\documentclass[12pt,preprint]{aastex}

\newcommand{\etal}{\mbox{et~al.}}

\shorttitle{VLA survey of the CDF-S}
\shortauthors{Mainieri et al.}

\begin{document}

%% LaTeX will automatically break titles if they run longer than
%% one line. However, you may use \\ to force a line break if
%% you desire.

\title{The VLA Survey of the Chandra Deep Field South. \\
    II. Identification and host galaxy properties of submillijansky sources.}

%% Use \author, \affil, and the \and command to format
%% author and affiliation information.
%% Note that \email has replaced the old \authoremail command
%% from AASTeX v4.0. You can use \email to mark an email address
%% anywhere in the paper, not just in the front matter.
%% As in the title, use \\ to force line breaks.

\author{
V. Mainieri\altaffilmark{1,2},
K.I. Kellermann\altaffilmark{3},
E.B. Fomalont\altaffilmark{3},
N. Miller\altaffilmark{3,4},
P. Padovani\altaffilmark{1},
P. Rosati\altaffilmark{1},
P. Shaver\altaffilmark{1},
J. Silverman\altaffilmark{2,5},
P. Tozzi\altaffilmark{6},
J. Bergeron\altaffilmark{7},
G. Hasinger\altaffilmark{2},
C. Norman\altaffilmark{4,8},
P. Popesso\altaffilmark{1,2}}
%% Notice that each of these authors has alternate affiliations, which
%% are identified by the \altaffilmark after each name.  Specify alternate
%% affiliation information with \altaffiltext, with one command per each
%% affiliation.

\altaffiltext{1}{ESO, Karl-Schwarschild-Strasse 2, D--85748 Garching, Germany}
\altaffiltext{2}{Max Planck Institut f\"ur extraterrestrische Physik,
  Giessenbachstrasse 1, D--85748 Garching, Germany}
\altaffiltext{3}{National Radio Astronomy Observatory, 520 Edgemont
  Road, Charlottesville, VA 22903-2475, U.S.A.}
\altaffiltext{4}{Department of Physics and Astronomy, Johns Hopkins
  University, Baltimore, MD 21210, U.S.A.}
\altaffiltext{5}{Institute of Astronomy, Department of Physics, Eidgen\"ossische Technische Hochschule, ETH Zurich, CH‐8093, Switzerland}
\altaffiltext{6}{INAF Osservatorio Astronomico di Trieste, via G.B.
  Tiepolo 11, I-34131, Trieste, Italy}
\altaffiltext{7}{Institut d'Astrophysique de Paris, 98bis Boulevard, 
	F-75014 Paris, France}  
\altaffiltext{8}{Space Telescope Science Institute, 3700 S. Martin
  Drive, Baltimore, MD 21210, U.S.A.}

%% Mark off your abstract in the ``abstract'' environment. In the manuscript
%% style, abstract will output a Received/Accepted line after the
%% title and affiliation information. No date will appear since the author
%% does not have this information. The dates will be filled in by the
%% editorial office after submission.

\begin{abstract}
  We present the optical and infrared identifications of the 266 radio
  sources detected at 20 cm with the Very Large Array in the Chandra
  Deep Field South \citep{kellermann07}. Using deep i-band Advanced
  Camera for Surveys, R-band Wide Field Imager, K-band SOFI/NTT,
  K-band ISAAC/VLT and Spitzer imaging data, we are able to find
  reliable counterparts for 254 ($\sim95\%$) VLA sources. Twelve radio
  sources remain unidentified and three of them are ``empty
  fields''. Using literature and our own data we are able to assign
  redshifts to 186 ($\sim70\%$) radio sources: 108 are spectroscopic
  redshifts and 78 reliable photometric redshifts. Based on the rest
  frame colors and morphological distributions of the host galaxies we
  find evidences for a change in the submillijansky radio source
  population: a) above $\approx 0.08$ mJy early-type galaxies are
  dominating; b) at flux densities below $\approx 0.08$ mJy, starburst
  galaxies become dominant.
\end{abstract}

%% Keywords should appear after the \end{abstract} command. The uncommented
%% example has been keyed in ApJ style. See the instructions to authors
%% for the journal to which you are submitting your paper to determine
%% what keyword punctuation is appropriate.


%% From the front matter, we move on to the body of the paper.
%% In the first two sections, notice the use of the natbib \citep
%% and \citet commands to identify citations.  The citations are
%% tied to the reference list via symbolic KEYs. The KEY corresponds
%% to the KEY in the \bibitem in the reference list below. We have
%% chosen the first three characters of the first author's name plus
%% the last two numeral of the year of publication as our KEY for
%% each reference.

%% Authors who wish to have the most important objects in their paper
%% linked in the electronic edition to a data center may do so by tagging
%% their objects with \objectname{} or \object{}.  Each macro takes the
%% object name as its required argument. The optional, square-bracket 
%% argument should be used in cases where the data center identification
%% differs from what is to be printed in the paper.  The text appearing 
%% in curly braces is what will appear in print in the published paper. 
%% If the object name is recognized by the data centers, it will be linked
%% in the electronic edition to the object data available at the data centers  
%%
%% Note that for sources with brackets in their names, e.g. [WEG2004] 14h-090,
%% the brackets must be escaped with backslashes when used in the first
%% square-bracket argument, for instance, \object[\[WEG2004\] 14h-090]{90}).
%%  Otherwise, LaTeX will issue an error. 

\section{Introduction}

Deep radio surveys have shown that there is an upturn in the
differential radio source counts below $\sim 1 mJy$ that cannot be
reproduced by evolutionary models of the millijansky radio populations
(e.g \citealt{condon84}; \citealt{windhorst85}; \citealt{ciliegi99};
\citealt{gruppioni99}; \citealt{richards00}; \citealt{seymour04}; etc.). 
Dedicated follow-up spectroscopic programs of many of these deep radio
surveys (e.g. \citealt{thuan87}; \citealt{prandoni01};
\citealt{afonso05}) have shown that the submillijansky radio sources are 
mainly identified with faint blue galaxies with disturbed morphologies
indicative of interactions and merging activity. However, due to the
faint magnitudes of a large fraction of these submillijansky sources,
all these works are based on a small percentages of
identifications. Interestingly, \cite{gruppioni99} has shown that even
at submillijansky level a large fraction of the radio sources are
identified with early-type galaxies once the spectroscopic follow-up
reaches fainter magnitude limit.

The aim of this paper is to identify the radio sources in one of the
best studied region of the sky, the Chandra Deep Field South (CDF-S)
and to study the properties of their host galaxies down to faint radio
flux densities. The CDF-S area has been imaged with the Very Large
Array (VLA) down to 8.5 $\mu Jy$ rms near the field center and
3.5\arcsec resolution at both 1.4 and 5 GHz.  The radio catalogue
include 266 sources. Details on the reduction process and the overall
VLA data are reported in \cite{kellermann07}, Paper I. The CDF-S
region is the area of the sky with the most extensive multi-wavelength
observations available, covering an impressively wide range of
wavelength. All of NASA's existing Great Observatories (HST, Chandra
and Spitzer), ESA's XMM-Newton and the Very Large Telescope at ESO
have devoted hundreds of hours each to obtain state-of-the-art
sensitivities in this field. The large amount of ancillary data made
the VLA survey in the CDF-S extremely valuable to understand better
the properties of the submillijansky radio population. At the same
time the radio data can help resolving uncertainties on the
identification or the nature of sources selected at other
wavelengths.\\ The paper is structured as follows: in Section 2 we
provide a general description of the ancillary data used in the
identification process; in Section 3 the identification methodology is
described; we present in Section 4 the identification catalogue; in
Section 5 we report on some particular sources with problematic
identification; Section 6 describes the redshift distribution of the
VLA sources; Section 7 is devoted to the discussion of the properties
of the host galaxies and finally we summarize our conclusions in
Section 8.\\ We use in this paper magnitudes in the AB system, if not
otherwise stated, and we assume a cosmology with H$_0 = 70$ km
s$^{-1}$ Mpc$^{-1}$, $\Omega _{\rm M} = 0.3$ and $\Omega _\Lambda =
0.7$.

\section{Optical and NIR data}
\label{data} 

\begin{figure}
   \centering \includegraphics[width=15cm]{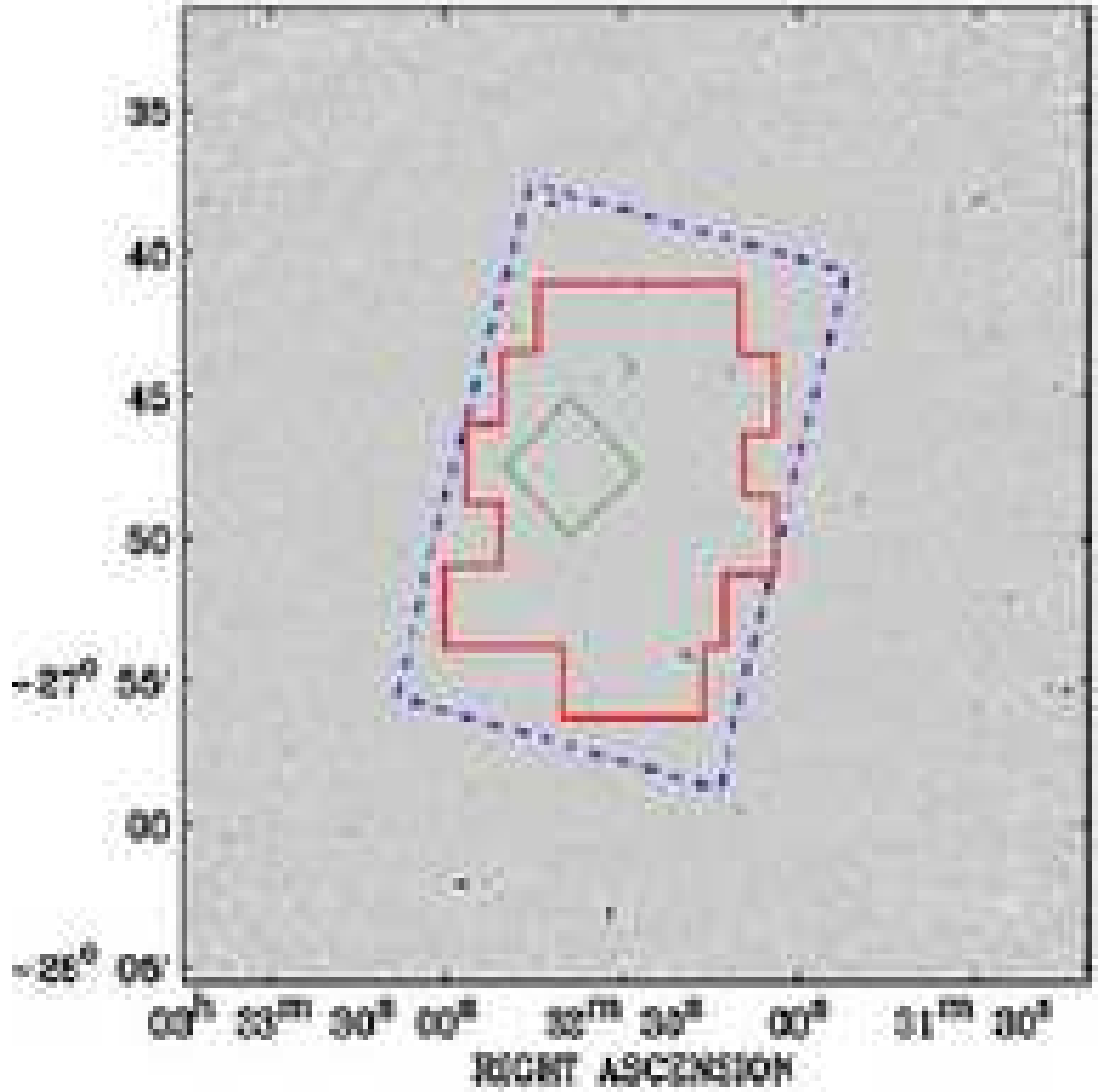}
   \caption{ Multiwavelength coverage of the VLA/CDF-S Survey. The
   background image is the 20 cm VLA observation. The area covered by
   ACS as part of the GOODS program is highlighted with a dashed line;
   the continuous line shows the overlay of the ISAAC/VLT deep Ks band
   imaging and finally the location of the UDF is indicated by the
   small square in the center.}
\label{multiwl}
\end{figure}

While we refer the reader to \cite{kellermann07} for an overview of
the multi-wavelength coverage of the CDF-S area, in this section we
describe the datasets used for the identification of the radio
sources.\\ In the optical band, we used data obtained respectively
with the Advanced Camera for Surveys (ACS) on board of HST and the
Wide Field Imager (WFI) installed at the 2.2m telescope in La Silla.
A central area of 160 arcmin$^2$ (dashed line in Fig. \ref{multiwl})
has been covered by ACS in four different filters: F435W (b), F606W
(v), F775W (i) and F850LP (z). These data have been taken as part of
'The Great Observatories Origins Deep Survey'
\citep{dickinson03,giavalisco04} down to the following limiting
magnitudes ($5\sigma$ in a 1 sq. arcsec area): 26.7, 26.7, 25.9,
25.6. For our identification purposes we used the F775W (i) band
catalogue, which is publicly available on the GOODS
web-site\footnote{http://www.stsci.edu/science/goods/}. For the outer
region of our VLA survey not overlapping with the GOODS-ACS area, we
used the WFI R (652 nm) band images from the ESO Imaging Survey
(EIS)\footnote{http://www.eso.org/science/goods/imaging/products.html}. These
data cover the whole VLA survey down to a limiting magnitude of 25.5
(AB, $5\sigma$ in 2 arcsec aperture).\\ In the NIR, we used the
ISAAC/VLT K$_{\rm S}$ (2162 nm) data from the GOODS survey on a
central area of $\sim 131$ arcmin$^2$ (continuous line in
Fig. \ref{multiwl}). The mean limiting magnitude of this observation
is 24.7 (AB, $5\sigma$ in 1 arcsec apertures; Retzlaff et al. in
preparation)\footnote{http://archive.eso.org/archive/adp/GOODS/ISAAC$\_$imaging$\_$v2.0/index.html}. Finally,
we have used shallower imaging in the K$_{\rm S}$ band obtained with
SOFI/NTT (21.4, AB, $5\sigma$ in 2 arcsec aperture; \citealt{olsen06})
which covers the whole VLA survey.

\begin{figure}
   \centering
   \includegraphics[width=8cm]{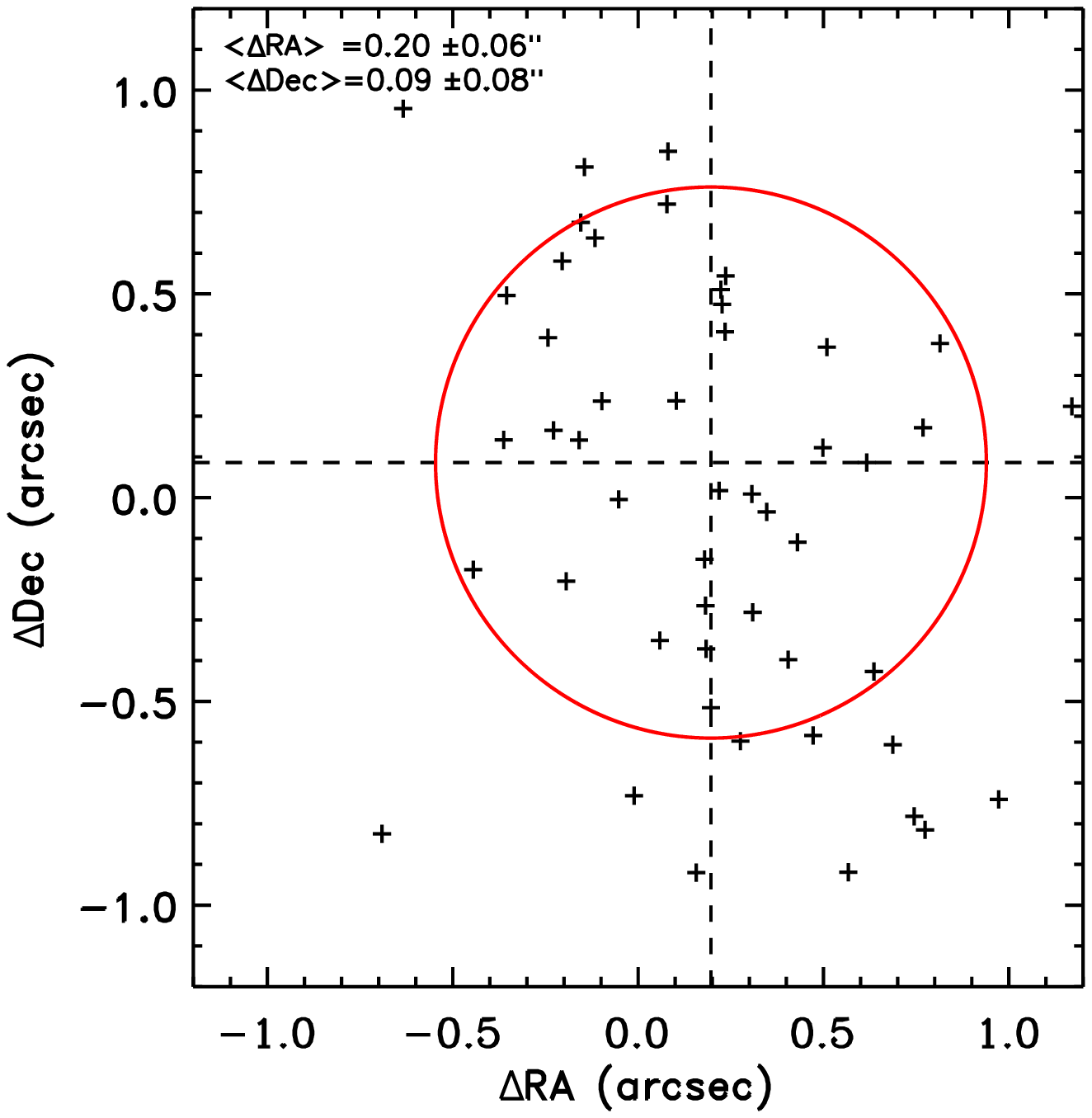}
   \includegraphics[width=8cm]{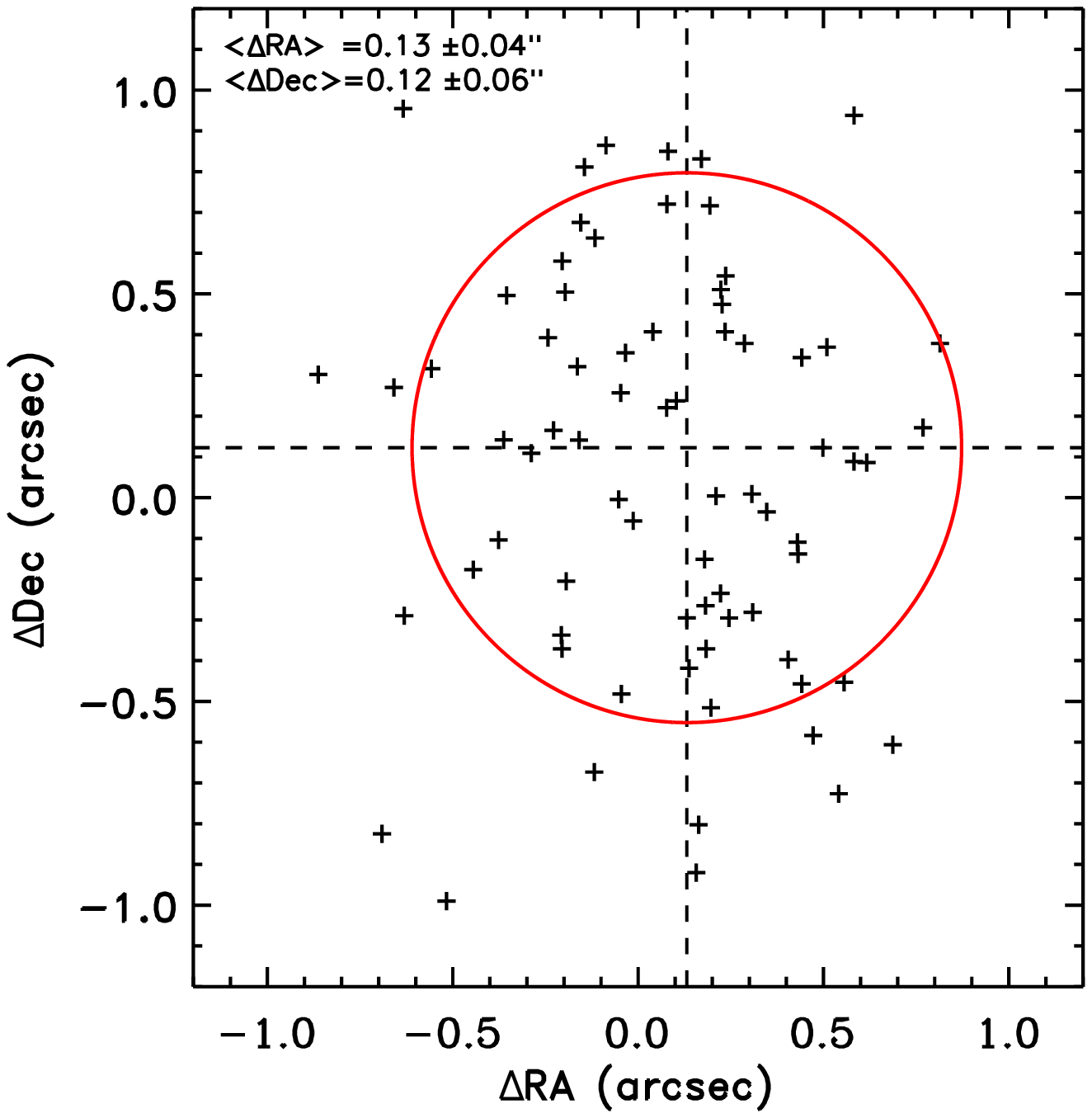}
   \caption{Offsets between the radio and the WFI-R (left
     panel) and SOFI-Ks (right panel) coordinates. We used only radio
     point-like sources with S/N$>5$ associated with a point-like counterpart. The circle is centered on the average offset found and its radius is equal to the average uncertainty in the radio coordinates.}
\label{dra_ddec}
\end{figure}

The offsets between the radio catalogue and each one of the
optical/NIR catalogs has been computed using a sample of
point-like radio sources with $S/N>5$ associated to a point-like
counterpart. In Fig. \ref{dra_ddec} we show the $\Delta {\rm
RA=RA(radio)-RA(optical/NIR}$) and $\Delta {\rm
Dec=Dec(radio)-Dec(optical/NIR}$) in the case of the WFI-R
catalogue (left panel) and SOFI-Ks catalogue (right panel). We found
mean offsets of $< \Delta {\rm RA} > = 0.20\pm0.06$ arcsec and $<
\Delta {\rm Dec} > = 0.09\pm0.08 $ arcsec for WFI-R, and $< \Delta
       {\rm RA} > = 0.13\pm0.04$ arcsec and $< \Delta {\rm Dec} > =
       0.12\pm0.06 $ arcsec for SOFI-Ks. Similar offsets have been
       computed for the other catalogs (see Table
       \ref{tab_offset}). Before proceeding in the identification
       process, we correct for these mean offsets each one of the
       optical/NIR catalogs, in order to have radio and optical/NIR
       positions in the same reference frame.

\section{Identification methodology}
\label{identification}

\begin{figure}
   \centering
   \includegraphics[width=10cm]{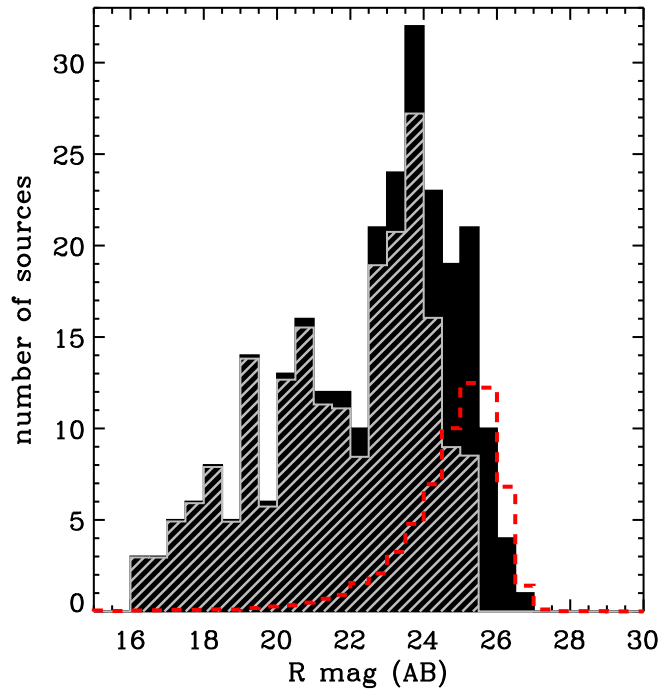}
   \caption{ The different component of Eq. (1) for the WFI-R
     catalogue. Magnitude distribution of possible counterparts found
     in circles of 2 arcsec radius around the radio sources (black
     filled histogram). The dashed histogram is the magnitude
     distribution of background objects that we expect to have in the
     area where the search for counterparts is performed (equal to
     the surface density as a function of magnitude of background
     objects, {\it n(m)}, times the area of a circle of 2 arcsec radius
     times the number of radio sources).  Finally, the magnitude
     distribution of optical counterparts once the background objects
     have been subtracted ( grey hatched histogram). }
\label{prob}
\end{figure}

Due to the faintness of both radio and optical samples, we could have
more than one candidate counterpart for a single radio source.
Therefore we used the likelihood ratio technique which provides a
reliability measure of an identification as a function of the distance
of the counterpart and of its magnitude.  This technique has been
first used in this context by \cite{richter75} and since by
\cite{deruiter77}, \cite{prestage83}, \cite{windhorst84},
\cite{wolstencroft86}, \cite{sutherland92}, \cite{ciliegi03},
\cite{ciliegi05}, \cite{brusa05} and \cite{brusa07}.\\ The likelihood
ratio (LR) is the probability that a given source with offset {\it r}
and measured magnitude {\it m} is the true counterpart, relative to
the probability that the same object is a chance background source:
\begin{equation}
LR = \frac{q(m)f(r)}{n(m)}
\end{equation}

\noindent where {\it q(m)} is the expected probability distribution as a
function of magnitude of the true counterparts, {\it f(r)} is the
probability distribution of the positional errors, and {\it n(m)} is
the surface density as a function of magnitude of background objects.
We refer the reader to \cite{ciliegi03} for a detailed discussion on
the procedure to calculate {\it q(m)}, {\it f(r)} and {\it n(m)}.  In
particular, to derive an estimate of {\it q(m)} we adopt a radius of 2
arcsec to maximize the over-density due to the presence of the
optical/NIR counterparts (we expect that the majority of the true
optical/NIR counterparts are inside a 2 arcsec circle around the radio
position). Inside {\it f(r)} both the radio and optical/NIR positional
uncertainties are included. For the optical/NIR positional error we
assume 0.05 arcsec while for the radio uncertainties we use the values
reported in Table 1 of
\cite{kellermann07}.\\ In Fig. \ref{prob} we show the distribution of
the different components of Eq. (1) for the WFI-R catalogue,
specifically the magnitude distribution of background objects that we
expect to have in the area where the search for counterparts is
performed (dashed histogram). This distribution is obtained
multiplying the surface density as a function of magnitude of
background objects, {\it n(m)}, times the area of a circle of 2 arcsec
radius times the number of radio sources. The black filled histogram
is the magnitude distribution of possible counterparts found in
circles of 2 arcsec radius around the radio sources and the grey
hatched histogram represents the expected magnitude distribution for
the true optical counterparts once the background objects have been
removed. We choose a threshold value for the likelihood ratio of
LR$_{\rm th}=0.2$ and we will consider objects below this value as
spurious counterparts. We find that this value is a good compromise
between having a low enough threshold to avoid missing many real
identifications and large enough to have few spurious ones. \\ Apart
from a LR value for each possible counterparts, we would like to have
a measure of the reliability of a particular source to be the true
counterpart. \cite{sutherland92} first pointed out that the presence
or absence of other candidate counterparts for a single radio source
provides {\it additional information} to that contained in LR and they
developed a self-consistent formula for the reliability {\it R$_j$} of
an object {\it j} to be the true counterpart:
\begin{equation}
R_j = \frac{LR_j}{\sum_i LR_i + (1-Q)}
\end{equation}

\noindent where {\it i} runs over the set of all candidate counterparts 
for that particular radio source, and Q$=\int_{-
\infty}^{m_{lim}}q(m)~dm$ is the probability that the counterpart of
the source is above the magnitude limit of the optical/NIR
catalogue. We choose Q$=0.8$ corresponding to the ratio of the
expected number of identification, the integral of {\it q(m)}, and the
total number of radio sources. We note that for values of Q in the
range 0.5-1.0 there is no significant difference in the result of the
identification process.\\ To summarize, we adopt the following
criterium to choose the most likely counterpart for a radio source: a)
if only one candidate counterpart with LR$>{\rm LR_{\rm th}}$ is
available, this will be chosen; b) if there are several candidate
counterparts with LR$>{\rm LR_{\rm th}}$, we will choose the one with
the highest reliability ({\it R$_j$}) value. We will call ``empty
field'' a radio source with no candidate counterpart within a distance
of 10 arcsec from the radio position.

\section{Identifications}
\label{counterparts}

\begin{figure}
   \centering
   \includegraphics[width=8cm]{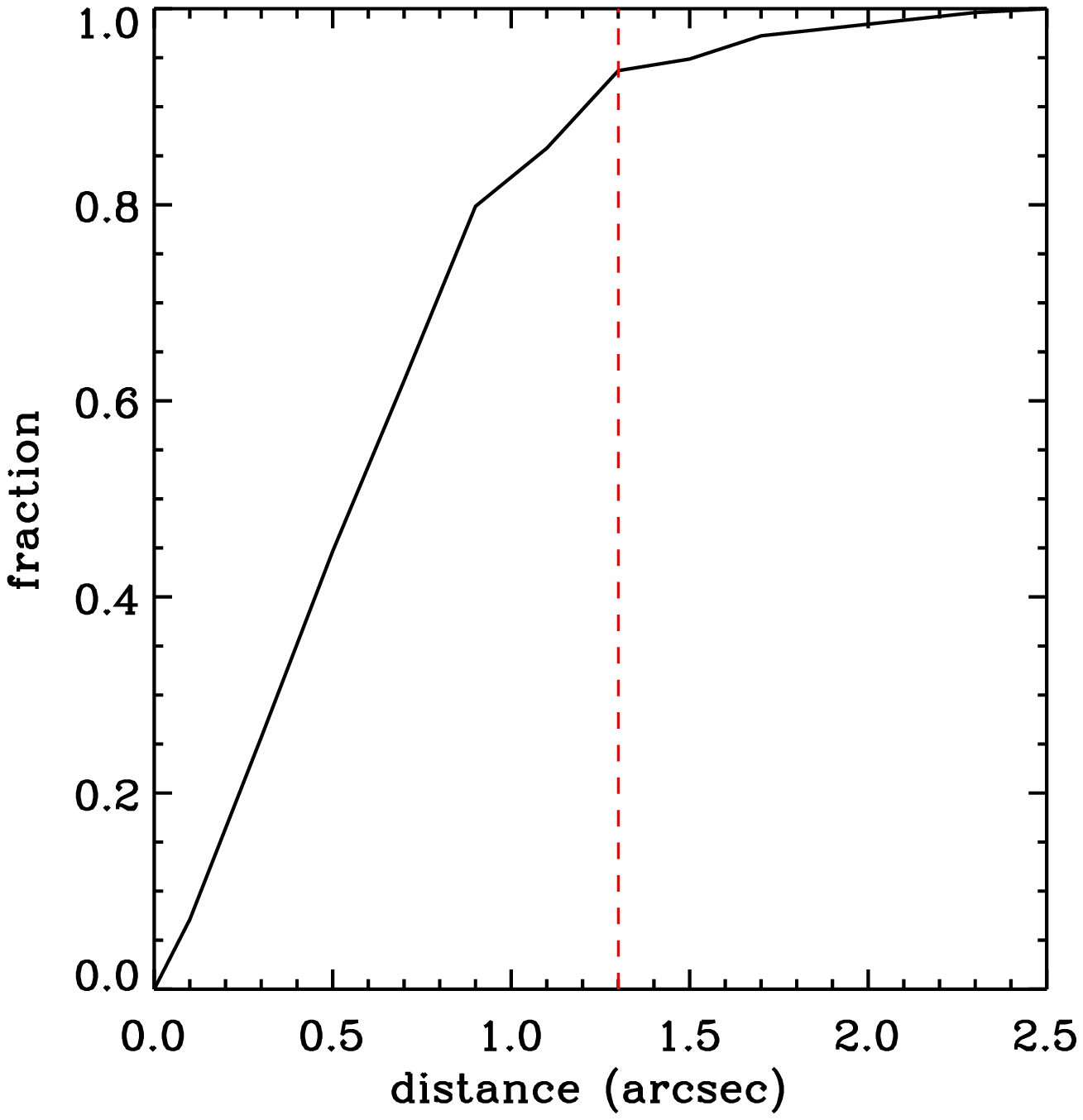}
   \includegraphics[width=8cm]{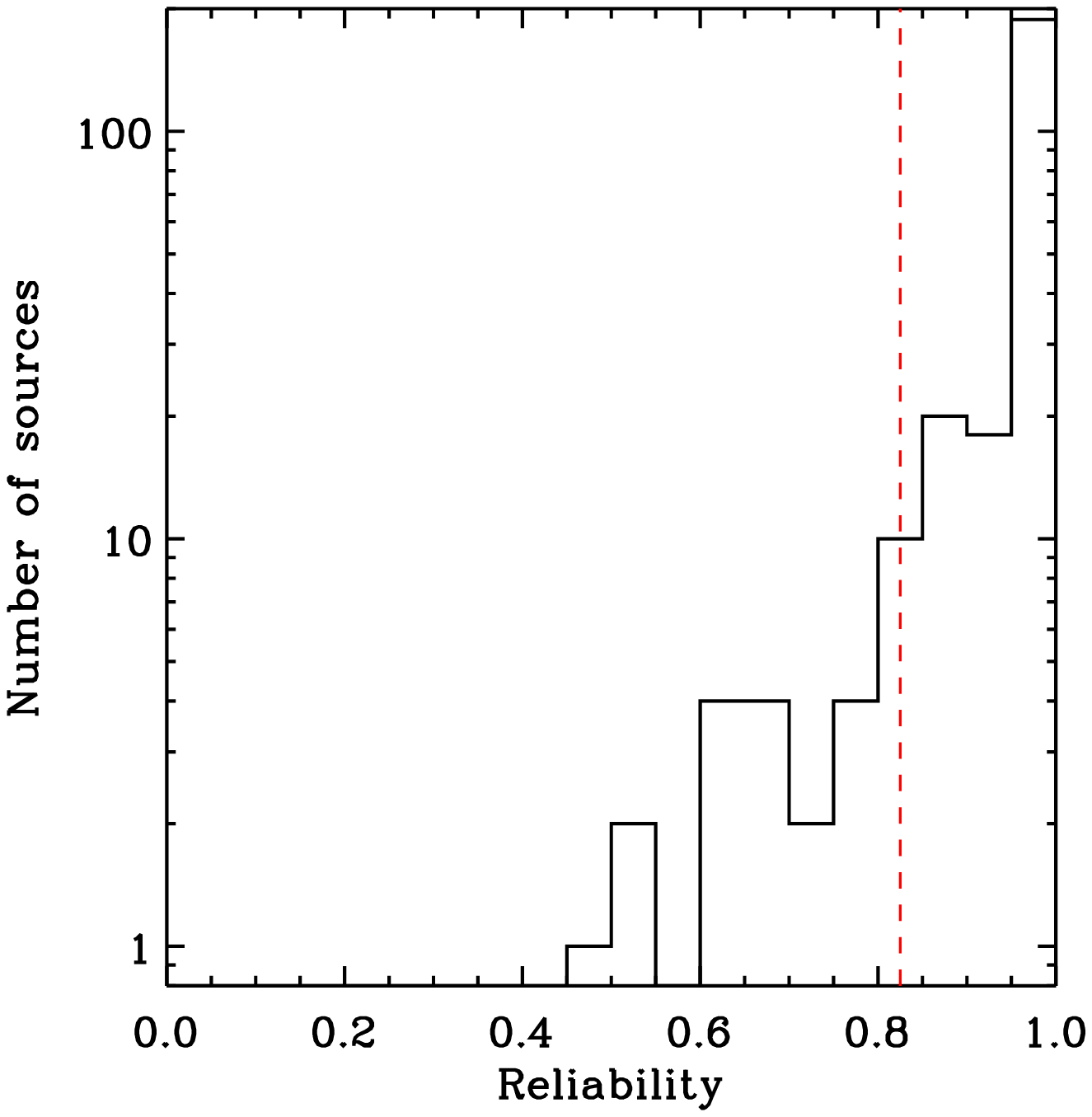}
   \caption{{\it Left:} the cumulative distribution of the radio to 
     optical/NIR separations. The dashed line indicates a separation
     of $1.3\arcsec$ within which $90\%$ of the counterparts are
     located. {\it Right:} the distribution of the reliability parameter,
     R, for the radio counterparts. The dashed line at R$=0.83$ marks
     the reliability value above which $90\%$ of the counterparts are
     located.}
\label{comulative}
\end{figure}

We use four different catalogs (ACS-i, WFI-R, ISAAC-Ks and SOFI-Ks)
and the procedure outlined in the previous section to identify the
counterparts of the VLA sources in the CDF-S area. In this section, we
first report on the identification success rate for each one of these
datasets and finally summarize the overall results of the
identification process.\\ We find reliable counterparts for 67 ($\sim
85\%$) of the 79 radio sources inside the area covered by
ACS/GOODS. The expected number of true counterparts, given by the sum
of the reliability (R) values for all sources with LR$>$LR$_{\rm th}$,
is $\sim 66.7$, therefore we expect that less than one of the optical
counterparts found from the ACS/GOODS catalogue could be spurious. We
are not able to assign an optical counterpart to twelve radio sources
of which four have a possible counterpart with LR$<$LR$_{\rm th}$
while eight appear like ``empty-field'' in the ACS images. The WFI-R
catalogue covers the all VLA image and we find reliable counterpart
for 230 ($\sim 86 \%$) of the 266 radio sources. Of these we expect a
maximum of three spurious identifications. The number of unidentified
sources is 36, twelve of which are ``empty-field''. From the NIR
catalogs, we are able to identify 56 ($92\%$) of the 61 radio sources
with ISAAC-Ks imaging and we expect that less than one of these should
be spurious. The shallower SOFI-Ks catalogue cover the all VLA area
and it provides reliable counterparts for 190 ($71\%$) of the 266
radio sources of which likely no more than six could be spurious. The
surface density of NIR sources is lower compared to that of optical
objects therefore we expect less spurious identifications due to
background sources: one for ISAAC and two for SOFI. Using the NIR
catalogs we are able to assign reliable counterpart to 18 radio
sources that were not identified in the optical.\\ Finally, to
maximize the number of radio sources identified, we consider two
additional datasets. The GEMS (Galaxy Evolution from Morphologies and
SEDs) program \cite{rix04} has imaged in two filters, F606W(v) and
F850LP(z), with ACS the whole area covered by our radio survey. Using
this data we are able to find reliable counterparts for five radio
sources previously unidentified (ID$= 19, 46, 78, 176, 213$). We also
use the Spitzer (3.6, 4.5, 5.8, 8.5, 24 micron) images obtained as
part of GOODS and the Spitzer/MIPS (24 and 70 micron) observations
performed as part of FIDEL\footnote{http://www.noao.edu/noao/fidel/}
(PI Mark Dickinson): from these mid-IR images we find reliable
counterparts for five additional radio sources (ID $= 70, 87, 100,
105, 216$).\\ Summarizing, from the optical, NIR catalogs, and
including the additional identifications from GEMS and Spitzer data,
we find reliable counterparts for 254 ($\sim 95\%$) of the 266 radio
sources.  The number of spurious identifications due to background
objects is expected to be lower than eight. Twelve radio sources
remain unidentified: three of them are '{\it empty fields}' while the
remaining nine have possible counterparts but with LR$<$LR$_{th}$. We
summarize the output of the likelihood ratio process in Table
\ref{tab_like}.  The left panel of Fig.
\ref{comulative} shows the cumulative distribution for the radio to
optical/NIR distances: $90\%$ of the counterparts are found inside a
circle of $1.3\arcsec$ radius and the maximum distance of a
counterparts is $2.5\arcsec$. We note that there are 20 radio sources
with an optical counterpart at more that $1.3\arcsec$ distance, eleven
of these are with extended radio images and this can explain the large
distance between the radio and the optical/NIR position. The other
nine counterparts of radio sources that are unresolved, generally have
low reliability values.  In the right panel of Fig.
\ref{comulative} we show the distribution of the reliability parameter
(R) for the 254 counterparts: $90\%$ of them have R$>0.83$.\\

We present in Table \ref{opt_catalogue} the catalogue\footnote{A
machine readable format of Table \ref{opt_catalogue} is available in
the electronic version of the paper.} for the optical/NIR
identification of the radio sources in the VLA/CDF-S area. The
catalogue is organized as follows:

\begin{itemize}

\item{Column 1~~~~ The radio source number used in this paper
    (see also \citealt{kellermann07}).}
%\item{Column 2~~~~    The GOODS identification number if available}
\item{Column 2-3~ The Right Ascension and Declination (J2000) of the radio source  (see also \citealt{kellermann07}).}
\item{Column 4-5~ The Right Ascension and Declination (J2000) of the
    primary counterpart. The positional error is 0.05 arcsec.}
\item{Column 6~~~~ The separation, in arcsec, between the radio source
    and its counterpart}
\item{Column 7~~~~    The likelihood ratio (LR) as defined in Sec. \ref{identification}}
\item{Column 8~~~~    The reliability parameter (R) as defined in Sec. \ref{identification}}
\item{Column 9~~~~ The catalogue from which the counterpart was selected: {\it
      ACS-i} (i band catalogue from ACS/GOODS), {\it WFI-R} (R band
      catalogue from WFI), {\it ISAAC-K} (K band catalogue from
      ISAAC), {\it SOFI-K} (K band catalogue from SOFI), {\it GEMS} (z
      band catalogue from ACS/GEMS), {\it SPITZER} (IRAC and MIPS
      images)}
\item{Column 10~~~   The R band magnitude (AB)}
\item{Column 11~~~   The K band magnitude (AB)}
\item{Column 12~~~   The spectroscopic redshift of the counterpart}
\item{Column 13~~~ Quality flag for the spectroscopic redshifts: '2'
    secure redshift (multiple spectral features), '1' tentative
    redshift (e.g. based on a single emission line)}
\item{Column 14~~~ The photometric redshift of the counterpart}
\end{itemize}

For some radio source we find more than one possible counterpart with
LR$>{\rm LR_{\rm th}}$. We call primary the counterpart with the
highest reliability value ( Table \ref{opt_catalogue}) and secondary
the remaining ones. Table \ref{opt_catalogue_sec} contains the
properties of these secondary counterparts.\\

\begin{figure}
\centering
\includegraphics[width=4cm]{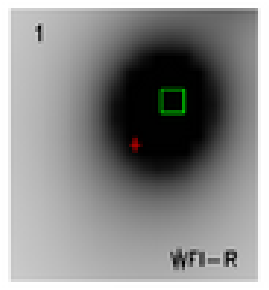}
\includegraphics[width=4cm]{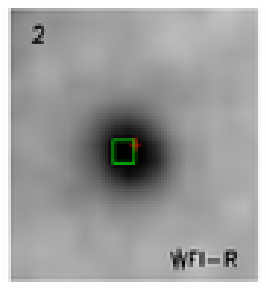}
\includegraphics[width=4cm]{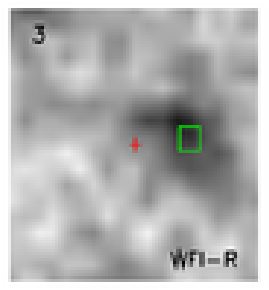}
\includegraphics[width=4cm]{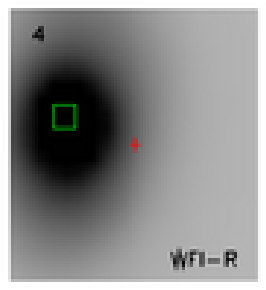}\\
\includegraphics[width=4cm]{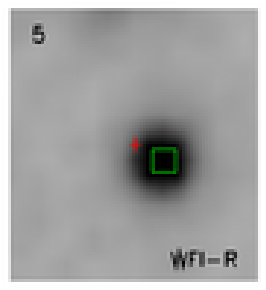}
\includegraphics[width=4cm]{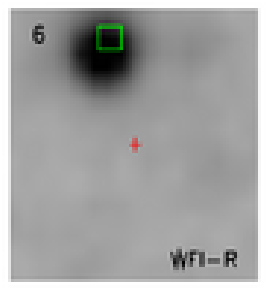}
\includegraphics[width=4cm]{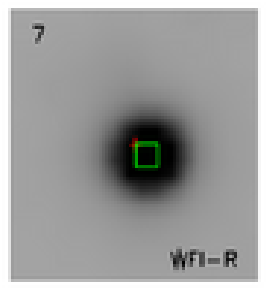}
\includegraphics[width=4cm]{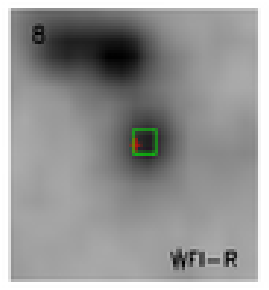}\\
\includegraphics[width=4cm]{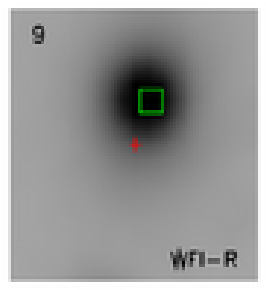}
\includegraphics[width=4cm]{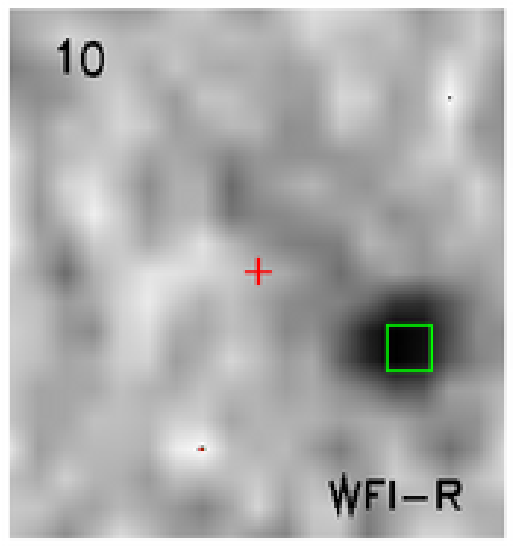}
\includegraphics[width=4cm]{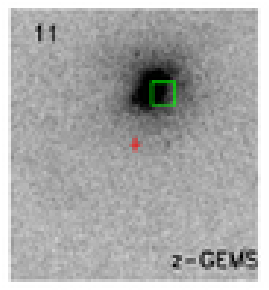}
\includegraphics[width=4cm]{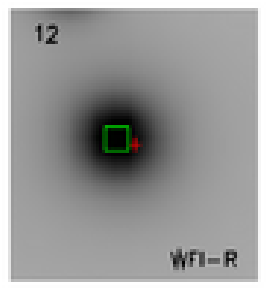}\\
\includegraphics[width=4cm]{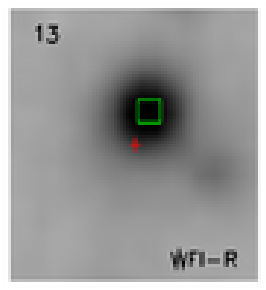}
\includegraphics[width=4cm]{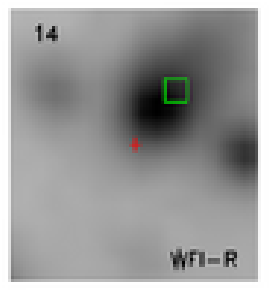}
\includegraphics[width=4cm]{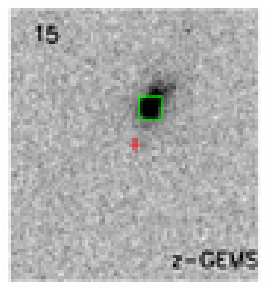}
\includegraphics[width=4cm]{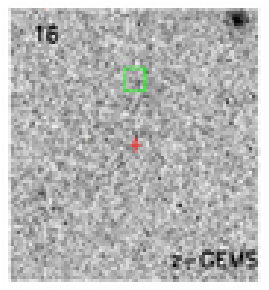}\\
\includegraphics[width=4cm]{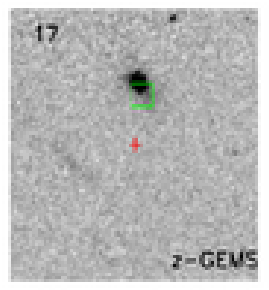}
\includegraphics[width=4cm]{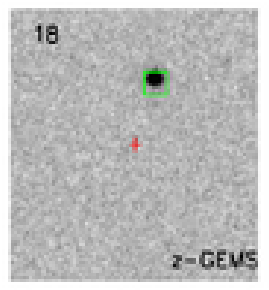}
\includegraphics[width=4cm]{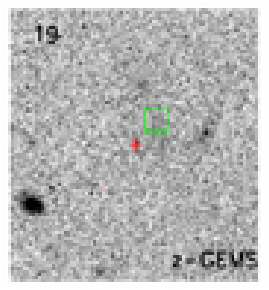}
\includegraphics[width=4cm]{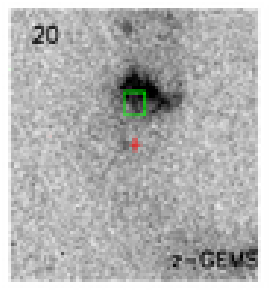}\\
\caption{Cutouts of the radio sources either from ACS (GOODS-i, GEMS-z) or WFI-R. Each one is 5 arcsec on a side. The red cross indicates the radio position, the green square the primary counterpart while the blue triangles are possible secondary counterparts. The complete series of cutouts is available in the electronic version of the paper.}
\label{radio_fc}
\end{figure}

\begin{figure}
\centering
\includegraphics[width=8cm]{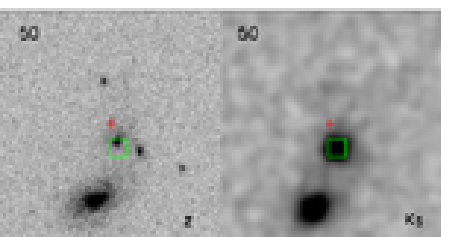}
\includegraphics[width=8cm]{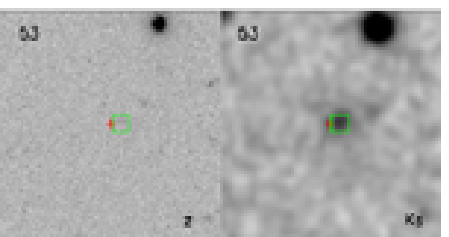}\\
\includegraphics[width=8cm]{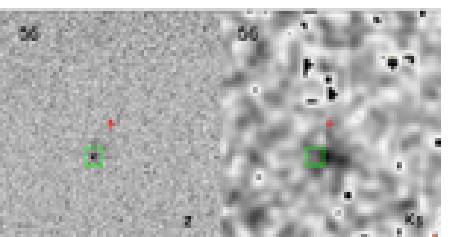}
\includegraphics[width=8cm]{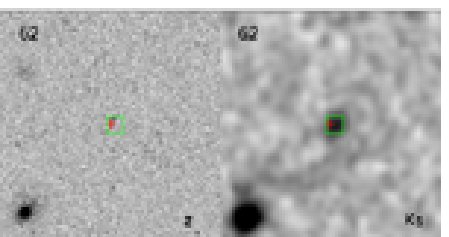}\\
\includegraphics[width=8cm]{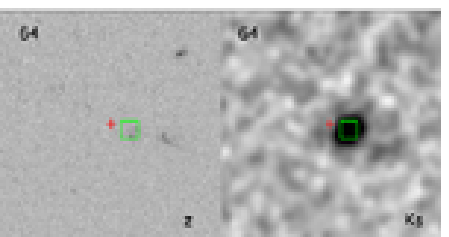}
\includegraphics[width=8cm]{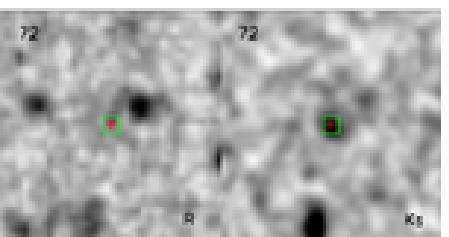}\\
\includegraphics[width=8cm]{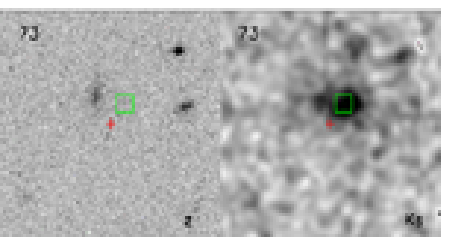}
\includegraphics[width=8cm]{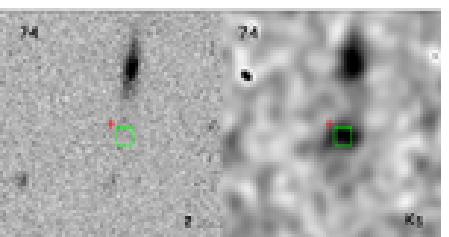}\\
\includegraphics[width=8cm]{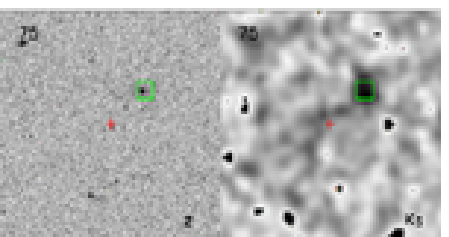}
\includegraphics[width=8cm]{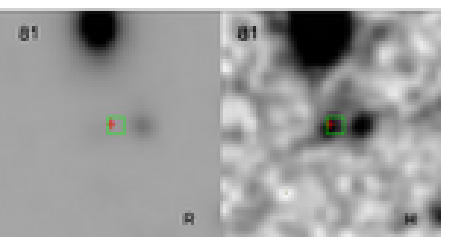}\\
\caption{Cutouts of the radio sources identified in the K band. For each of these sources, we show the optical (left) and K band(right) cutouts. Each one is 10 arcsec on a side. The red cross indicated the radio position, the green square the primary counterpart while the blue triangles are possible secondary counterparts.}
\label{radio_fc_IR}
\end{figure}

\begin{figure}
\centering
\includegraphics[width=8cm]{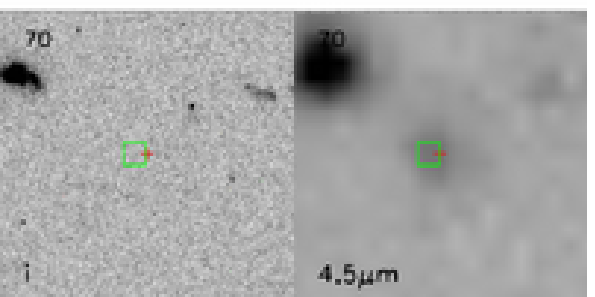}
\includegraphics[width=8cm]{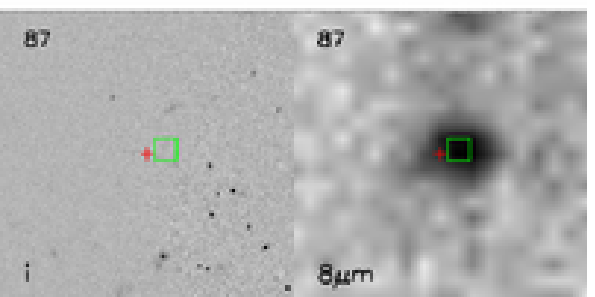}\\
\includegraphics[width=8cm]{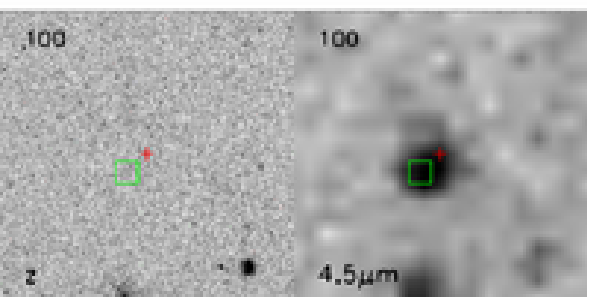}
\includegraphics[width=8cm]{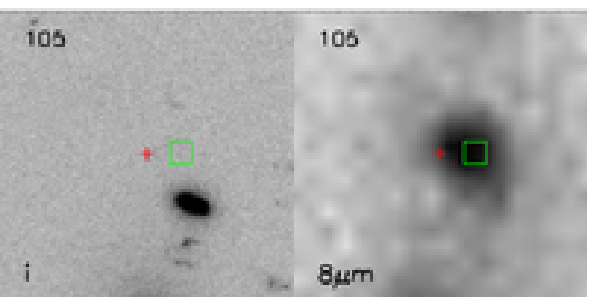}\\
\includegraphics[width=8cm]{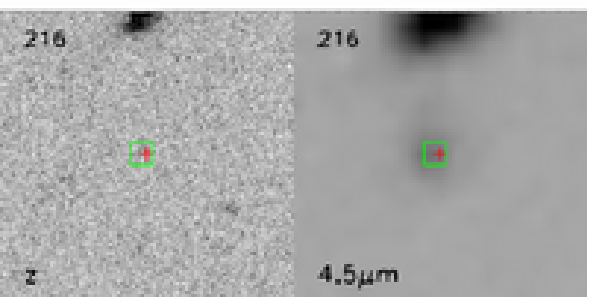}
\caption{Cutouts of the radio sources identified in the Spitzer
  bands. For each of these sources, we show the optical (left) and IR
  (right) cutouts. Each one is 10 arcsec on a side. The red cross
  indicated the radio position, the green square the primary
  counterparts while the blue triangles are possible secondary
  counterparts.}
\label{radio_fc_Spitzer}
\end{figure}

For each one of the 266 radio sources we have produced an optical
cutout (see Fig. \ref{radio_fc}) using the ACS-{\it i} GOODS images,
the ACS-{\it z} data from GEMS or finally, for those radio sources not
covered by ACS, the WFI-R mosaic. The default size of the cutouts is
$10 \times 10$ arcsec.  We indicate the radio position (red cross),
the primary counterpart (green square), and secondary counterpart
(blue circle).\\ As already mentioned, for 18 radio sources
unidentified in the optical, a reliable counterpart is found in the
NIR bands. Optical and K band cutouts for these objects are shown in
Fig. \ref{radio_fc_IR}.\\ Five radio sources are correctly identified
only from the Spitzer observations in the CDF-S region. We show in
Fig. \ref{radio_fc_Spitzer} their optical and Spitzer cutouts. The
details of these five sources are the following:

\noindent{\bf RID 70 - J033159.82-274540.3} The most likely counterpart
becomes visible only at 4.5 $\mu m$. This radio source is also in the ATCA
1.4GHz survey \citep{afonso06}, ATCDFS J033159.86-274541.3, and the
distance between the two radio coordinates is $\sim 1$ arcsec. This
source was unidentified in \cite{afonso06} since they limited the
search for counterparts to the optical bands.\\
{\bf RID 87 - J033209.90-275015.7} The counterpart is clearly visible in
all the four IRAC bands. \\
{\bf RID 100 - J033213.31-273934.1} The counterpart is visible only in the
4.5 $\mu m$ Spitzer image.\\
{\bf RID 105 - J033215.39-275037.5} The counterpart of this radio source
become visible only in the Spitzer bands. This object is also detected
with Chandra ( XID$=587$) and it was identified in \cite{giacconi02}
with a R$=21.0$ galaxy located at $\sim 1.8$ arcsec south-west of the
X-ray position. We believe that this X-ray source has the same
counterpart as the one chosen for the the radio source. The spectroscopic
redshift for XID$=587$ of z$=0.246$ reported by \cite{szokoly04}
belongs to the old counterpart, while for the newly proposed
counterpart we currently have only a photometric redshift estimate of
$1.80\pm0.08$ \citep{grazian06}.\\
{\bf RID 216 - J033303.30-275328.0} The most likely counterpart is visible
only in the 3.6 $\mu m$ Spitzer image. .

\begin{figure}
  \centering \includegraphics[width=10cm]{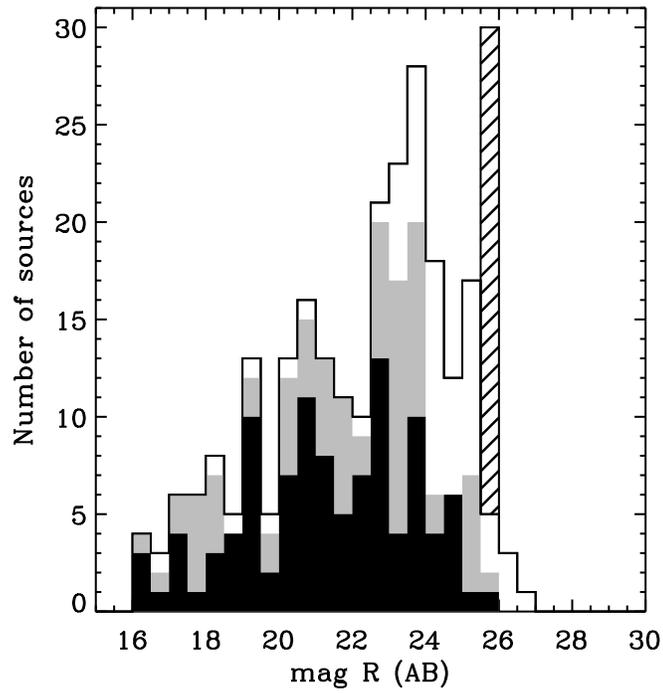} \caption{R
  magnitudes distribution for the primary counterparts (the total
  histogram). Black shading refers to sources with spectroscopic
  redshifts, the grey shading denotes objects with photometric
  redshifts and finally radio sources with no counterparts in the R
  band and for which we can only provide lower limit to their
  magnitudes are represented by the hatched box.  }
\label{Rmag}
\end{figure}

We show in Fig. \ref{Rmag} the R magnitude distribution for the
primary counterparts (the total histogram). We highlight sources with a
spectroscopic redshift (black shading) and the ones with a
photometric redshift (grey shading). It is clear from this plot how
the ability to obtain a redshift for the optical counterparts
decreases going to fainter magnitudes: $\sim 88\%$ of the objects with
R$_{\rm AB}<24$ have a redshift (spectroscopic or photometric) while
this fraction decreases to $\sim 30\%$ for sources with R$_{\rm
  AB}>24$.

\section{Notes on particular sources}
\label{particular}

\begin{figure}
\centering
\includegraphics[width=5cm]{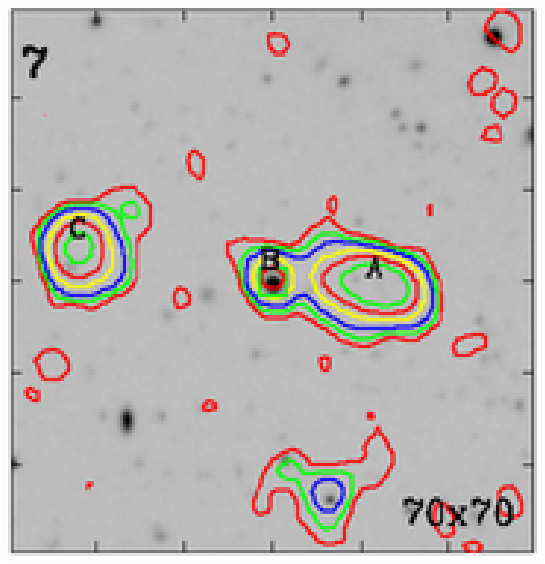}
\includegraphics[width=5cm]{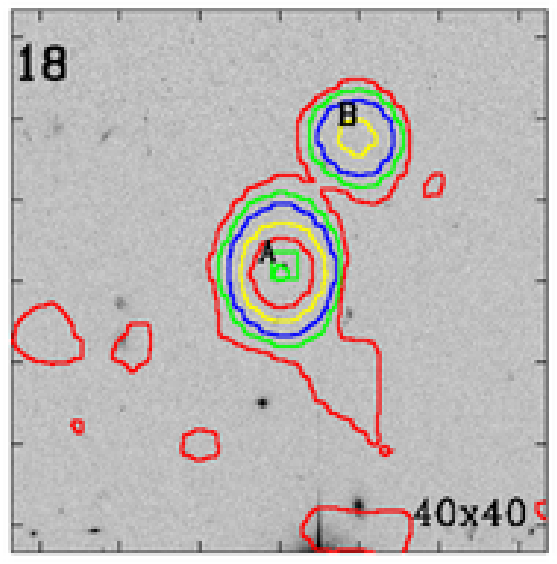}
\includegraphics[width=5cm]{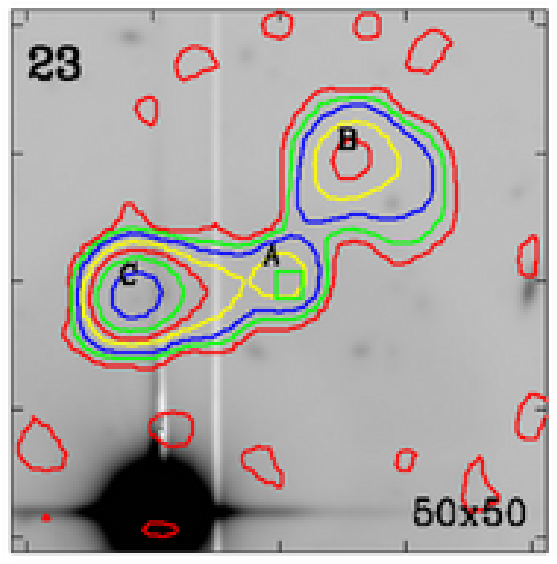}\\
\includegraphics[width=5cm]{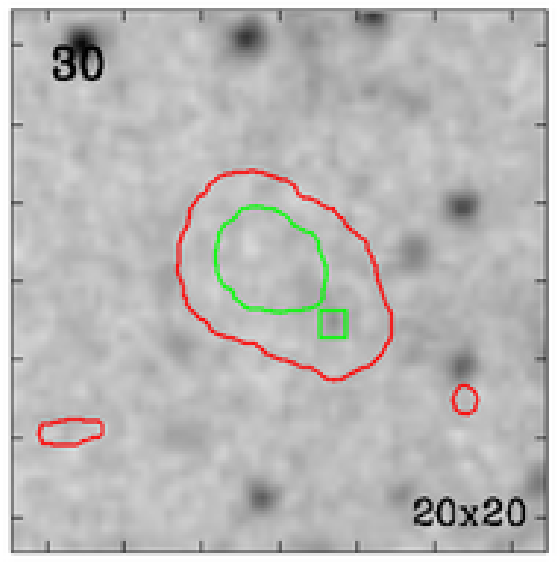}
\includegraphics[width=5cm]{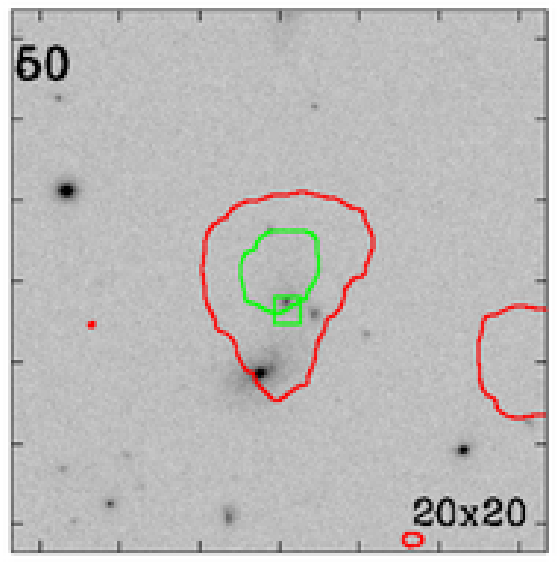}
\includegraphics[width=5cm]{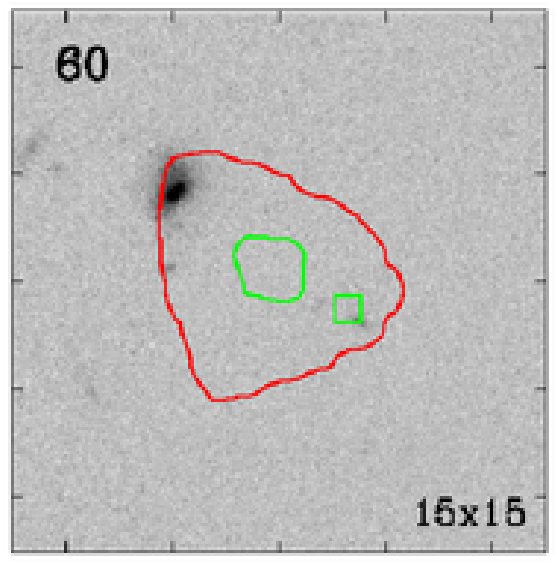}\\
\includegraphics[width=5cm]{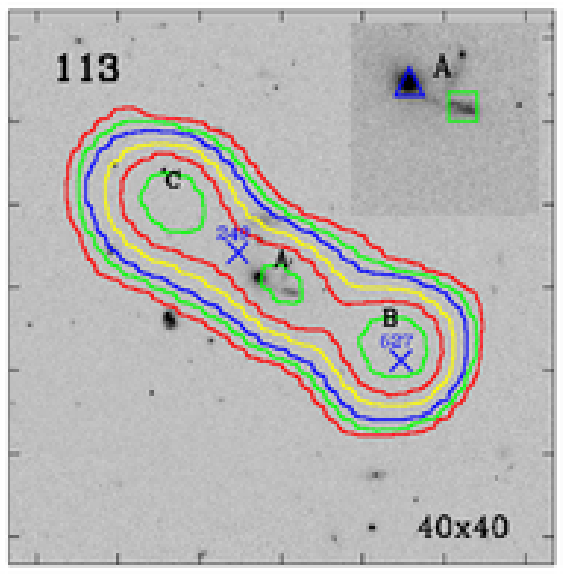}
\includegraphics[width=5cm]{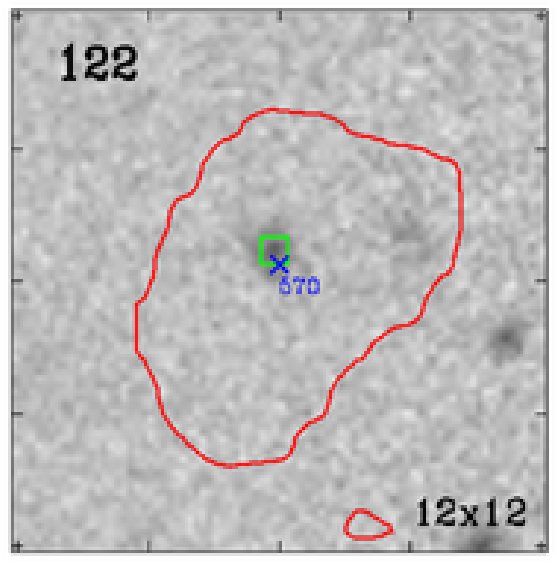}
\includegraphics[width=5cm]{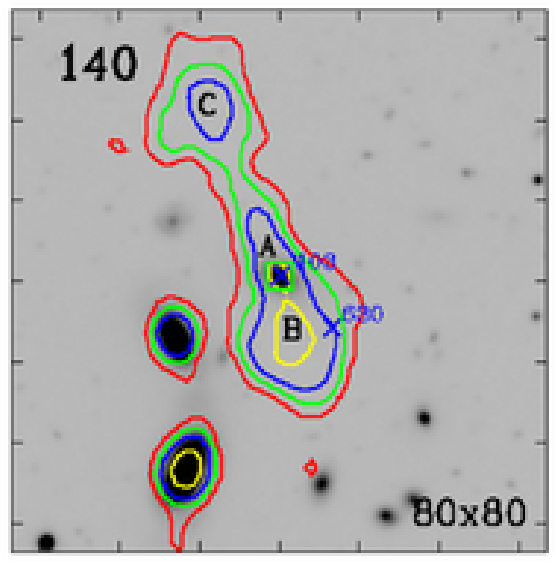}\\
\includegraphics[width=5cm]{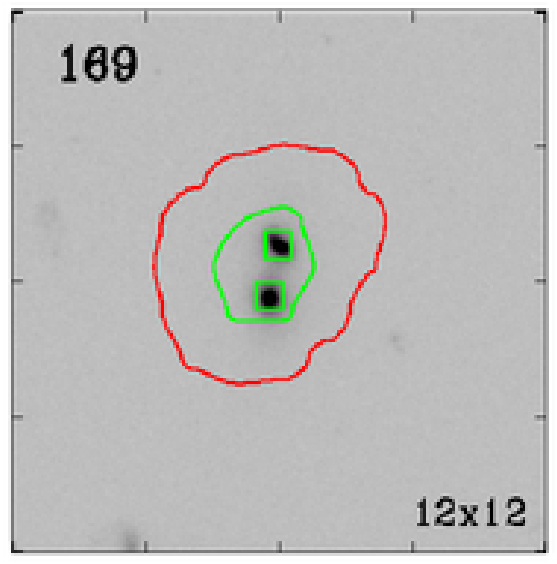}
\includegraphics[width=5cm]{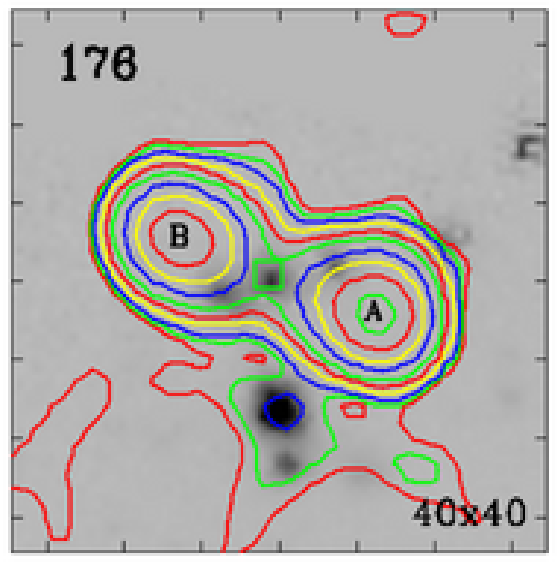}
\includegraphics[width=5cm]{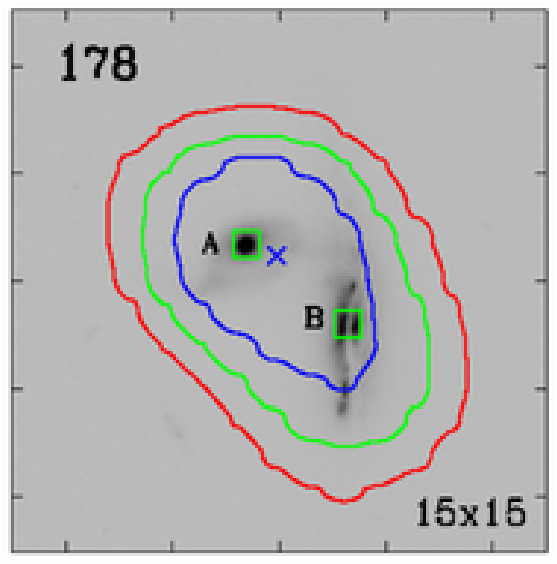}\\
\caption{Cutouts of some particular radio source discussed in Section \ref{particular} with overlaid radio contours. The green square indicates the primary counterpart while the blue x are X-ray detections. The id of the source is on the top-left corner, while the dimension (in arcsec) of the cutout is reported on the bottom-right corner.}
\label{radio_special}
\end{figure}

Some of the identifications presented need further discussion due to a
complex appearance either in the optical/NIR or in the radio
band. Cutouts of each of these source are shown in
Fig. \ref{radio_special}. The component letters in this section refer
to radio source components as given in Table 1 of
\cite{kellermann07}.\\

\noindent{\bf RID 7 - J033115.06-275518.4} This complex multi-component
radio source is identified with an R$\sim 21$ optical/NIR galaxy coincident with component B.\\
{\bf RID 18 - J033127.23-274247.6} A possible blending of two sources, the
counterpart, a R$=24$ galaxy, coincident with the component A. \\
{\bf RID 23 - J033130.05-273814.2} Assuming that the central component
is the core of a double lobed objects, we tentatively identified the
radio source with a faint (R$\approx 26$) optical source.\\
{\bf RID 30 - J033138.56-273808.8} The faint radio source ( S$_{1.4}=175
 \mu Jy$) is extended towards the faint galaxy to the south-west. It is
likely a blend of two faint sources with the south-west one identified
with the counterpart we have given in our catalogue. \\
{\bf RID 50 - J033150.35-274119.1} The counterpart ( R$=23.9$) listed in
Table \ref{opt_catalogue} is close to the peak of the radio emission
and is relatively red ($K_S=20.6$), while an extension of the radio
emission points south to a $\sim 3$ arcsec distant spiral galaxy (
R$=21.5$). The photometric redshift from COMBO-17 reported in
Table \ref{opt_catalogue} could be affected by the contamination in the
optical bands of a nearby ($\approx 1.3$ arcsec) source.\\
{\bf RID 60 - J033154.90-275340.5} Probably the blend of two radio
sources. Currently, in Table \ref{opt_catalogue} is listed the
possible counterparts , R$=24.5$, for one of the two sources; the
other blended radio source could be identified with a R$=21.3$ galaxy
with an offset of $\sim 3$ arcsec from the radio centroid.\\
{\bf RID 113 - J033219.17-275407.7} This is a powerful extended double
lobed radio source (FRII). We choose as the most likely counterpart
the secondary according to our method based on LR (blue triangle in
the inset of the figure). The same counterpart was also chosen for
the X-ray source 249
\citep{giacconi02}. Higher resolution is needed to be certain of the 
optical counterpart. Another X-ray source from
\cite{giacconi02}, XID$=527$ is coincident with the bright peak on the
southwest radio lobe ( component C); not surprisingly this X-ray
source was unidentified.\\
{\bf RID 122 - J033222.49-274805.4} This radio source is identified with a
K$=22.8$ galaxy located at $\sim 0.5$ arcsec from the VLA
position. This counterpart is not visible in the optical bands but is
brigther in the NIR and in all the four IRAC/Spitzer bands. This
object is also detected with Chandra ( XID$=570$ from
\citealt{giacconi02}) and it  was identified with a
R$=24.6$ galaxy located $\sim 1.2$ arcsec north-east of the X-ray
position. We propose that this X-ray sources is identified with the
same counterpart of the radio source. This radio source is also part
of the ATCA 1.4 GHz catalogue presented by
\cite{afonso06}, ATCDFS J033222.36-274807.3, and the distance between
the two radio positions is $\sim 2.6$ arcsec. In their catalogue, this
source remains unidentified, we note that our K band counterpart
candidate is $\sim 4$ arcsec from the
\cite{afonso06} radio position.\\
{\bf RID 140 - J033228.83-274356.4} This source is identified with an $I
\sim 18.3$ galaxy $0.6$ arcsec from the central radio component (RID 140A). 
This identification is also supported by an X-ray detection (XID=103
from \citealt{giacconi02}). We note that \cite{afonso06} reported a
radio source, ATCDFS J033228.71-274402.3, $\sim 7$ arcsec away from
our radio position. This ATCA source remained unidentified in their
catalogue, and we believe this was due to the less accurate position
provided by the ATCA survey compared to our. Finally, another X-ray
detection, XID=630 from \cite{giacconi02}, was tentatively identified
with a R$\approx 23$ galaxy, we suggest that is instead be associated
with the south component of this radio jet (B).\\ {\bf RID 169 -
J033239.47-275301.3}. This radio source is associated with a pair of
interacting galaxies. One of the galaxies has a spectroscopic
redshifts of z=0.686 \citep{vanzella05} while the other as a
photometric redshifts estimate of z$_{\rm phot}=0.65 \pm 0.01$ from
COMBO-17.\\ {\bf RID 176 - J033242.57-273816.4} A double lobe radio
galaxy.  The likely core is clearly visible at 4.5 $\mu m$ with
Spitzer.\\ {\bf RID 178 - J033244.20-275142.1} This radio source is
probably the blend of two objects (A and B) that correspond to a
couple of interacting galaxies at z=0.279 \citep{szokoly04}. The
counterpart of 178A has high excitation emission lines from the
optical spectrum typical of an AGN and is also an X-ray source (
XID$=98$;
\citealt{giacconi02}).\\
%{\bf 184 - J033245.96-275316.2} We identified this radio source with a
%faint (R$\sim 25.5$) counterpart $0.4$ arcsec away. The identification
%is supported by a detection also at 3.6 micron with IRAC/Spitzer. This
%radio source corresponds to ATCDFS J033246.03-275318.2 ($\sim 2.6$
%arcsec is the difference between the two radio positions) which
%remained unidentified in \cite{afonso06}.\\
{\bf RID 186 - J033246.83-274215.1} This S$_{20cm} \sim 122 \mu Jy$ radio
source is offsetted from the center of a bright ( R$\approx 16.3$)
galaxy. An X-ray source ( XID$=84$; \citealt{giacconi02}) is
coincident with the centroid of the galaxy.\\
{\bf RID 207 - J033257.08-280209.7} The optical counterpart of this bright
extended double radio source coincides also with an X-ray detection (
XID$=508$; \citealt{lehmer05}). The galaxy could be the central CD galaxy of a
group/cluster of galaxies at z$=0.664\pm0.001$ (photometric redshift
from COMBO-17) and several arc-like structures can be clearly seen in
the ACS image (see inset in the cutout).\\
{\bf RID 215 - J033303.22-275306.1} The radio emission is coming from two
interacting galaxies with consistent estimates of the photometric
redshifts from COMBO-17, z$_{\rm phot}=0.73 \pm 0.10$ and z$_{\rm
phot}=0.82 \pm 0.06$.\\
%{\bf 237 - J033314.85-280431.8} The radio contours (S$_{\rm 20 cm} =
%590 \mu$Jy of this source are extended to the north-east. It is
%tentatively identified with a magR$\sim 23.6$ galaxy 3.4 arcsec away,
%but the likelihood ratio of this being the right counterpart is below
%the threshold LR$_{\rm th}$.
{\bf RID 244-245-248} These three radio sources could be one single
sources with two lobes. We note that RID=248 has a possible
counterpart (mag R$\sim 24$, distance$\sim 0.8$ arcsec) with
LR$>$LR$_{\rm th}$ which is not consistent with it being a lobe of the
radio source, but it could be the base of a jet which includes RID 244
and 245.\\

\begin{figure}
\centering
\includegraphics[width=18cm]{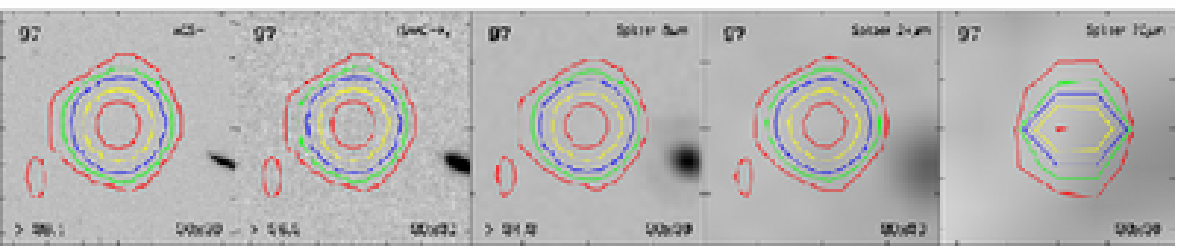}
\includegraphics[width=18cm]{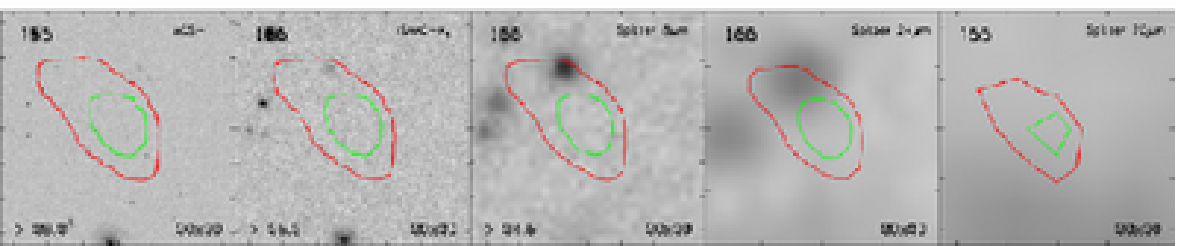}
\includegraphics[width=18cm]{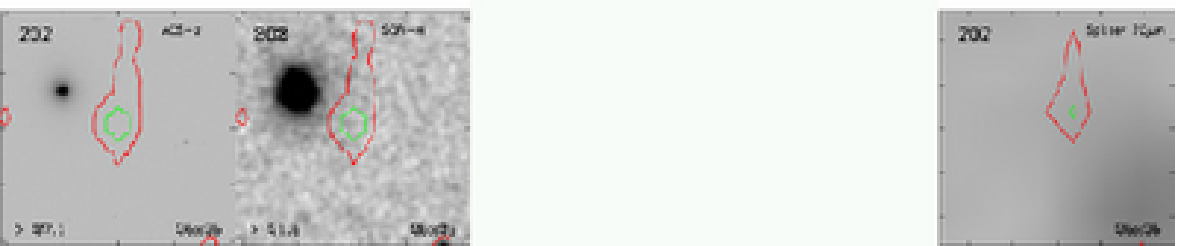}
\caption{ Cutouts of the three sources without any counterpart. In each 
cutout is indicated the sources identification number (top-left), the
band (top-right), the $1 \sigma$ magnitude limit (bottom-left) ad the
dimensions of the cutout (bottom-right).}
\label{radio_unid}
\end{figure}

We are not able to find any counterpart for the following radio sources (see
Fig. \ref{radio_unid}):\\

\noindent{\bf RID 97 - J033213.09-274350.7} This bright radio source
(S$_{20cm}=1.42$ mJy) remains unidentified although is inside the GOODS
area and therefore covered by deep ACS, ISAAC and Spitzer imaging. This
radio source corresponds to ATCDFS J033213.08-274351.0 of
\cite{afonso06}. We considered the possibility that RID 92 and RID 97 
might be lobes of a double radio galaxy. RID 92 lies 38 arcsec
away. However, RID 92 appears to be firmly identified with an
R$\sim$22 galaxy, 0.2 arcsec from the radio position and coincide with
an X-ray source, so this interpretation is unlikely.  RID 97 is
slightly resolved, but about half of the flux density appears to be in
an unresolved component, suggesting that it is an isolated source.
However, we cannot exclude the possibility, that RID 97 is part of RID
92, although there is no evidence of a connection between the two
sources.\\
\noindent{\bf RID 155 - J033235.02-275455.2} This source (S$_{20cm}=0.25$ mJy) 
remains unidentified although is inside the GOODS
area and therefore covered by deep ACS, ISAAC and Spitzer imaging.\\
\noindent{\bf RID 202 - J033252.54-275942.9} This radio source 
(S$_{20cm}=0.1$ mJy) remains unidentified. It is outside the GOODS
area therefore is not covered by the deep IR imaging by ISAAC and
Spitzer.\\

\section{Redshift distribution}

\begin{figure}
   \centering 
   \includegraphics[width=7cm]{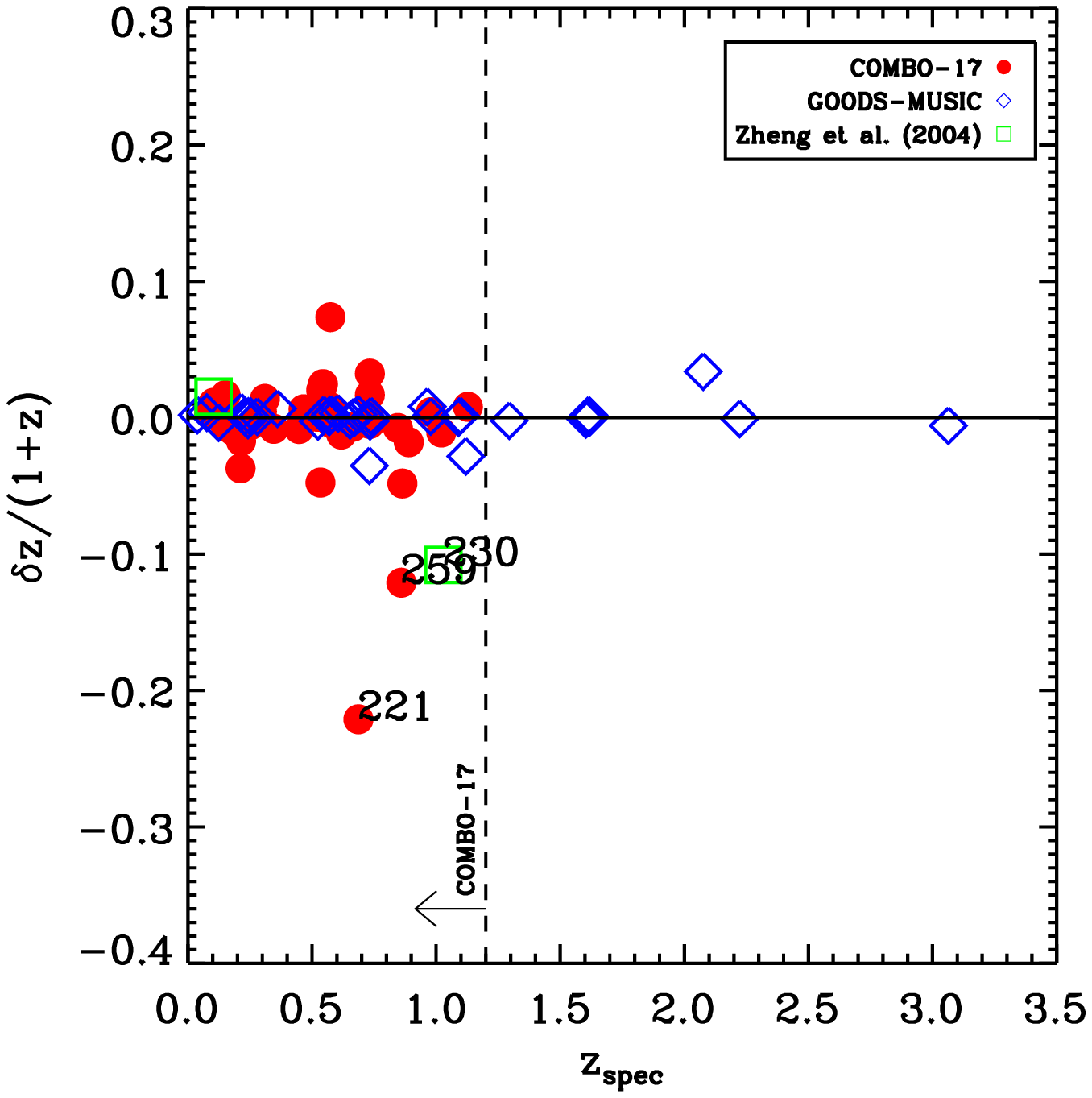}
   \includegraphics[width=7cm]{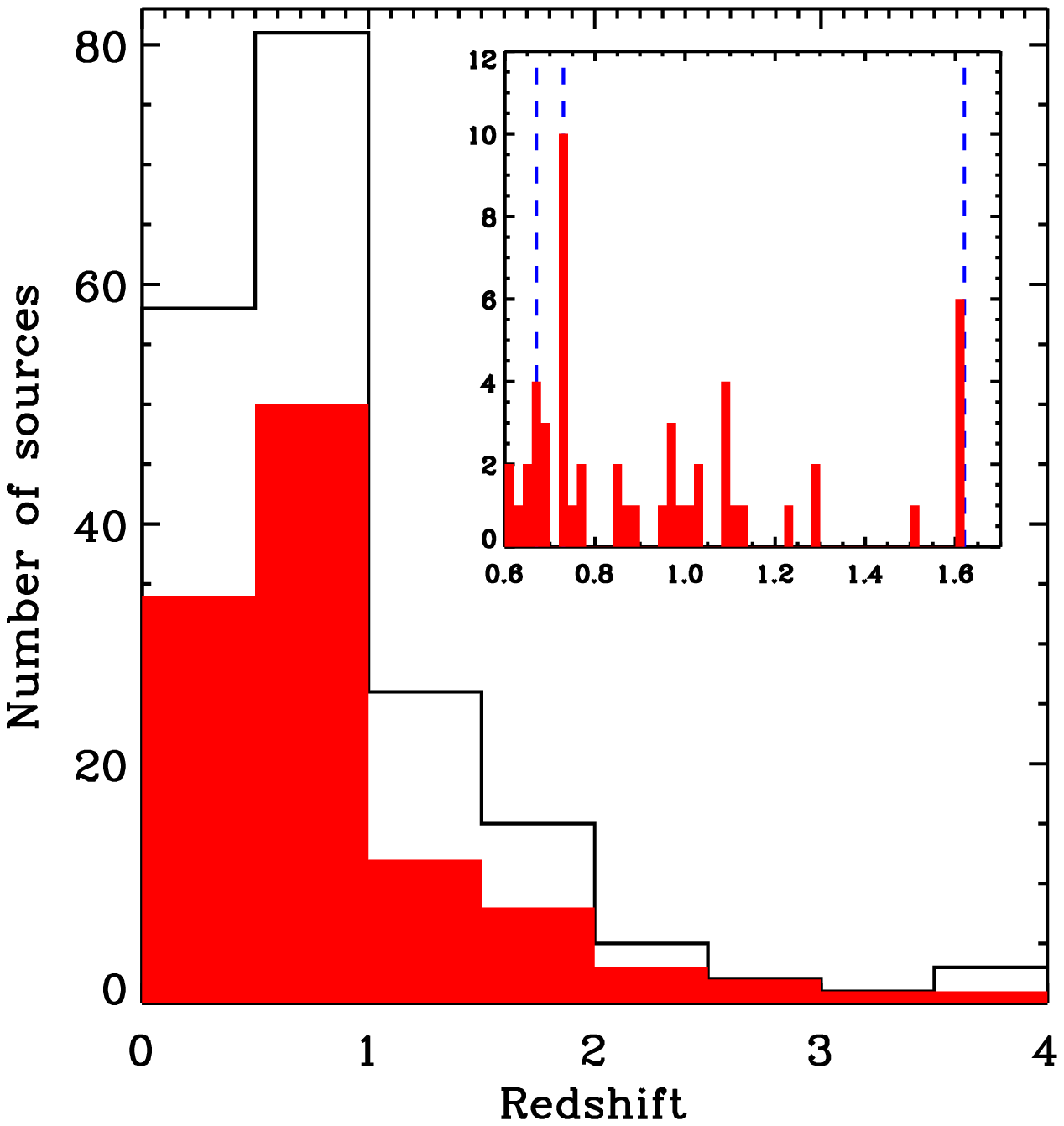} 
   \caption{{\it Left:} Photometric redshift errors as a function of redshift. The dashed
   line indicates the value of z$=1.2$ to which we limit our use of
   this dataset. {\it Right:} Redshift distribution for the VLA/CDF-S
   sample.  The total histogram is obtained using both spectroscopic
   (49) and photometric (110) redshifts, while the shaded one refers
   only to the spectroscopic one. The inset is a zoom in the region
   0.6$<$z$<$1.7 with a redshift bin of 0.02. The vertical dashed
   lines shows the location of three redshift spikes reported by
   \cite{gilli03} at z$=$0.67,0.73,1.61.}
\label{zspec_phot}
\end{figure}

We have compared the primary counterparts found in
Sec. \ref{counterparts} with the spectroscopic catalogs available in
the CDF-S area. A total of 108 ($\sim 41\%$) radio sources have a
spectroscopic redshift. Thirty-six redshifts are provided by the
follow-up campaign of the X-ray sources in the CDF-S
\citep{szokoly04}, seven from the FORS-2/GOODS program
\citep{vanzella05, vanzella06, vanzella08}, 25 from the VIMOS/GOODS program 
\citep{popesso08}, six from the VVDS survey
\citep{lefevre04}, two from the spectroscopic follow-up of the K20
survey \citep{mignoli05}, 31 from the optical follow-up of the X-ray
sources in the E-CDF-S (Silverman et al. in preparation) and one from
\cite{ravikumar07}. We report in Col. 12 of Table  \ref{opt_catalogue}
these spectroscopic redshifts and in Col. 13 a quality flag for these
measures. Since each spectroscopic survey has its own redshift quality
flag system, we try to homogenize and simplify these different
definitions using the following scheme: '2' secure redshift obtained
using multiple spectral features, '1' tentative redshift (e.g. based
on a single emission line).\\ For radio sources with no
spectroscopic information available, we collected photometric estimates
of their redshifts. Due to the deep and wide photometry available in
this region of the sky, reliable photometric redshifts have been
produced. Almost the entire area of the VLA observations is covered by
the COMBO-17 survey which provides extremely accurate photometric
redshifts using photometry in 17 pass-bands from 350 to 930 nm
\citep{wolf04}. We use this dataset with two limitations. We
consider only COMBO-17 sources with R$<24$ (Vega): at these magnitudes
the errors on the photometric redshift estimates are expected to be
less than $|z_{\rm phot}-z_{\rm spec}|/(1+z_{\rm spec})\approx
0.06$. The COMBO-17 data for galaxies fainter than R$=24$ (Vega) are
too shallow. Further, we limit the use of COMBO-17 photometric
redshifts to z$<1.2$ because at higher redshifts the COMBO-17
estimates become increasingly inaccurate due to the lack of NIR
coverage ( see Sec.  4.6 of
\citealt{wolf04}). We have waived this last constraint for objects
best fitted with a QSO templates ( MC$_{class} = {\rm 'QSO'}$ in
\citealt{wolf04}) for which the photometric redshifts are accurate
at least to z$\approx 4$ (see Fig. 18 of \citealt{wolf04}). We compare
our catalogue with \cite{zheng04} that has derived photometric
redshifts for a large fraction of the X-ray sources in the CDF-S area
and the GOODS-MUSIC catalogue \citep{grazian06} in which are published
photometric redshifts obtained using the excellent optical and NIR
data in the GOODS region. In total we are able to assign photometric
redshifts to 78 radio sources without spectroscopic measures: 66 from
COMBO-17, nine from GOODS-MUSIC and three from
\cite{zheng04}. In order to check the accuracy of these photometric
redshifts, we select only sources with a secure spectroscopic redshift
(Col. 13 of Table \ref{opt_catalogue} equal to 2) leading to a sample
of 80 objects. In Fig. \ref{zspec_phot} (left panel) we plot the
difference between the photometric and spectroscopic redshift. The
photometric redshifts are extremely precise with a $|z_{\rm
phot}-z_{\rm spec}|/(1+z_{\rm spec}) < 0.1$ for almost all the
objects. Only three sources have larger errors: RID $ 221, 230,
259$. The spectra of RID 221 does not leave any doubt on the correctness
of the spectroscopic redshift (clear Ca H\&K lines) and its photometry
does not seem to be affected by near-by sources. Similarly, a further
inspection of the optical spectra of RID $230$ and RID $259$ confirms
the spectroscopic redshift z$=1.029$ (broad MgII and CIII] emission
lines) and z=0.860 ([OII] and [OII] emission lines)
respectively. Nevertheless, for two of these sources the disagreement
between photometric and spectroscopic redshift is small ($\sim
0.1$). Clipping these three ``outliers'' out, we obtain an rms$=0.013$
for $|z_{\rm phot}-z_{\rm spec}|/(1+z_{\rm spec})$. \\ The redshift
distribution of the radio sample is plotted in Fig. \ref{zspec_phot}
(right panel): spectroscopic redshifts are shown as a filled
histogram, the total histogram instead includes also photometric
redshifts. In an inset of the same figure we show a zoom of the region
0.6$<$z$<1.7$ using a finer binning in the histogram ($\Delta z
=0.02$). We note that radio sources are good tracers of large scale
structures already detected at other wavebands in this region of the
sky (NIR,
\citealt{cimatti02}; optical, \citealt{lefevre04},
\citealt{vanzella05}; X-ray \citealt{gilli03},
\citealt{szokoly04}). The more prominent spikes in our redshifts
distribution are at z$=0.735 \pm 0.004$ (ten objects, four of these
are also X-ray sources from \citealt{giacconi02}) and z$=1.614 \pm
0.011$ (six objects, two of these are also X-ray sources from
\citealt{giacconi02}). Both these structure were traced in the X-rays
\citep{gilli03}, respectively with 19 and 5 X-ray sources. Of the two
prominent large scale structures discovered in the X-ray survey, at
z=0.67 and z=0.73, the radio sources are tracing only the second. As
noted by
\cite{gilli03} while the structure at z=0.73 is dominated by a
standard galaxy cluster with a significant concentration around a
central CD galaxy, the one at z=0.67 appears to be a rather loose
structure uniformly distributed in the field.

\section{Host galaxy properties}
\label{host}

\begin{figure}
  \centering \includegraphics[width=7cm]{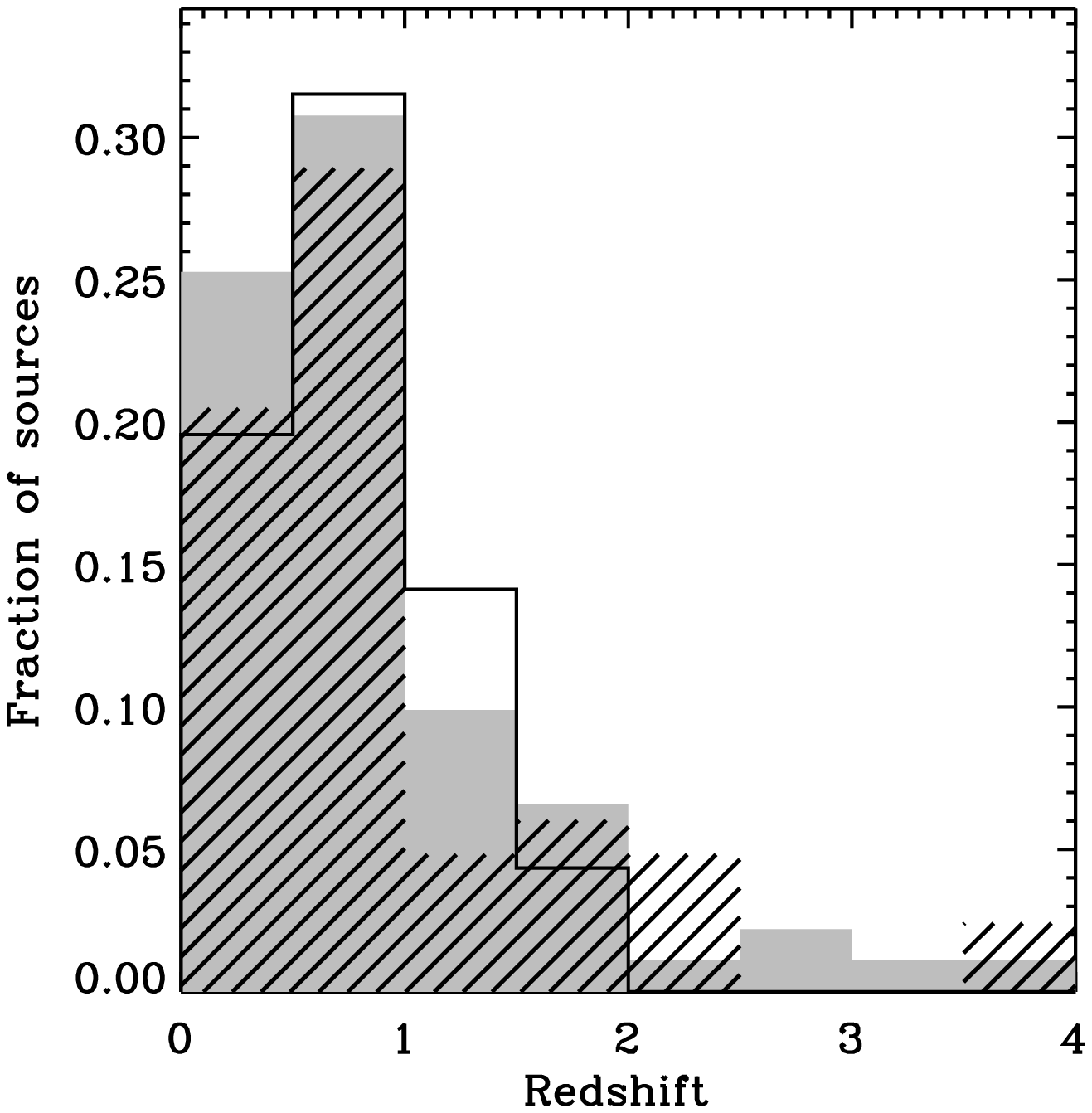}
  \centering \includegraphics[width=7cm]{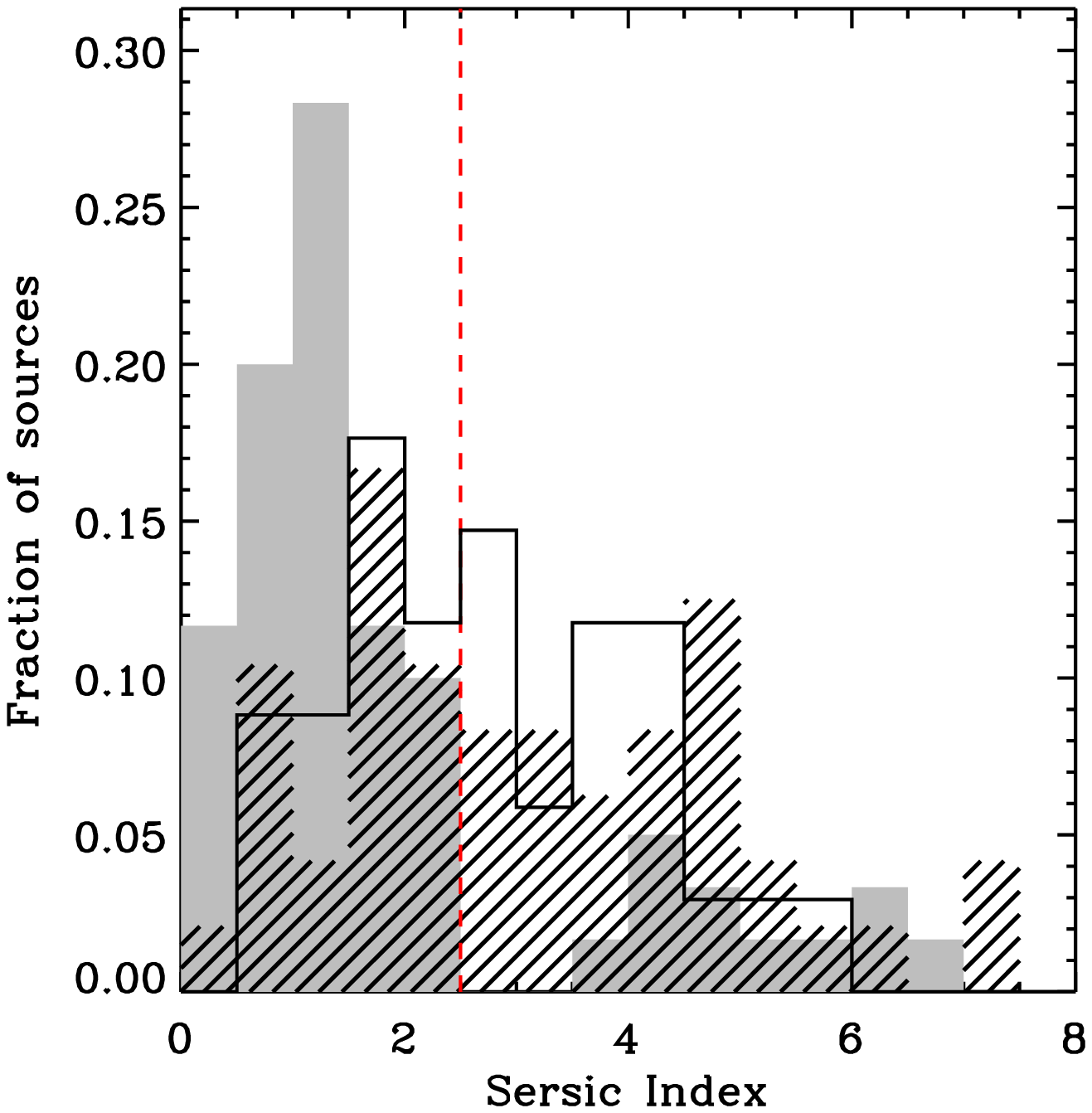}
  \caption{Left panel: redshift distribution for radio sources with
  S(1.4 GHz)$>0.2$ mJy (continuous line), $0.08<$S(1.4 GHz)$<0.2$
  (hatched histogram) and S(1.4 GHz)$<0.08$ mJy (grey shaded
  histogram). Right panel: distribution of the Sersic index values for
  radio sources with S(1.4 GHz)$>0.2$ mJy (continuous line),
  $0.08<$S(1.4 GHz)$<0.2$ (hatched histogram) and S(1.4 GHz)$<0.08$
  mJy (grey shaded histogram). The vertical line mark the value
  $n=2.5$, empirical dividing value between early and late type
  galaxies. }
\label{Fradio_z}
\end{figure}

\begin{figure}
\centering
\includegraphics[width=15cm]{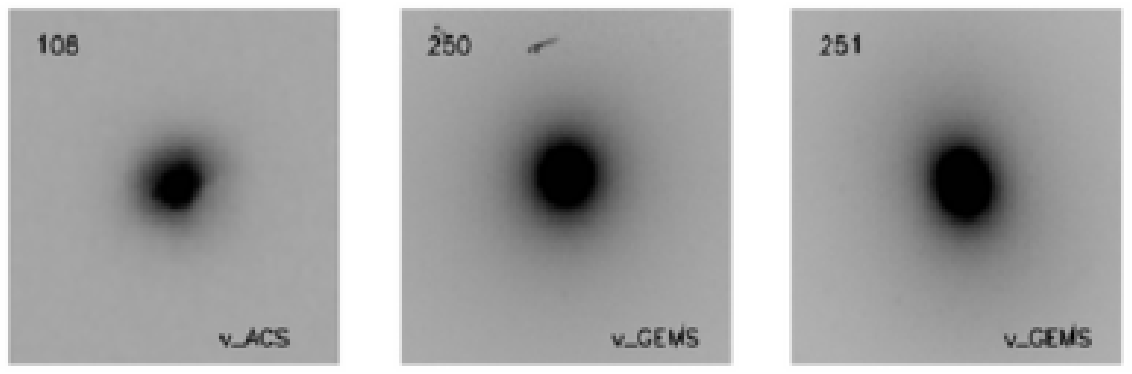}\\
\includegraphics[width=15cm]{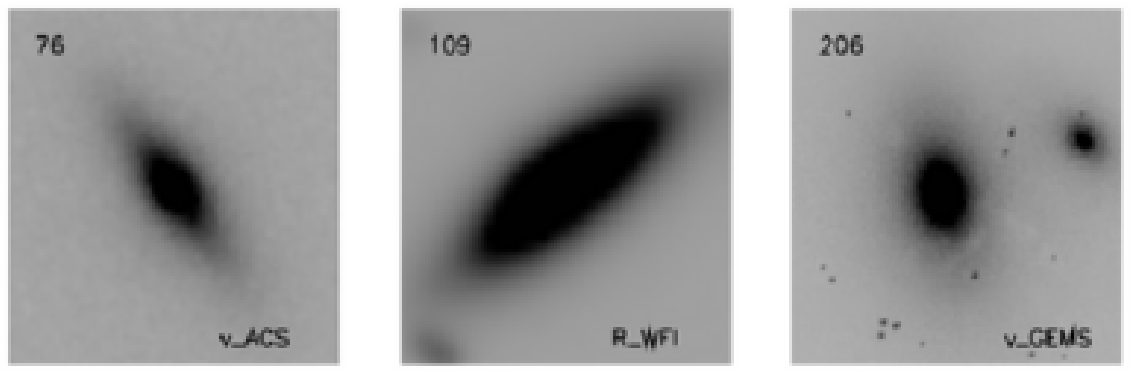}\\
\includegraphics[width=15cm]{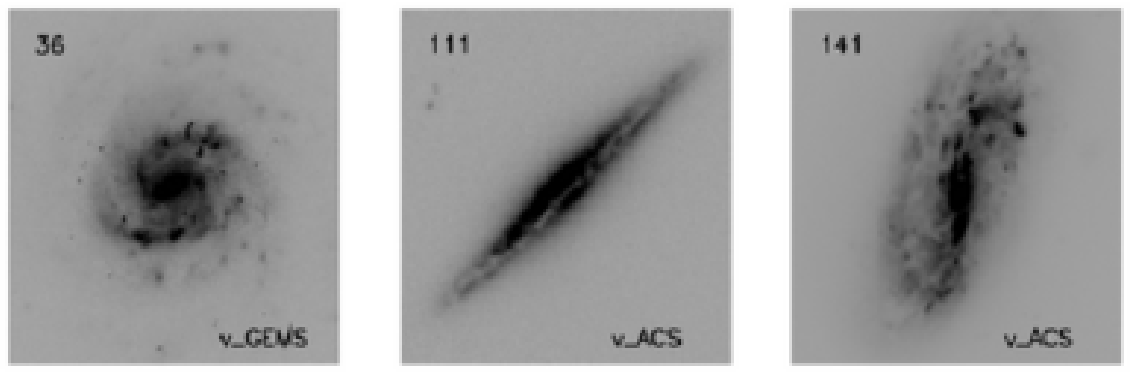}\\
\includegraphics[width=15cm]{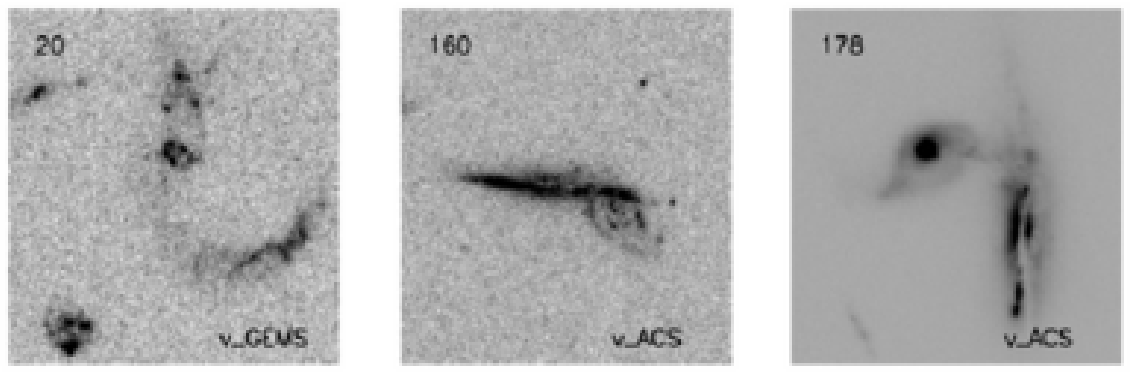}\\
\caption{Image cutouts ($10''\times10''$) of sources belonging 
to the four morphological classes defined in Sec. \ref{host}: E (first
row), S0 (second row), S (third row), I (fourth row).}
\label{morph_template}
\end{figure}

\begin{figure}
  \centering \includegraphics[width=7cm]{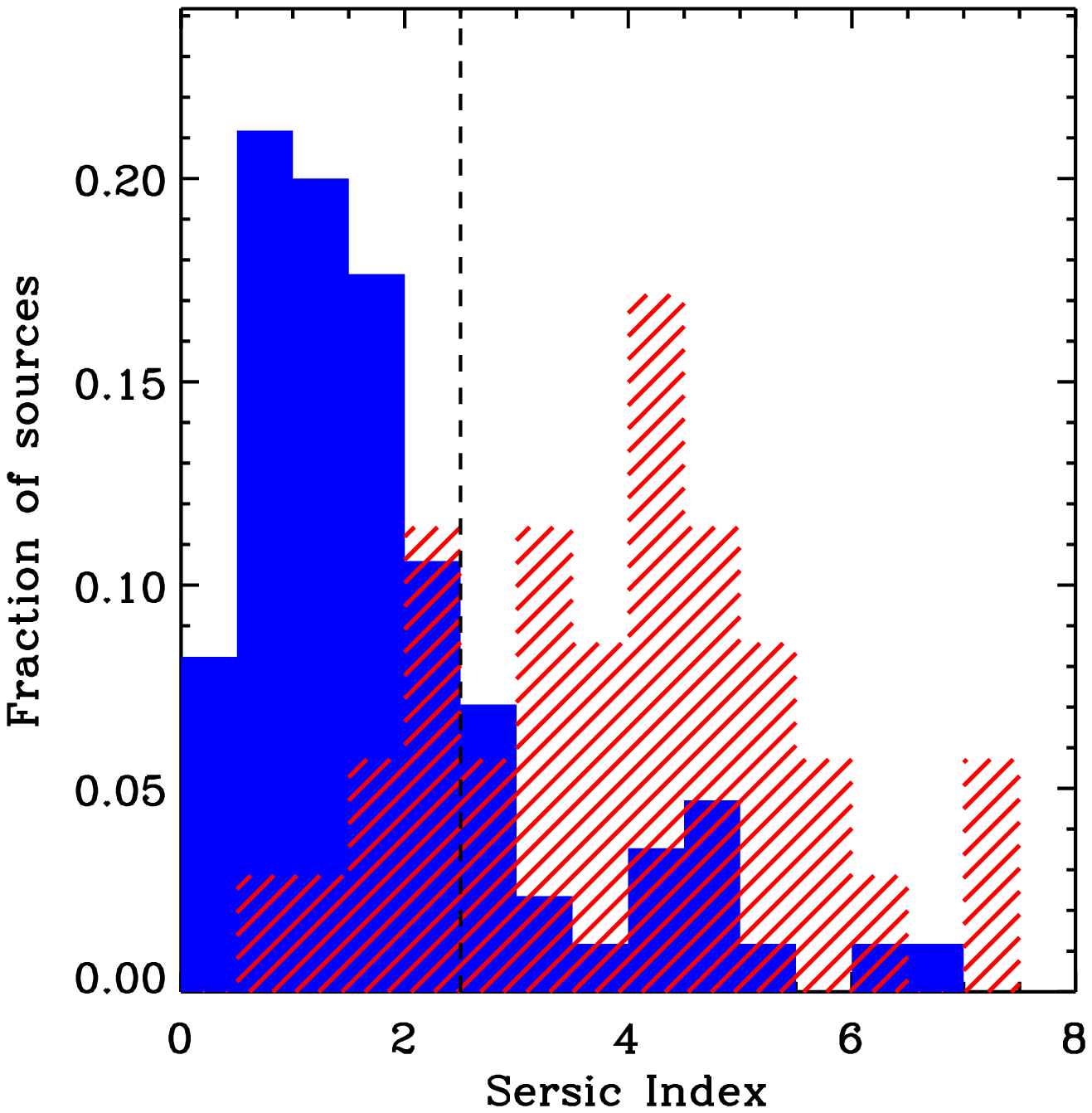}
  \centering \includegraphics[width=7cm]{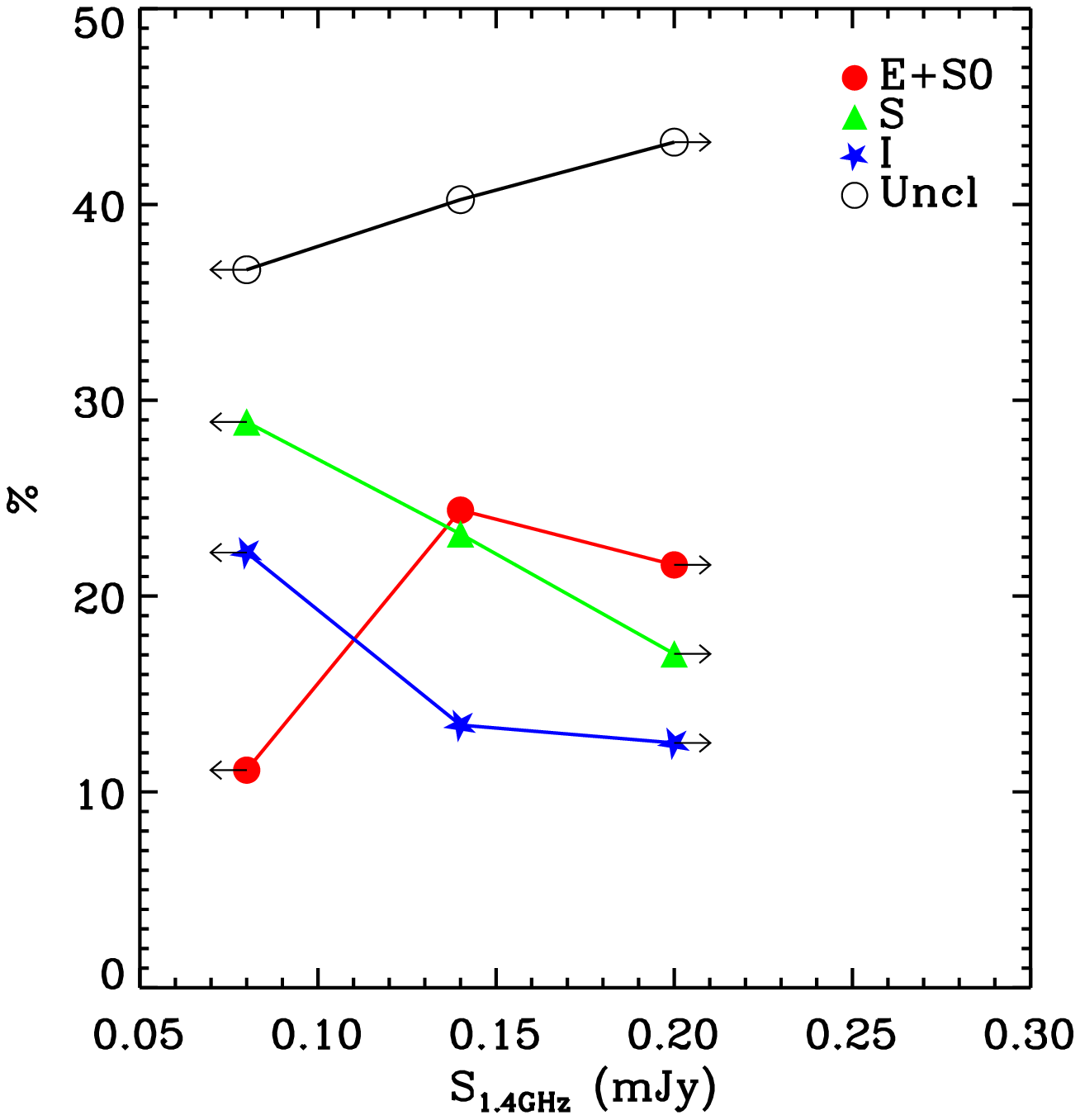}
  \caption {Left panel: distribution of Sersic index values for
  sources with visual morphological classification 'E'or 'S0' (hatched
  histogram) and 'S' or 'I' ( shaded histogram). Right panel:
  percentages of E$+$S0, S, I and unclassified sources in the three
  radio flux density intervals defined in Sec. \ref{host}. }
\label{Fradio_morph}
\end{figure}

\begin{figure}
  \centering \includegraphics[width=8cm]{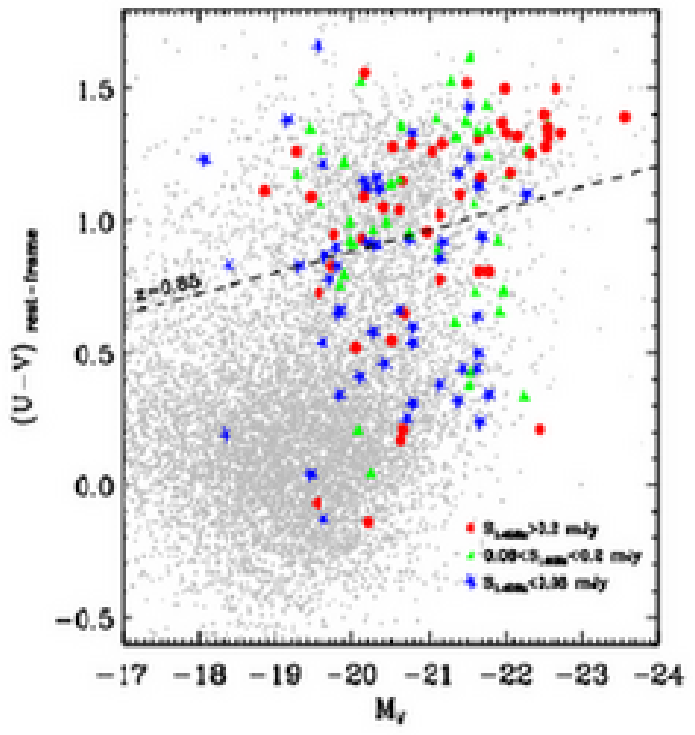}
  \centering \includegraphics[width=8cm]{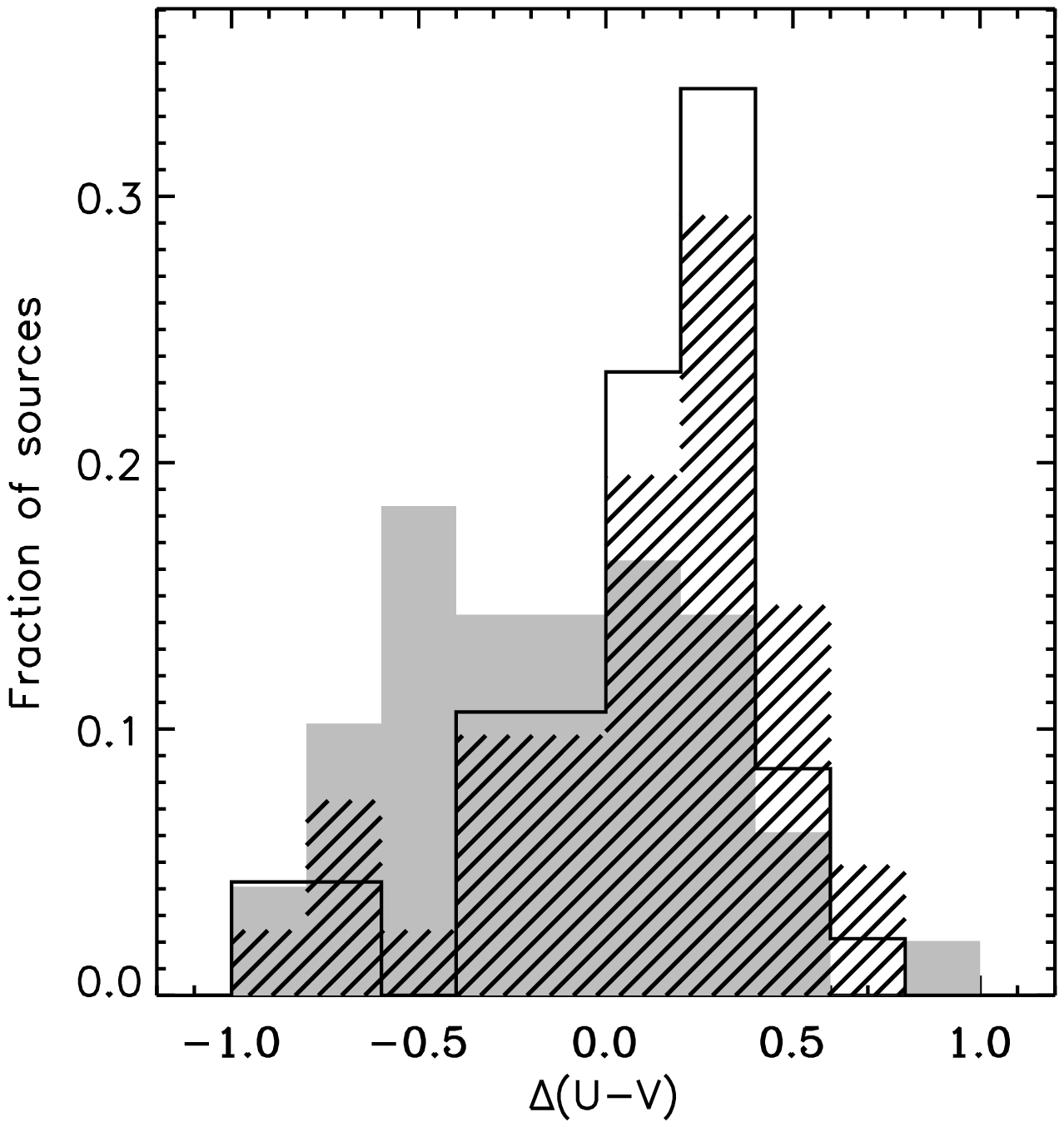}
  \caption{{\it Left panel}: Rest-frame U-V colors versus the
  absolute magnitude in the V band for radio sources with S$>0.2$ mJy
  (circles), $0.08<$S$<0.2$ mJy (triangles), S$<0.08$ mJy (stars) and
  optically selected galaxies from COMBO-17 (grey dots). The dashed
  line is the lower edge of red-sequence galaxies as defined by Bell
  et al. (2004) with z=0.85 (average redshift of our radio
  sources). {\it Right panel:} distribution of the difference between
  the rest-frame U-V color and the value of the lower edge of
  red-sequence galaxies (see left panel) for a given M$_{\rm V}$
  value. The continuous line refers to radio sources with S$>0.2$ mJy,
  the hatched histogram to $0.08<$S$<0.2$ mJy and finally sources with
  S$<0.08$ mJy are represented by the shaded histogram.  }
\label{Fradio_colors}
\end{figure}

In this section we investigate the properties of the host galaxy as a
function of the 1.4 GHz flux density of the radio source. We want in
particular, to study the behavior of three observables as a function
of the radio flux density: a) redshift, b) morphology and c)
rest-frame colors of the host galaxies. We are motivated by the known
upturn in the normalized source count at 1.4 GHz below 3 mJy compared
to the extrapolation from brighter flux densities (e.g. Fig. 7 in
Paper I). This is generally modeled by the mix of two populations of
sources: AGN and starburst galaxies. Several models
(e.g. \citealt{seymour04}) predict that the starburst galaxies become
dominant at 1.4 GHz flux densities below $\sim 0.2$ mJy. Further
discussion on the source population properties will be presented by
Padovani et al. (2008, Paper IV).\\ We divide our sample in three flux
density bins with approximately the same number of sources: S$<0.08$
mJy ( 91 objects), $0.08<$S$<0.2$ mJy ( 83 objects) and S$>0.2$ mJy (
92 objects).

\noindent a) {\it redshift}\\ In the three flux density intervals defined above 
respectively 71 ($\sim 78\%$), 56 ($\sim 67\%$) and 64 ($\sim 69\%$)
objects have a redshift (spectroscopic or photometric). We show in
Fig. \ref{Fradio_z} (left panel) the redshift distributions of the
three sub-samples. The brighter objects (continuous line) are almost
all at z$<1.5$. Due to the small area of our VLA survey, we are
missing the population of radio-loud QSOs that dominate this regime,
as we know from shallower large-area radio surveys. The intermediate
flux density bin (hatched histogram) shows a redshift distribution
similar to the above one with a few objects at z$\approx 4$. A K-S test
confirms with a probability of $46\%$ that the two distributions are
drawn from the same one.\\ Finally, the redshift distribution for the
sources in the faintest bin (shaded histogram) has a tail at higher
redshift ($\sim 16\%$ have z$>1.5$). Previous surveys
(e.g. \citealt{windhorst95},
\citealt{richards98}, \citealt{richards99}, \citealt{ciliegi05}) 
found that the majority of the $\mu$Jy radio sources are at low
(z$<1$) redshifts. If we consider our faintest flux density bin,
$22\%$ of the objects are at z$>1$, and for another $22\%$ of sources
we do not have any redshift information. The fraction of sources at
$z>1$ could grow up to $44\%$ if all the unidentified objects are at
high redshift. A K-S test gives a probability of $99\%$ that the
redshift distributions of the faintest and intermediate flux density
bins are drawn from the same one.

\noindent b) {\it morphology}\\
In this section we investigate the morphological properties of the
host galaxies. The area of our VLA observations is almost entirely
covered by the GEMS survey (\citealt{rix04}). The GEMS team has made
publicly available\footnote{http://www.mpia.de/GEMS/gems.htm}
\citep{haeussler07} a catalogue containing the Sersic index ($n$)
parameter (\citealt{sersic68}) obtained using GALFIT
(\citealt{peng02}). We consider only objects in this catalogue with
CONSTR\_FLAG$=1$ and SCIENCE\_FLAG$=1$ (see Haeussler et
al. 2007). Applying these constraints, we are left with 60 (S$<0.08$
mJy), 48 ($0.08<$S$<0.2$ mJy) and 34 (S$>0.2$ mJy) objects
respectively. We are therefore able to study the morphological
properties of $\sim 66\%$, $\sim 58\%$ and $\sim 37\%$ of the objects
in each flux density bin. The distributions of Sersic indexes are
shown in Fig.\ref{Fradio_z} (right panel). It has been empirically
found that a Sersic index n$=2.5$ (vertical line in
Fig. \ref{Fradio_z}) can roughly discriminate between early-type
galaxies and late-type galaxies (e.g., Blanton et al. 2003; Shen et
al. 2003; Hogg et al. 2004). Galaxies with n$>2.5$ are generally
early-type, while the majority of galaxies with n$<2.5$ are
late-type. We find that for the high and intermediate flux density
bins (continuum and hatched histograms) the distribution of Sersic
index is relatively flat with $\approx 55 \%$ of the objects with
$n>2.5$. According to a Kolmogorov-Smirnov (K-S) test these two
distributions are drawn from the same one. Instead, the distribution
of Sersic indexes appears completely different for the faintest flux
density bin (shaded histogram) with only $\approx 18 \%$ of the
sources with $n>2.5$. A K-S test gives a probability as low as $\sim
0.05\%$ that the distribution for the faintest flux density bin and
any of the other two are drawn from the same sample. This is
suggesting that at $\approx 0.08$ mJy there is a significant change in
the radio population with late-type galaxies becoming dominant. We are
aware of the fact that using only the Sersic index to divide galaxies
in morphological classes could produce a mixture of different galaxy
populations (e.g. \citealt{sargent07}). Therefore we visually
classified our radio sources in four broad morphological categories:
elliptical (E), lenticular (S0), spiral (S) and irregular (I). We show
examples for these four categories in Fig. \ref{morph_template}. We
are able to classify 161 ($60\%$) of the 266 radio sources. In Fig.
\ref{Fradio_morph} (left panel) we show the distribution of Sersic
index values for E+S0 (hatched histogram) and S+I (shaded
histogram). The distributions of Sersic indexes are clearly different
for these two classes of objects (K-S probability $\approx
10^{-6}\%$): E+S0 tend to have high value of n, while S+I peak at
n$<$2.5. This plot confirms that our visual morphological
classification is robust and at the same time is reassuring about the
use of Sersic index to separate early and late type galaxies. We can
now use this visual morphological classification to study the
evolution of the host galaxy properties as a function of the S$_{\rm
1.4 GHz}$ flux density. In Fig. \ref {Fradio_morph} (right panel) we
plot the percentages of early-type (E+S0), spiral and irregular
galaxies in the three flux density bins. While a large fraction of
radio sources with S$_{\rm 1.4 GHZ} > 0.08$ mJy are early-type, at
lower flux densities late-type galaxies become five times more
numerous than elliptical or lenticular galaxies.

\noindent c) {\it rest-frame colors}\\
We finally consider the rest-frame colors of our radio sources. We
compare the list of optical/NIR counterparts presented in this work
with the COMBO-17 photometric catalogue
\citep{wolf04}. We include only COMBO-17 sources with R magnitude
brighter that 24 (Vega) and we exclude sources classified as 'QSO'
from the SED fitting. We further exclude sources with broad ($> 2000$
km s$^{-1}$) emission lines in their optical spectra to avoid
contamination by the central AGN. Removing any other potential AGN
according to the criterium L$_{\rm X}[0.5-10 {\rm keV}]>10^{42}$ erg
s$^{-1}$ does not change the result. We finally exclude sources for
which the COMBO-17 redshift (that has been used to compute the
rest-frame colors by
\citealt{wolf04}) is not in agreement with an available spectroscopic
one. We obtain rest-frame colors for 47 ( S$>0.2$ mJy), 41 (
$0.08<$S$<0.2$ mJy) and 49 ( S$<0.08$ mJy) objects respectively. In
Fig. \ref{Fradio_colors} (left panel), we plot the rest-frame U-V
colors as a function of the absolute magnitude M$_{\rm V}$ for radio
sources with S$>0.2$ mJy (circles), $0.08<$S$<0.2$ mJy (triangles),
S$<0.08$ mJy (stars) and optically selected galaxies from COMBO-17
(grey dots). The distribution of rest-frame colors for the faintest
radio flux density bin appears clearly different from the other two: while
sources with S$>0.2$ mJy are mainly concentrated in the top part of
the plot, below 0.08 mJy the host galaxies are widely spread in the
color-magnitude diagram. The dashed line in Fig.\ref{Fradio_colors}
(left panel) is the empirical redshift-dependent color divisions that
separate blue and red galaxies population. Following Bell et
al. (2004), we define red-sequence/early-type galaxies as being redder
than $<U-V>=1.15-0.31\times z-0.08(M_V -5 logH+20)$, where z$=0.85$,
the average redshift of our radio sample. We find that $\sim 70\%$ of
the radio sources in the two brighter flux density bins are red-sequence
galaxies while this percentage decreases to $\sim 40\%$ for sources
with S$<0.08$ mJy. Again, we have an indication that early-type
galaxies are the dominant population of the intermediate radio flux density
bin, while late-type galaxies becomes dominant at the faintest radio
flux densities. A K-S test on the distributions of the U-V colors (see right
panel of Fig. \ref{Fradio_colors}) provides a probability as low as
$0.5 \%$ for the S$<0.08$ mJy colors being similar to the ones of
brighter radio sources.

%This supports our finding in Paper IV (Padovani et al. 2007) that the
%bulk ($>1/2$) of our radio sources are faint radio galaxies with
%star-forming galaxies making up a minority of the sample.

%These results are in good agreement with spectroscopic identifications
%at the submillijansky level, which find $\approx 60-70\%$ of elliptical
%galaxies and broad line AGN among the optical counterparts of S$_{1.4
%GHz}<0.2$ mJy radio sources \citep{gruppioni99}. 

Recently, \cite{bondi07} found that between 0.15 and 0.5 mJy the
median radio spectral index is significantly flatter compared to
brighter or fainter sources and they interpreted this as evidence that
a population of flat spectrum low luminosity compact AGNs and radio
quiet QSOs is dominating the radio emission in this flux density
range, while a starburst population is expected to have a steeper
spectral index. At the same time, they confirm that early-type
galaxies have spectral index significantly flatter than starburst
ones. The
\cite{bondi07} results are consistent with our finding that the
majority of radio sources with flux densities $0.08<$S$<0.2$ mJy are
hosted in bulge dominated early-type host galaxies while at lower
radio flux densities (S$\lesssim 0.08$ mJy) the starburst galaxies
become dominant.

\section{Conclusions}

We have used a likelihood ratio technique to identify the 266 radio
sources from the VLA survey in the CDF-S. Using the available imaging
in {\it i}, R, K$_{\rm S}$, 3.6 $\mu m$, 4.5 $\mu m$, 5.8 $\mu m$, 8.5
$\mu m$, 24 $\mu m$ and 70 $\mu m$ bands we were able to find a
reliable counterpart for 254 ($\sim 95 \%$) radio sources. Using
literature data and our own follow-up, a total of 186 ( $\sim70\%$)
sources have a redshift: 108 are spectroscopic redshifts and 78
reliable photometric redshifts. The ability of obtaining a redshift
(either spectroscopic or photometric) is a strong function of the
magnitude of the counterparts: $\sim 88\%$ of the objects with R$_{\rm
AB}<24$ have a redshift while this fraction decreases to $\sim 30\%$
for sources with R$_{\rm AB}>24$. The redshift distribution of the VLA
sources peaks around z$\approx 0.9$. The radio sources are good
tracers of large scale structures already detected at other wavebands
in this region of the sky (NIR, optical, X-ray). In particular,
the main peaks of our redshifts distribution are at z$=0.735 \pm
0.004$ (10 objects) and z$=1.614 \pm 0.011$ (6 objects).

We have studied the properties of the host galaxies of the radio
sources dividing the sample in three flux density bins equally
populated: S$>0.2$ mJy, $0.08<$S$<0.2$ mJy and S$<0.08$ mJy. While the
properties of the host galaxies in the two brighter flux density bins
look similar, we find evidences for a change in the dominant radio
population at S$\approx 0.08$ mJy. The radio sources in the
intermediate and bright flux density bins:
\begin{itemize}
\item show a Sersic indexes distribution that resembles that of 
early-type galaxies with a tail of disk dominated galaxies;
\item according to a  visual morphological classification are 
mainly elliptical or lenticular galaxies;
\item in a rest-frame color-magnitude diagram (U$-V$ versus M$_{\rm V}$) 
are preferentially ($70\%$) located between the
early-type/red-sequence galaxies;
\end{itemize}
while sources with S$< 0.08$ mJy:
\begin{itemize}
\item have a Sersic indexes distribution that peaks at low values of n, 
indicating a low value for the bulge to disk ratio, with only $\approx
18 \%$ of the sources with $n>2.5$;
\item is five times more likely to be hosted in a late-type 
galaxy instead of an elliptical or lenticular one;
\item are widely spread in the color-magnitude diagram, with 
$\approx 60\%$ of them not being an early-type/red-sequence galaxy.
\end{itemize}

Summarizing, we suggest that the flux density bin S $\gtrsim 0.08$ mJy
is dominated by a population of early-type galaxies while at flux
densities below $\approx 0.08$ mJy starburst galaxies become dominant.

\acknowledgements
  This work is based on observations with the National Radio Astronomy
  Observatory which is a facility of the National Science Foundation
  operated under cooperative agreement by Associated Universities,
  Inc.  We thank the referee for constructive comments that improved
  the quality of the paper. We thank P. Ciliegi for enlightening
  discussions on the likelihood ratio technique. John Kelly assisted
  with the early data reduction. We acknowledge the ESO/GOODS project
  for the ISAAC and FORS2 data obtained using the Very Large Telescope
  at the ESO Paranal Observatory under Program ID(s): LP168.A-0485,
  170.A-0788, 074.A-0709, and 275.A-5060. We acknowledge the GOODS
  team, the GOODS-MUSIC team, the COMBO-17 team, the GEMS team and the
  FIDEL team for making images and catalogs publicly available. PT
  acknowledges financial contribution from contract ASI--INAF
  I/023/05/0.
% We are grateful to K. Jahnke for providing us the GEMS PSF.

\clearpage

\begin{table*}
       \caption{Positional offsets between the radio and the optical/nearIR catalogs.}
       \label{tab_offset}
%       \vspace{0.2cm}
\begin{minipage}{0.99\textwidth}
\footnotesize
\begin{tabular}{lcccc}
\hline\hline\noalign{\smallskip}   
 & \multicolumn{2}{c}{Optical} &  \multicolumn{2}{c}{NIR} \\     
      &                             
ACS-i &
WFI-R &
ISAAC-K &
SOFI-K \\

\noalign{\smallskip}\hline\noalign{\smallskip}  
$< \Delta {\rm RA} >$:  & 0.15\arcsec & 0.20\arcsec & 0.15\arcsec  & 0.13\arcsec\\
$< \Delta {\rm Dec} >$: & 0.12\arcsec & 0.09\arcsec & -0.23\arcsec & 0.12\arcsec \\

\hline

\noalign{\smallskip}
\end{tabular}

\end{minipage}
\end{table*}                                                                         

\clearpage

\begin{table*}
       \caption{Output of the likelihood ratio technique for the radio 
sources identification.}
       \label{tab_like}
%       \vspace{0.2cm}
\begin{minipage}{0.99\textwidth}
\footnotesize
\begin{tabular}{rccccc}
\hline\hline\noalign{\smallskip}   
 & \multicolumn{2}{c}{Optical} &  \multicolumn{2}{c}{NIR} & Total \\     
      &                             
ACS-i &
WFI-R &
ISAAC-K &
SOFI-K &
       \\

\noalign{\smallskip}\hline\noalign{\smallskip}  
radio sources$^a$: & 79 & 266 & 61 & 266  & 266 \\

LR$>$LR$_{\rm th}$ counterparts$^b$: & 67 & 230 & 56 & 190  & 254 \\

LR$<$LR$_{\rm th}$ counterparts$^c$: & 4 & 24 & 1 & 12 & 9 \\

frac. identified$^d$: & 85$\%$ & 86$\%$ & 92$\%$ & 71$\%$ & 95$\%$ \\

possible spurious$^e$:  & $<1$ & 3 & $<1$ & 6 & 8 \\

unidentified$^f$: & 12 & 36 & 5 & 76 & 12 \\

empty fields$^g$: & 8 & 12 & 4 & 64 & 3 \\

\hline

\noalign{\smallskip}
\end{tabular}

$^a$ Number of radio sources inside the area imaged in the different bands\\
$^b$ Number of radio sources with a counterpart with LR above the threshold LR$>$LR$_{\rm th}$ and therefore considered reliable.\\
$^c$ Number of radio sources only with a counterpart with LR below the threshold LR$<$LR$_{\rm th}$ and therefore considered unreliable.\\
$^d$ Percentage of identified sources (considering only counterparts with LR$>$LR$_{\rm th}$).\\
$^e$ Maximum number of spurious sources expected between the counterparts with  LR$>$LR$_{\rm th}$.\\
$^f$ Number of radio sources without any counterparts or with counterparts with LR$<$LR$_{\rm th}$.\\
$^g$ Number of radio sources without any possible counterpart.\\
\end{minipage}
\end{table*}

%\begin{figure}
%  \centering \includegraphics[width=8cm]{figures/RK_z.ps}
%   \caption{(R-K) color versus the spectroscopic (or photometric) redshift for the radio sources in the CDF-S area. }
%\label{RK_z}
%\end{figure}

\clearpage
\pagestyle{empty}

\begin{deluxetable}{rrrrrcrrcccccc}
\tabletypesize{\scriptsize}
\rotate
\tablecaption{Optical, near infrared primary counterparts of the 20 cm sources in the E-CDF-S \label{opt_catalogue}}
\tablewidth{0pt}
\tablehead{
 & \multicolumn{2}{c}{RADIO} & \multicolumn{2}{c}{OPTICAL}  &  &  &  &  &  &  &  &  & \\
(1)  &  \multicolumn{1}{c}{(2)} &  \multicolumn{1}{c}{(3)} &  \multicolumn{1}{c}{(4)} &  \multicolumn{1}{c}{(5)} &  \multicolumn{1}{c}{(6)} &  \multicolumn{1}{c}{(7)} &  \multicolumn{1}{c}{(8)} &  \multicolumn{1}{c}{(9)} &  \multicolumn{1}{c}{(10)} &  \multicolumn{1}{c}{(11)} &  \multicolumn{1}{c}{(12)} & \multicolumn{1}{c}{(13)} & \multicolumn{1}{c}{(14)}  \\
RID  &  \multicolumn{1}{c}{RA} &  \multicolumn{1}{c}{Dec} &  \multicolumn{1}{c}{RA} &  \multicolumn{1}{c}{Dec} &  \multicolumn{1}{c}{dist} &  \multicolumn{1}{c}{LR} &  \multicolumn{1}{c}{Rel} &  \multicolumn{1}{c}{catalogue} &  \multicolumn{1}{c}{R} &  \multicolumn{1}{c}{K} &  \multicolumn{1}{c}{z} & \multicolumn{1}{c}{Qual-z} & \multicolumn{1}{c}{z$_{phot}$}    \\
 & \multicolumn{2}{c}{(J2000)} & \multicolumn{2}{c}{(J2000)}  & ($^{\prime\prime}$) &  &  &  & (AB) & (AB) &  &  & \\
}
\startdata
1 & 3:31:11.70 & -27:31:44.3 & 3:31:11.64 & -27:31:43.5 & 0.90 & 101.25 & 1.00 & WFI-R & 16.79$\pm$0.01 & $>$21.4 &  &  &  \\
2 & 3:31:13.96 & -27:39:10.3 & 3:31:13.98 & -27:39:10.4 & 0.40 & 29.79 & 0.99 & WFI-R & 23.13$\pm$0.04 & $>$21.4 &  &  &  \\
3 & 3:31:14.20 & -27:48:44.3 & 3:31:14.12 & -27:48:44.2 & 1.00 & 0.85 & 0.81 & WFI-R & 25.12$\pm$0.19 & $>$21.4 &  &  &  \\
4 & 3:31:14.42 & -27:39:07.1 & 3:31:14.53 & -27:39:06.6 & 1.60 & 35.25 & 0.99 & WFI-R & 18.04$\pm$0.01 & $>$21.4 &  &  &  \\
5 & 3:31:14.50 & -27:55:46.3 & 3:31:14.46 & -27:55:46.6 & 0.60 & 19.06 & 0.99 & WFI-R & 23.40$\pm$0.03 & $>$21.4 &  &  &  \\
6 & 3:31:14.87 & -27:55:43.4 & 3:31:14.91 & -27:55:41.5 & 2.00 & 1.60 & 0.89 & WFI-R & 23.28$\pm$0.03 & $>$21.4 &  &  &  \\
7 & 3:31:15.06 & -27:55:18.4 & 3:31:15.04 & -27:55:18.6 & 0.30 & 67.91 & 1.00 & WFI-R & 20.84$\pm$0.01 & $>$21.4 &  &  & 0.49$\pm$0.03$^1$ \\
8 & 3:31:16.00 & -27:44:43.2 & 3:31:15.98 & -27:44:43.2 & 0.10 & 21.47 & 0.98 & WFI-R & 23.98$\pm$0.04 & $>$21.4 &  &  & 0.86$\pm$0.04$^1$ \\
9 & 3:31:17.36 & -28:01:47.8 & 3:31:17.34 & -28:01:47.0 & 0.70 & 63.25 & 1.00 & WFI-R & 20.98$\pm$0.01 & $>$21.4 &  &  & 0.67$\pm$0.02$^1$ \\
10 & 3:31:18.74 & -27:49:02.4 & 3:31:18.62 & -27:49:03.1 & 1.70 & 0.30 & 0.60 & WFI-R & 26.00$\pm$0.10 & $>$21.4 &  &  &  \\
11 & 3:31:19.92 & -27:35:50.2 & 3:31:19.88 & -27:35:49.3 & 1.00 & 21.29 & 0.99 & WFI-R & 21.62$\pm$0.01 & $>$21.4 &  &  & 0.90$\pm$0.01$^1$ \\
12 & 3:31:20.12 & -27:39:01.3 & 3:31:20.15 & -27:39:01.2 & 0.50 & 128.81 & 1.00 & WFI-R & 20.30$\pm$0.01 & $>$21.4 &  &  & 0.55$\pm$0.02$^1$ \\
13 & 3:31:23.32 & -27:49:06.4 & 3:31:23.30 & -27:49:05.8 & 0.50 & 31.86 & 0.99 & WFI-R & 22.70$\pm$0.02 & $>$21.4 &  &  & 1.00$\pm$0.04$^1$ \\
14 & 3:31:24.94 & -27:52:08.5 & 3:31:24.88 & -27:52:07.5 & 1.10 & 33.52 & 0.99 & WFI-R & 20.11$\pm$0.01 & $>$21.4 &  &  &  \\
15 & 3:31:25.31 & -27:59:59.4 & 3:31:25.29 & -27:59:58.7 & 0.60 & 19.22 & 0.99 & WFI-R & 22.96$\pm$0.02 & $>$21.4 &  &  & 0.79$\pm$0.08$^1$ \\
16 & 3:31:27.07 & -27:59:58.6 & 3:31:27.07 & -27:59:57.4 & 1.10 & 3.46 & 0.95 & WFI-R & 23.90$\pm$0.09 & $>$21.4 &  &  &  \\
17 & 3:31:27.07 & -27:44:10.4 & 3:31:27.06 & -27:44:09.5 & 0.80 & 14.83 & 0.99 & WFI-R & 23.10$\pm$0.02 & $>$21.4 &  &  & 0.55$\pm$0.00$^1$ \\
18 & 3:31:27.23 & -27:42:47.6 & 3:31:27.20 & -27:42:46.5 & 1.10 & 2.68 & 0.93 & WFI-R & 24.23$\pm$0.05 & $>$21.4 &  &  &  \\
19 & 3:31:27.58 & -27:44:39.5 & 3:31:27.55 & -27:44:39.1 & 0.60 & 0.30 & 0.60 & GEMS-z & $>$25.5 & $>$21.4 &  &  &  \\
20 & 3:31:28.60 & -27:49:35.6 & 3:31:28.60 & -27:49:34.9 & 0.70 & 25.43 & 0.99 & WFI-R & 22.77$\pm$0.01 & $>$21.4 & 0.846$^b$ & 2 & 0.83$\pm$0.03$^1$ \\
21 & 3:31:29.80 & -27:32:18.7 & 3:31:29.78 & -27:32:18.4 & 0.30 & 42.28 & 1.00 & WFI-R & 22.75$\pm$0.02 & $>$21.4 &  &  &  \\
22 & 3:31:29.88 & -27:57:22.7 & 3:31:29.84 & -27:57:22.5 & 0.50 & 0.37 & 0.65 & WFI-R & 26.11$\pm$0.29 & $>$21.4 &  &  &  \\
23 & 3:31:30.05 & -27:38:14.2 & 3:31:29.98 & -27:38:14.6 & 0.90 & 0.30 & 0.60 & WFI-R & 26.00$\pm$0.10$^\star$ & $>$21.4 &  &  &  \\
24 & 3:31:30.15 & -27:56:01.9 & 3:31:30.06 & -27:56:02.4 & 1.20 & 29.96 & 0.99 & WFI-R & 19.91$\pm$0.01 & $>$21.4 &  &  & 0.69$\pm$0.01$^1$ \\
25 & 3:31:30.79 & -27:57:34.7 & 3:31:30.73 & -27:57:34.9 & 0.70 & 11.12 & 0.98 & WFI-R & 23.86$\pm$0.04 & $>$21.4 &  &  & 1.06$\pm$0.17$^1$ \\
26 & 3:31:32.87 & -28:01:16.6 & 3:31:32.82 & -28:01:15.9 & 0.90 & 142.48 & 1.00 & WFI-R & 17.71$\pm$0.01 & $>$21.4 &  &  & 0.15$\pm$0.01$^1$ \\
27 & 3:31:34.26 & -27:38:28.7 & 3:31:34.21 & -27:38:28.5 & 0.50 & 27.83 & 0.99 & WFI-R & 22.93$\pm$0.03 & $>$21.4 &  &  & 1.08$\pm$0.06$^1$ \\
28 & 3:31:37.77 & -27:51:41.8 & 3:31:37.73 & -27:51:40.9 & 0.90 & 30.62 & 0.99 & WFI-R & 21.24$\pm$0.01 & $>$21.4 &  &  & 0.36$\pm$0.04$^1$ \\
29 & 3:31:38.47 & -27:59:41.9 & 3:31:38.67 & -27:59:41.1 & 2.80 & 0.01 & 0.01 & WFI-R & 25.20$\pm$0.12 & $>$21.4 &  &  &  \\
30 & 3:31:38.56 & -27:38:08.8 & 3:31:38.41 & -27:38:10.4 & 2.60 & 0.01 & 0.01 & WFI-R & 25.00$\pm$0.10 & $>$21.4 &  &  &  \\
31 & 3:31:39.52 & -27:41:19.1 & 3:31:39.50 & -27:41:19.6 & 0.60 & 19.20 & 0.99 & WFI-R & 23.04$\pm$0.02 & $>$21.4 &  &  &  \\
32 & 3:31:40.10 & -27:36:48.1 & 3:31:40.05 & -27:36:47.7 & 0.60 & 30.25 & 0.99 & WFI-R & 21.78$\pm$0.01 & $>$21.4 &  &  & 0.69$\pm$0.03$^1$ \\
33 & 3:31:43.24 & -27:54:05.1 & 3:31:43.22 & -27:54:05.5 & 0.50 & 2.40 & 0.92 & WFI-R & 25.15$\pm$0.12 & $>$21.4 &  &  &  \\
34 & 3:31:43.37 & -27:51:01.9 & 3:31:43.33 & -27:51:02.0 & 0.50 & 215.34 & 1.00 & WFI-R & 19.06$\pm$0.01 & $>$21.4 &  &  & 0.26$\pm$0.03$^1$ \\
35 & 3:31:43.42 & -27:42:48.9 & 3:31:43.43 & -27:42:48.6 & 0.20 & 64.55 & 1.00 & WFI-R & 21.34$\pm$0.01 & $>$21.4 & 0.466$^d$ & 2 & 0.47$\pm$0.01$^1$ \\
36 & 3:31:43.56 & -27:55:28.0 & 3:31:43.59 & -27:55:28.1 & 0.50 & 294.56 & 1.00 & WFI-R & 17.80$\pm$0.01 & 17.78$\pm$0.07 &  &  & 0.11$\pm$0.01$^1$ \\
37 & 3:31:44.05 & -27:38:35.0 & 3:31:43.96 & -27:38:35.2 & 1.10 & 5.28 & 0.96 & WFI-R & 23.41$\pm$0.03 & $>$21.4 &  &  & 0.06$\pm$0.04$^1$ \\
38 & 3:31:44.48 & -27:42:11.0 & 3:31:44.45 & -27:42:12.1 & 1.20 & 5.90 & 0.85 & WFI-R & 23.49$\pm$0.03 & 20.93$\pm$0.36 &  &  & 0.83$\pm$0.05$^1$ \\
39 & 3:31:46.00 & -27:51:30.0 & 3:31:45.97 & -27:51:30.1 & 0.40 & 39.28 & 0.99 & WFI-R & 22.25$\pm$0.01 & 19.50$\pm$0.05 & 0.681$^a$ & 0 & 0.65$\pm$0.04$^1$ \\
40 & 3:31:46.09 & -28:00:26.0 & 3:31:46.09 & -28:00:26.4 & 0.40 & 9.08 & 0.97 & WFI-R & 24.45$\pm$0.07 & $>$21.4 &  &  &  \\
41 & 3:31:46.56 & -27:57:34.0 & 3:31:46.57 & -27:57:34.7 & 0.80 & 83.38 & 1.00 & WFI-R & 20.07$\pm$0.01 & 17.96$\pm$0.03 &  &  & 0.36$\pm$0.02$^1$ \\
42 & 3:31:47.39 & -27:45:42.0 & 3:31:47.38 & -27:45:41.7 & 0.20 & 2.54 & 0.93 & WFI-R & 25.31$\pm$0.17 & 21.18$\pm$0.20 &  &  &  \\
43 & 3:31:47.66 & -27:50:14.0 & 3:31:47.66 & -27:50:13.7 & 0.30 & 64.36 & 1.00 & WFI-R & 21.25$\pm$0.01 & 18.98$\pm$0.03 &  &  & 0.54$\pm$0.02$^1$ \\
44 & 3:31:47.86 & -27:54:52.0 & 3:31:47.77 & -27:54:52.0 & 1.20 & 1.82 & 0.90 & WFI-R & 24.49$\pm$0.08 & 20.82$\pm$0.13 &  &  &  \\
45 & 3:31:49.70 & -27:43:25.6 & 3:31:49.69 & -27:43:26.3 & 0.80 & 60.74 & 1.00 & WFI-R & 20.80$\pm$0.01 & 18.49$\pm$0.02 & 0.618$^d$ & 2 & 0.60$\pm$0.02$^1$ \\
46 & 3:31:49.86 & -27:48:38.0 & 3:31:49.87 & -27:48:38.6 & 0.70 & 2.33 & 0.92 & GEMS-z & $>$25.5 & 19.13$\pm$0.02 &  &  &  \\
47 & 3:31:50.03 & -27:58:06.1 & 3:31:50.01 & -27:58:06.2 & 0.30 & 8.77 & 0.97 & WFI-R & 24.43$\pm$0.06 & $>$21.4 &  &  &  \\
48 & 3:31:50.08 & -27:39:47.2 & 3:31:50.15 & -27:39:48.1 & 1.40 & 1.00 & 0.83 & WFI-R & 23.71$\pm$0.05 & 20.11$\pm$0.09 &  &  & 1.13$\pm$0.12$^1$ \\
49 & 3:31:50.33 & -27:58:18.1 & 3:31:50.30 & -27:58:18.9 & 1.00 & 0.44 & 0.69 & WFI-R & 25.61$\pm$0.20 & $>$21.4 &  &  &  \\
50 & 3:31:50.35 & -27:41:19.1 & 3:31:50.33 & -27:41:20.2 & 1.10 & 25.47 & 0.99 & SOFI-K & 23.94$\pm$0.03 & 20.64$\pm$0.07 & 1.791$^d$ & 1 & 0.70$\pm$0.09$^1_\ast$ \\
51 & 3:31:50.78 & -27:47:03.1 & 3:31:50.78 & -27:47:03.5 & 0.50 & 21.99 & 0.99 & WFI-R & 23.60$\pm$0.03 & 21.07$\pm$0.13 &  &  &  \\
52 & 3:31:51.09 & -27:44:36.4 & 3:31:51.10 & -27:44:37.3 & 1.00 & 7.65 & 0.97 & WFI-R & 23.66$\pm$0.04 & 20.37$\pm$0.08 &  &  & 0.85$\pm$0.19$^1$ \\
53 & 3:31:51.32 & -27:50:56.0 & 3:31:51.28 & -27:50:56.0 & 0.50 & 79.54 & 1.00 & SOFI-K & 24.26$\pm$0.07 & 20.64$\pm$0.20 &  &  &  \\
54 & 3:31:51.94 & -27:53:26.1 & 3:31:51.94 & -27:53:27.0 & 1.00 & 1.41 & 0.88 & WFI-R & 24.74$\pm$0.10 & 22.24$\pm$0.42 & 2.940$^a$ & 2 &  \\
55 & 3:31:52.08 & -27:43:21.1 & 3:31:52.00 & -27:43:21.3 & 1.00 & 10.66 & 0.97 & WFI-R & 22.94$\pm$0.03 & 22.25$\pm$0.22 & 1.501$^d$ & 2 &  \\
56 & 3:31:52.10 & -27:39:25.1 & 3:31:52.16 & -27:39:26.6 & 1.60 & 1.10 & 0.85 & SOFI-K & 24.55$\pm$0.06 & 21.81$\pm$0.23 &  &  &  \\
57 & 3:31:52.82 & -27:44:29.1 & 3:31:52.77 & -27:44:30.4 & 1.50 & 0.14 & 0.40 & WFI-R & 25.19$\pm$0.14 & $>$21.4 &  &  &  \\
58 & 3:31:53.41 & -28:02:21.0 & 3:31:53.42 & -28:02:21.1 & 0.30 & 15.33 & 0.99 & WFI-R & 23.96$\pm$0.09 & $>$21.4 &  &  & 0.96$\pm$0.11$^1$ \\
59 & 3:31:54.08 & -27:50:03.9 & 3:31:54.05 & -27:50:05.6 & 1.80 & 3.08 & 0.94 & WFI-R & 21.85$\pm$0.01 & 19.35$\pm$0.05 &  &  & 0.71$\pm$0.04$^1$ \\
60 & 3:31:54.90 & -27:53:40.5 & 3:31:54.75 & -27:53:41.3 & 2.00 & 0.09 & 0.32 & WFI-R & 24.46$\pm$0.10 & 21.67$\pm$0.34 &  &  &  \\
61 & 3:31:55.00 & -27:44:10.1 & 3:31:54.98 & -27:44:10.5 & 0.50 & 15.21 & 0.98 & WFI-R & 23.93$\pm$0.05 & 20.02$\pm$0.06 & 0.758$^e$ & 1 &  \\
62 & 3:31:56.31 & -27:40:00.5 & 3:31:56.29 & -27:40:00.6 & 0.20 & 41.88 & 1.00 & SOFI-K & $>$25.5 & 21.78$\pm$0.21 &  &  &  \\
63 & 3:31:56.45 & -27:52:36.1 & 3:31:56.43 & -27:52:36.6 & 0.60 & 12.51 & 0.98 & WFI-R & 23.86$\pm$0.04 & 20.06$\pm$0.08 &  &  & 0.77$\pm$0.08$^1$ \\
64 & 3:31:56.77 & -27:52:25.2 & 3:31:56.71 & -27:52:25.5 & 0.90 & 32.09 & 0.99 & SOFI-K & $>$25.5 & 21.13$\pm$0.11 &  &  &  \\
65 & 3:31:56.97 & -27:59:39.2 & 3:31:56.87 & -27:59:39.2 & 1.20 & 0.41 & 0.67 & WFI-R & 25.22$\pm$0.12 & $>$21.4 &  &  &  \\
66 & 3:31:57.80 & -27:42:07.4 & 3:31:57.79 & -27:42:08.7 & 1.40 & 4.44 & 0.96 & WFI-R & 22.32$\pm$0.01 & 19.34$\pm$0.04 & 0.665$^d$ & 2 & 0.65$\pm$0.03$^1$ \\
67 & 3:31:58.21 & -27:38:15.2 & 3:31:58.20 & -27:38:16.5 & 1.40 & 2.37 & 0.86 & WFI-R & 23.91$\pm$0.04 & $>$21.4 &  &  &  \\
68 & 3:31:58.29 & -27:50:41.4 & 3:31:58.27 & -27:50:41.9 & 0.60 & 1.18 & 0.83 & WFI-R & 25.16$\pm$0.08 & 21.40$\pm$0.25 &  &  & 1.99$\pm$0.23$^3$ \\
69 & 3:31:58.93 & -27:43:03.2 & 3:31:58.91 & -27:43:04.1 & 1.10 & 33.20 & 0.99 & ACS-i & 23.19$\pm$0.03 & 20.15$\pm$0.08 &  &  & 0.81$\pm$0.08$^1$ \\
70 & 3:31:59.82 & -27:45:40.3 & 3:31:59.85 & -27:45:40.3 & 0.00 & 10.00 & 0.90 & SPITZER & $>$27.5 & $>$24.7 &  &  &  \\
71 & 3:32:00.84 & -27:35:56.4 & 3:32:00.84 & -27:35:56.8 & 0.50 & 32.57 & 0.96 & WFI-R & 22.90$\pm$0.02 & $>$21.4 &  &  & 0.95$\pm$0.04$^1$ \\
72 & 3:32:01.29 & -27:53:36.4 & 3:32:01.29 & -27:53:36.5 & 0.10 & 12.97 & 0.98 & SOFI-K & $>$25.5 & 22.35$\pm$0.29 &  &  &  \\
73 & 3:32:01.47 & -27:46:47.9 & 3:32:01.42 & -27:46:47.0 & 1.10 & 35.88 & 0.99 & SOFI-K & $>$25.5 & 19.95$\pm$0.09 &  &  &  \\
74 & 3:32:02.86 & -27:56:12.3 & 3:32:02.81 & -27:56:12.8 & 0.80 & 31.56 & 0.99 & SOFI-K & $>$25.5 & 21.36$\pm$0.19 &  &  &  \\
75 & 3:32:03.34 & -27:53:14.3 & 3:32:03.22 & -27:53:12.8 & 2.20 & 0.22 & 0.52 & SOFI-K & 24.95$\pm$0.12 & 21.91$\pm$0.33 &  &  &  \\
76 & 3:32:03.65 & -27:46:04.0 & 3:32:03.65 & -27:46:03.7 & 0.30 & 324.86 & 1.00 & ACS-i & 20.99$\pm$0.01 & 18.51$\pm$0.02 & 0.574$^a$ & 2 & 0.59$\pm$0.00$^1$ \\
77 & 3:32:03.85 & -27:58:04.4 & 3:32:03.87 & -27:58:05.4 & 1.20 & 0.66 & 0.77 & WFI-R & 24.98$\pm$0.14 & $>$21.4 &  &  &  \\
78 & 3:32:04.67 & -28:00:57.3 & 3:32:04.69 & -28:00:57.7 & 0.50 & 6.56 & 0.90 & GEMS-z & 24.34$\pm$0.06 & $>$21.4 &  &  &  \\
79 & 3:32:04.80 & -27:46:48.3 & 3:32:04.84 & -27:46:47.6 & 0.90 & 2.06 & 0.87 & ACS-i & 26.10$\pm$0.10 & 21.26$\pm$0.19 &  &  &  \\
80 & 3:32:06.12 & -27:32:36.1 & 3:32:06.09 & -27:32:35.5 & 0.60 & 32.17 & 0.99 & WFI-R & 22.44$\pm$0.02 & $>$21.4 &  &  &  \\
81 & 3:32:06.34 & -27:56:26.3 & 3:32:06.32 & -27:56:26.4 & 0.30 & 75.80 & 1.00 & SOFI-K & $>$25.5 & 20.96$\pm$0.14 &  &  & 0.50$\pm$0.01$^1$ \\
82 & 3:32:06.45 & -27:47:29.3 & 3:32:06.44 & -27:47:28.8 & 0.40 & 182.38 & 1.00 & ACS-i & 22.71$\pm$0.01 & 19.56$\pm$0.03 & 1.021$^b$ & 2 & 1.00$\pm$0.04$^1$ \\
83 & 3:32:07.27 & -27:51:21.2 & 3:32:07.32 & -27:51:20.4 & 1.10 & 4.31 & 0.95 & WFI-R & 23.98$\pm$0.03 & 20.81$\pm$0.07 & 0.034$^d$ & 1 & 0.80$\pm$0.18$^1$ \\
84 & 3:32:08.56 & -27:46:48.4 & 3:32:08.53 & -27:46:48.3 & 0.20 & 1422.87 & 1.00 & ACS-i & 19.45$\pm$0.01 & 17.06$\pm$0.01 & 0.310$^a$ & 2 & 0.33$\pm$0.00$^1$ \\
85 & 3:32:08.70 & -27:47:34.4 & 3:32:08.66 & -27:47:34.4 & 0.40 & 887.95 & 1.00 & ACS-i & 18.98$\pm$0.01 & 17.59$\pm$0.01 & 0.544$^a$ & 2 & 0.58$\pm$0.02$^1$ \\
86 & 3:32:09.70 & -27:42:48.2 & 3:32:09.71 & -27:42:48.1 & 0.20 & 397.46 & 1.00 & ACS-i & 21.96$\pm$0.01 & 19.71$\pm$0.04 & 0.733$^a$ & 2 & 0.76$\pm$0.01$^1$ \\
87 & 3:32:09.90 & -27:50:15.7 & 3:32:09.85 & -27:50:15.5 & 1.00 & 10.00 & 0.90 & SPITZER & $>$27.5 & $>$24.7 &  &  &  \\
88 & 3:32:10.16 & -27:59:38.4 & 3:32:10.14 & -27:59:38.3 & 0.10 & 77.83 & 1.00 & WFI-R & 21.27$\pm$0.01 & $>$21.4 &  &  & 0.65$\pm$0.02$^1$ \\
89 & 3:32:10.73 & -27:48:07.4 & 3:32:10.72 & -27:48:07.1 & 0.20 & 125.28 & 1.00 & ACS-i & 23.04$\pm$0.03 & 19.97$\pm$0.01 & 0.653$^b$ & 2 & 0.65$\pm$0.08$^2$ \\
90 & 3:32:10.81 & -27:59:26.4 & 3:32:10.82 & -27:59:26.8 & 0.50 & 104.84 & 1.00 & WFI-R & 20.85$\pm$0.01 & $>$21.4 & 0.651$^g$ & 1 & 0.66$\pm$0.01$^1$ \\
91 & 3:32:10.82 & -27:46:28.2 & 3:32:10.79 & -27:46:27.8 & 0.30 & 34.13 & 0.99 & ACS-i & 24.59$\pm$0.08 & 20.01$\pm$0.02 & 1.610$^f$ & 2 & 1.61$\pm$0.08$^2$ \\
92 & 3:32:10.91 & -27:44:15.2 & 3:32:10.91 & -27:44:14.9 & 0.20 & 150.27 & 1.00 & ACS-i & 22.50$\pm$0.01 & 20.75$\pm$0.01 & 1.615$^a$ & 2 & 1.62$\pm$0.08$^2$ \\
93 & 3:32:11.00 & -27:40:53.3 & 3:32:10.99 & -27:40:53.7 & 0.50 & 31.37 & 0.99 & ACS-i & 24.60$\pm$0.10 & 21.35$\pm$0.18 & 0.181$^a$ & 2 &  \\
94 & 3:32:11.50 & -27:48:16.1 & 3:32:11.49 & -27:48:15.6 & 0.30 & 627.69 & 1.00 & ACS-i & 20.42$\pm$0.01 & 18.22$\pm$0.01 & 0.547$^b$ & 2 & 0.55$\pm$0.08$^2$ \\
95 & 3:32:11.53 & -27:47:13.5 & 3:32:11.51 & -27:47:13.1 & 0.30 & 282.46 & 1.00 & ACS-i & 22.02$\pm$0.01 & 19.44$\pm$0.01 & 0.576$^f$ & 2 & 0.58$\pm$0.08$^2$ \\
96 & 3:32:11.67 & -27:37:26.3 & 3:32:11.64 & -27:37:26.0 & 0.30 & 406.92 & 1.00 & WFI-R & 18.88$\pm$0.01 & $>$21.4 &  &  & 1.57$\pm$0.09$^1$ \\
97 & 3:32:13.09 & -27:43:50.7  &  &  &  &  &  &  &  &  &  &  &   \\
98 & 3:32:13.19 & -27:57:44.4 & 3:32:13.15 & -27:57:44.5 & 0.50 & 8.26 & 0.86 & WFI-R & 24.36$\pm$0.04 & $>$21.4 &  &  & 1.05$\pm$0.17$^1$ \\
99 & 3:32:13.26 & -27:42:40.9 & 3:32:13.25 & -27:42:40.9 & 0.10 & 667.01 & 1.00 & ACS-i & 20.40$\pm$0.01 & 18.54$\pm$0.01 & 0.605$^a$ & 2 & 0.61$\pm$0.08$^2$ \\
100 & 3:32:13.31 & -27:39:34.1 & 3:32:13.36 & -27:39:34.8 & 1.00 & 10.00 & 0.90 & SPITZER & $>$25.5 & $>$21.4 &  &  &  \\
101 & 3:32:13.52 & -27:49:52.5 & 3:32:13.48 & -27:49:52.8 & 0.60 & 63.55 & 0.82 & ACS-i & 22.78$\pm$0.02 & 19.80$\pm$0.01 &  &  & 0.05$\pm$0.08$^2$ \\
102 & 3:32:14.14 & -27:49:10.4 & 3:32:14.13 & -27:49:10.1 & 0.20 & 22.03 & 0.99 & ACS-i & 24.44$\pm$0.06 & 22.47$\pm$0.05 & 2.076$^b$ & 2 & 2.18$\pm$0.08$^2$ \\
103 & 3:32:14.80 & -27:56:40.3 & 3:32:14.82 & -27:56:40.5 & 0.40 & 119.04 & 1.00 & WFI-R & 20.90$\pm$0.01 & 17.81$\pm$0.02 & 0.733$^d$ & 2 & 0.73$\pm$0.03$^1$ \\
104 & 3:32:15.37 & -27:37:06.4 & 3:32:15.41 & -27:37:06.3 & 0.70 & 3.61 & 0.95 & WFI-R & 24.63$\pm$0.11 & $>$21.4 &  &  &  \\
105 & 3:32:15.39 & -27:50:37.5 & 3:32:15.30 & -27:50:37.5 & 1.00 & 10.00 & 0.90 & SPITZER & $>$27.5 & $>$24.7 &  &  &  \\
106 & 3:32:15.96 & -27:34:38.5 & 3:32:15.96 & -27:34:38.2 & 0.30 & 39.53 & 0.99 & WFI-R & 23.01$\pm$0.03 & $>$21.4 &  &  & 0.76$\pm$0.01$^1$ \\
107 & 3:32:17.06 & -27:58:46.4 & 3:32:17.10 & -27:58:47.4 & 1.20 & 0.90 & 0.82 & WFI-R & 24.59$\pm$0.10 & $>$21.4 &  &  &  \\
108 & 3:32:17.09 & -27:43:03.4 & 3:32:17.14 & -27:43:03.3 & 0.90 & 150.78 & 1.00 & ACS-i & 21.29$\pm$0.01 & 19.48$\pm$0.01 & 0.569$^a$ & 2 & 0.57$\pm$0.08$^2$ \\
109 & 3:32:17.13 & -27:59:17.3 & 3:32:17.06 & -27:59:16.6 & 1.10 & 76.23 & 1.00 & WFI-R & 18.10$\pm$0.01 & $>$21.4 &  &  & 0.14$\pm$0.00$^1$ \\
110 & 3:32:17.22 & -27:52:21.3 & 3:32:17.17 & -27:52:20.8 & 0.60 & 80.24 & 0.99 & ACS-i & 23.57$\pm$0.03 & 20.10$\pm$0.01 & 1.097$^a$ & 2 & 1.10$\pm$0.08$^2$ \\
111 & 3:32:17.89 & -27:50:59.4 & 3:32:17.87 & -27:50:59.3 & 0.10 & 510.88 & 1.00 & ACS-i & 18.92$\pm$0.01 & 16.97$\pm$0.01 & 0.124$^e$ & 2 & 0.12$\pm$0.08$^2$ \\
112 & 3:32:18.05 & -27:47:19.3 & 3:32:18.02 & -27:47:18.5 & 0.70 & 379.13 & 1.00 & ACS-i & 21.08$\pm$0.01 & 17.77$\pm$0.01 & 0.734$^a$ & 2 & 0.73$\pm$0.08$^2$ \\
113 & 3:32:19.17 & -27:54:07.7 & 3:32:19.29 & -27:54:06.1 & 2.20 & 0.39 & 0.01 & ACS-i & 22.17$\pm$0.01 & 18.58$\pm$0.01 & 0.964$^a$ & 2 & 1.12$\pm$0.08$^2$ \\
114 & 3:32:19.42 & -27:52:17.8 & 3:32:19.52 & -27:52:17.7 & 1.40 & 4.76 & 0.96 & ACS-i & 23.70$\pm$0.04 & 20.10$\pm$0.01 &  &  & 1.00$\pm$0.08$^2$ \\
115 & 3:32:19.79 & -27:41:22.3 & 3:32:19.81 & -27:41:22.7 & 0.70 & 680.55 & 1.00 & ACS-i & 19.41$\pm$0.01 & 17.94$\pm$0.01 & 0.229$^a$ & 2 & 0.23$\pm$0.08$^2$ \\
116 & 3:32:19.91 & -27:57:21.3 & 3:32:19.90 & -27:57:20.4 & 0.80 & 16.66 & 0.99 & WFI-R & 22.72$\pm$0.02 & 19.70$\pm$0.10 &  &  & 0.70$\pm$0.03$^1$ \\
117 & 3:32:20.24 & -27:52:22.5 & 3:32:20.26 & -27:52:22.0 & 0.50 & 317.01 & 1.00 & ACS-i & 20.63$\pm$0.01 & 18.73$\pm$0.01 & 0.343$^a$ & 2 & 0.46$\pm$0.08$^2$ \\
118 & 3:32:21.00 & -27:47:06.3 & 3:32:20.92 & -27:47:05.4 & 1.20 & 83.16 & 0.91 & ACS-i & 20.94$\pm$0.01 & 18.81$\pm$0.01 & 0.670$^b$ & 2 & 0.67$\pm$0.08$^2$ \\
119 & 3:32:21.08 & -27:35:30.4 & 3:32:21.08 & -27:35:30.0 & 0.30 & 50.82 & 1.00 & WFI-R & 22.14$\pm$0.01 & $>$21.4 &  &  & 0.71$\pm$0.03$^1$ \\
120 & 3:32:21.29 & -27:44:35.6 & 3:32:21.28 & -27:44:35.6 & 0.20 & 443.73 & 0.86 & ACS-i & 20.53$\pm$0.01 & 18.22$\pm$0.01 & 0.524$^b$ & 2 & 0.52$\pm$0.08$^2$ \\
121 & 3:32:22.04 & -27:42:44.0 & 3:32:22.01 & -27:42:43.3 & 0.60 & 10.94 & 0.79 & ACS-i & 25.56$\pm$0.11 & 21.61$\pm$0.04 &  &  & 2.12$\pm$0.08$^2$ \\
122 & 3:32:22.49 & -27:48:05.4 & 3:32:22.50 & -27:48:04.7 & 0.50 & 9.37 & 0.98 & ISAAC-K & 25.07$\pm$0.14 & 22.77$\pm$0.07 &  &  &  \\
123 & 3:32:22.52 & -27:49:34.1 & 3:32:22.48 & -27:49:35.0 & 1.10 & 39.15 & 0.99 & ACS-i & 22.78$\pm$0.02 & 19.57$\pm$0.01 & 0.731$^b$ & 2 & 0.67$\pm$0.08$^2$ \\
124 & 3:32:22.58 & -28:00:24.3 & 3:32:22.59 & -28:00:23.4 & 0.80 & 139.35 & 1.00 & WFI-R & 17.47$\pm$0.01 & $>$21.4 &  &  & 0.13$\pm$0.01$^1$ \\
125 & 3:32:22.59 & -27:44:26.3 & 3:32:22.59 & -27:44:25.8 & 0.30 & 424.99 & 1.00 & ACS-i & 21.34$\pm$0.01 & 19.01$\pm$0.01 & 0.738$^b$ & 2 & 0.74$\pm$0.08$^2$ \\
126 & 3:32:22.68 & -27:41:26.1 & 3:32:22.69 & -27:41:26.2 & 0.40 & 17.92 & 0.96 & ISAAC-K & 27.34$\pm$0.28 & 22.56$\pm$0.08 &  &  & 1.71$\pm$0.08$^2$ \\
127 & 3:32:23.71 & -27:36:48.5 & 3:32:23.67 & -27:36:48.2 & 0.40 & 301.36 & 1.00 & WFI-R & 17.71$\pm$0.01 & $>$21.4 &  &  & 0.12$\pm$0.01$^1$ \\
128 & 3:32:23.84 & -27:58:45.5 & 3:32:23.79 & -27:58:45.6 & 0.50 & 316.19 & 1.00 & WFI-R & 18.76$\pm$0.01 & $>$21.4 & 0.122$^d$ & 2 & 0.12$\pm$0.01$^1$ \\
129 & 3:32:24.32 & -28:01:14.3 & 3:32:24.29 & -28:01:14.3 & 0.30 & 74.27 & 1.00 & WFI-R & 21.16$\pm$0.01 & $>$21.4 &  &  & 0.54$\pm$0.02$^1$ \\
130 & 3:32:24.60 & -27:54:42.3 & 3:32:24.53 & -27:54:43.0 & 1.10 & 313.80 & 1.00 & ACS-i & 19.45$\pm$0.01 & 18.17$\pm$0.01 & 0.126$^a$ & 1 & 0.12$\pm$0.08$^2$ \\
131 & 3:32:25.07 & -27:38:22.4 & 3:32:25.06 & -27:38:22.6 & 0.20 & 23.48 & 0.99 & WFI-R & 23.63$\pm$0.03 & $>$21.4 &  &  & 0.61$\pm$0.08$^1$ \\
132 & 3:32:25.10 & -27:54:50.4 & 3:32:25.16 & -27:54:50.1 & 1.00 & 43.66 & 1.00 & ACS-i & 23.17$\pm$0.03 & 19.63$\pm$0.01 & 1.090$^a$ & 2 & 1.09$\pm$0.08$^2$ \\
133 & 3:32:25.22 & -27:42:18.5 & 3:32:25.17 & -27:42:18.8 & 0.70 & 71.55 & 1.00 & ACS-i & 23.21$\pm$0.02 & 20.68$\pm$0.01 & 1.617$^a$ & 2 & 1.62$\pm$0.08$^2$ \\
134 & 3:32:26.95 & -27:41:06.5 & 3:32:27.01 & -27:41:05.0 & 1.60 & 82.27 & 1.00 & ACS-i & 19.28$\pm$0.01 & 18.19$\pm$0.01 & 0.734$^a$ & 2 & 0.73$\pm$0.08$^2$ \\
135 & 3:32:27.79 & -27:37:52.1 & 3:32:27.80 & -27:37:49.7 & 2.30 & 1.33 & 0.87 & SOFI-K & 23.39$\pm$0.03 & $>$21.4 &  &  & 1.12$\pm$0.22$^1$ \\
136 & 3:32:27.97 & -27:46:39.4 & 3:32:27.99 & -27:46:39.2 & 0.40 & 800.59 & 1.00 & ACS-i & 20.18$\pm$0.01 & 17.97$\pm$0.01 & 0.248$^b$ & 2 & 0.25$\pm$0.08$^2$ \\
137 & 3:32:28.38 & -27:38:41.7 & 3:32:28.35 & -27:38:41.7 & 0.30 & 652.96 & 1.00 & ACS-i & 20.15$\pm$0.01 & 18.18$\pm$0.03 &  &  & 0.39$\pm$0.01$^1$ \\
138 & 3:32:28.46 & -27:58:09.9 & 3:32:28.40 & -27:58:10.1 & 0.70 & 115.37 & 1.00 & WFI-R & 19.18$\pm$0.01 & $>$21.4 & 0.212$^d$ & 2 & 0.17$\pm$0.04$^1$ \\
139 & 3:32:28.75 & -27:46:21.3 & 3:32:28.74 & -27:46:20.4 & 0.70 & 128.03 & 1.00 & ACS-i & 22.37$\pm$0.01 & 19.66$\pm$0.01 & 0.738$^a$ & 2 & 0.74$\pm$0.08$^2$ \\
140 & 3:32:28.83 & -27:43:56.4 & 3:32:28.82 & -27:43:55.6 & 0.60 & 235.10 & 1.00 & ACS-i & 18.44$\pm$0.01 & 17.18$\pm$0.01 & 0.215$^a$ & 2 & 0.22$\pm$0.08$^2$ \\
141 & 3:32:29.88 & -27:44:24.5 & 3:32:29.88 & -27:44:24.4 & 0.20 & 685.54 & 1.00 & ACS-i & 16.74$\pm$0.01 & 15.93$\pm$0.01 & 0.076$^a$ & 2 & 0.08$\pm$0.08$^2$ \\
142 & 3:32:29.97 & -27:44:05.2 & 3:32:29.99 & -27:44:04.8 & 0.50 & 316.93 & 1.00 & ACS-i & 17.24$\pm$0.01 & 16.31$\pm$0.01 & 0.076$^a$ & 2 & 0.08$\pm$0.08$^2$ \\
143 & 3:32:30.57 & -27:59:11.7 & 3:32:30.55 & -27:59:11.5 & 0.20 & 329.15 & 1.00 & WFI-R & 17.91$\pm$0.01 & $>$21.4 &  &  & 0.13$\pm$0.01$^1$ \\
144 & 3:32:31.46 & -27:58:52.0 & 3:32:31.42 & -27:58:51.7 & 0.60 & 10.00 & 0.90 & SOFI-K & $>$25.5 & $>$21.4 &  &  &  \\
145 & 3:32:31.47 & -27:46:23.3 & 3:32:31.46 & -27:46:23.2 & 0.00 & 46.41 & 0.98 & ACS-i & 23.53$\pm$0.03 & 20.80$\pm$0.01 & 2.223$^a$ & 2 & 2.22$\pm$0.08$^2$ \\
146 & 3:32:31.54 & -27:50:29.3 & 3:32:31.55 & -27:50:28.8 & 0.50 & 19.60 & 0.99 & ACS-i & 25.02$\pm$0.13 & 20.75$\pm$0.01 & 1.613$^c$ & 2 & 1.56$\pm$0.08$^2$ \\
147 & 3:32:31.58 & -28:04:35.7 & 3:32:31.25 & -28:04:36.1 & 4.40 & 0.04 & 0.17 & WFI-R & 23.12$\pm$0.03 & $>$21.4 &  &  &  \\
148 & 3:32:32.00 & -28:03:10.2 & 3:32:32.00 & -28:03:09.8 & 0.30 & 82.51 & 1.00 & WFI-R & 19.92$\pm$0.01 & $>$21.4 &  &  & 1.97$\pm$0.01$^1$ \\
149 & 3:32:32.83 & -27:46:08.4 & 3:32:32.81 & -27:46:07.8 & 0.40 & 363.82 & 1.00 & ACS-i & 20.80$\pm$0.01 & 18.74$\pm$0.01 & 0.361$^b$ & 2 & 0.37$\pm$0.08$^2$ \\
150 & 3:32:33.03 & -27:50:30.5 & 3:32:33.00 & -27:50:30.2 & 0.30 & 175.37 & 1.00 & ACS-i & 22.04$\pm$0.01 & 19.28$\pm$0.01 & 0.669$^c$ & 2 & 0.67$\pm$0.08$^2$ \\
151 & 3:32:33.45 & -27:52:28.4 & 3:32:33.44 & -27:52:28.1 & 0.10 & 15.20 & 0.99 & ISAAC-K & $>$27.5 & 23.05$\pm$0.08 &  &  & 3.62$\pm$0.08$^2$ \\
152 & 3:32:34.78 & -27:39:57.2 & 3:32:34.74 & -27:39:57.5 & 0.50 & 588.91 & 1.00 & ACS-i & 19.17$\pm$0.01 & 17.61$\pm$0.01 & 0.214$^d$ & 2 & 0.19$\pm$0.01$^1$ \\
153 & 3:32:34.87 & -28:00:46.7 & 3:32:34.89 & -28:00:45.7 & 1.00 & 31.73 & 0.99 & WFI-R & 20.96$\pm$0.01 & $>$21.4 &  &  & 0.68$\pm$0.03$^1$ \\
154 & 3:32:34.94 & -27:55:33.1 & 3:32:35.08 & -27:55:33.0 & 2.10 & 38.31 & 0.99 & ACS-i & 16.38$\pm$0.01 & 14.82$\pm$0.01 & 0.038$^a$ & 2 & 0.04$\pm$0.08$^2$ \\
155 & 3:32:35.02 & -27:54:55.2  &  &  &  &  &  &  &  &  &  &  &   \\
156 & 3:32:35.75 & -27:49:16.4 & 3:32:35.71 & -27:49:16.0 & 0.40 & 13.89 & 0.99 & ACS-i & 24.96$\pm$0.11 & 21.85$\pm$0.02 & 2.579$^b$ & 1 & 2.58$\pm$0.08$^2$ \\
157 & 3:32:36.20 & -27:49:32.5 & 3:32:36.17 & -27:49:31.8 & 0.50 & 318.14 & 1.00 & ACS-i & 21.19$\pm$0.01 & 18.92$\pm$0.01 & 0.547$^a$ & 2 & 0.55$\pm$0.08$^2$ \\
158 & 3:32:36.51 & -27:34:53.8 & 3:32:36.53 & -27:34:53.5 & 0.40 & 11.06 & 0.98 & WFI-R & 23.99$\pm$0.05 & $>$21.4 &  &  &  \\
159 & 3:32:37.24 & -27:57:49.2 & 3:32:37.22 & -27:57:48.1 & 1.00 & 104.07 & 1.00 & WFI-R & 18.43$\pm$0.01 & 17.20$\pm$0.03 &  &  & 0.13$\pm$0.01$^1$ \\
160 & 3:32:37.29 & -27:51:27.4 & 3:32:37.34 & -27:51:27.4 & 0.90 & 75.51 & 0.81 & ACS-i & 21.59$\pm$0.01 & 19.71$\pm$0.01 &  &  & 0.52$\pm$0.08$^2$ \\
161 & 3:32:37.79 & -27:50:00.4 & 3:32:37.74 & -27:50:00.4 & 0.60 & 22.61 & 0.77 & ACS-i & 23.95$\pm$0.05 & 21.24$\pm$0.02 & 1.619$^c$ & 2 & 1.62$\pm$0.08$^2$ \\
162 & 3:32:37.80 & -27:52:12.4 & 3:32:37.76 & -27:52:12.3 & 0.30 & 28.48 & 0.98 & ACS-i & 24.68$\pm$0.05 & 20.60$\pm$0.01 & 1.603$^a$ & 2 & 1.60$\pm$0.08$^2$ \\
163 & 3:32:38.79 & -27:49:56.5 & 3:32:38.82 & -27:49:56.3 & 0.50 & 453.77 & 1.00 & ACS-i & 20.39$\pm$0.01 & 18.39$\pm$0.01 & 0.242$^a$ & 2 & 0.24$\pm$0.08$^2$ \\
164 & 3:32:38.80 & -27:44:49.2 & 3:32:38.79 & -27:44:48.9 & 0.10 & 111.16 & 1.00 & ACS-i & 22.92$\pm$0.01 & 19.98$\pm$0.01 & 0.736$^d$ & 2 & 0.59$\pm$0.08$^2$ \\
165 & 3:32:38.94 & -27:57:00.4 & 3:32:38.94 & -27:57:00.7 & 0.40 & 281.10 & 1.00 & WFI-R & 18.96$\pm$0.01 & 17.33$\pm$0.02 & 0.298$^d$ & 2 & 0.30$\pm$0.00$^1$ \\
166 & 3:32:38.95 & -27:59:19.4 & 3:32:38.88 & -27:59:18.7 & 1.10 & 10.67 & 0.98 & WFI-R & 22.35$\pm$0.01 & $>$21.4 & 0.652$^e$ & 1 &  \\
167 & 3:32:39.19 & -27:53:57.4 & 3:32:39.19 & -27:53:57.7 & 0.50 & 185.08 & 1.00 & ACS-i & 21.94$\pm$0.01 & 20.05$\pm$0.01 &  &  & 0.61$\pm$0.08$^2$ \\
168 & 3:32:39.47 & -27:54:11.3 & 3:32:39.41 & -27:54:11.8 & 0.90 & 15.64 & 0.99 & ACS-i & 23.72$\pm$0.07 & 19.98$\pm$0.01 & 1.295$^c$ & 2 & 1.29$\pm$0.08$^2$ \\
169 & 3:32:39.47 & -27:53:01.3 & 3:32:39.48 & -27:53:01.6 & 0.60 & 230.83 & 0.54 & ACS-i & 21.25$\pm$0.01 & 18.91$\pm$0.01 & 0.686$^c$ & 2 & 0.69$\pm$0.08$^2$ \\
170 & 3:32:39.68 & -27:48:51.4 & 3:32:39.67 & -27:48:50.6 & 0.60 & 13.19 & 0.86 & ACS-i & 25.18$\pm$0.09 & 21.61$\pm$0.03 & 3.064$^a$ & 2 & 3.04$\pm$0.08$^2$ \\
171 & 3:32:39.74 & -28:03:11.9 & 3:32:39.70 & -28:03:11.4 & 0.80 & 10.00 & 0.90 & WFI-R & 24.00$\pm$0.01 & $>$21.4 &  &  &  \\
172 & 3:32:40.76 & -27:55:05.3 & 3:32:40.76 & -27:55:05.4 & 0.30 & 219.30 & 1.00 & ACS-i & 22.27$\pm$0.01 & 19.63$\pm$0.05 & 0.978$^c$ & 2 & 0.98$\pm$0.08$^2$ \\
173 & 3:32:40.87 & -27:55:46.2 & 3:32:40.84 & -27:55:46.7 & 0.70 & 2.55 & 0.93 & ACS-i & 25.60$\pm$0.13 & 21.38$\pm$0.26 & 0.625$^a$ & 1 & 0.62$\pm$0.08$^2$ \\
174 & 3:32:41.65 & -28:01:28.3 & 3:32:41.61 & -28:01:27.7 & 0.60 & 23.22 & 0.99 & WFI-R & 23.13$\pm$0.02 & $>$21.4 &  &  & 0.61$\pm$0.07$^1$ \\
175 & 3:32:42.00 & -27:39:49.3 & 3:32:41.99 & -27:39:49.3 & 0.10 & 309.27 & 1.00 & WFI-R & 19.01$\pm$0.01 & 17.28$\pm$0.01 & 0.152$^b$ & 2 & 0.15$\pm$0.01$^1$ \\
176 & 3:32:42.57 & -27:38:16.4 & 3:32:42.63 & -27:38:16.0 & 0.90 & 0.52 & 0.72 & GEMS-z & $>$25.5 & $>$21.4 &  &  &  \\
177 & 3:32:43.16 & -27:55:14.3 & 3:32:43.17 & -27:55:14.7 & 0.60 & 19.16 & 0.98 & ACS-i & 23.83$\pm$0.05 & 21.44$\pm$0.31 & 0.579$^d$ & 1 &  \\
178A & 3:32:44.20 & -27:51:42.1 & 3:32:44.27 & -27:51:41.1 & 1.30 & 78.66 & 0.99 & ACS-i & 19.99$\pm$0.01 & 18.09$\pm$0.01 & 0.279$^a$ & 2 & 0.28$\pm$0.08$^2$ \\
178B & 3:32:44.20 & -27:51:42.1 & 3:32:44.05 & -27:51:43.3 & 2.20 & 0.31 & 0.00 & ACS-i & 19.34$\pm$0.01 & 17.12$\pm$0.01 & 0.279$^a$ & 2 & 0.28$\pm$0.08$^2$ \\
179 & 3:32:44.22 & -27:57:45.2 & 3:32:44.23 & -27:57:44.5 & 0.60 & 19.43 & 0.99 & WFI-R & 23.12$\pm$0.03 & 20.19$\pm$0.26 &  &  & 1.09$\pm$0.04$^1$ \\
180 & 3:32:44.88 & -27:47:28.0 & 3:32:44.87 & -27:47:27.6 & 0.30 & 396.27 & 1.00 & ACS-i & 19.07$\pm$0.01 & 17.57$\pm$0.01 &  &  & 0.19$\pm$0.03$^1$ \\
181 & 3:32:44.98 & -27:54:39.3 & 3:32:45.02 & -27:54:39.6 & 0.80 & 818.33 & 1.00 & ACS-i & 19.69$\pm$0.01 & 18.00$\pm$0.01 & 0.459$^a$ & 2 & 0.46$\pm$0.08$^2$ \\
182 & 3:32:45.38 & -28:04:50.3 & 3:32:45.34 & -28:04:49.8 & 0.60 & 9.06 & 0.97 & WFI-R & 24.17$\pm$0.09 & $>$21.4 &  &  &  \\
183 & 3:32:45.96 & -27:57:44.4 & 3:32:45.96 & -27:57:45.4 & 1.10 & 72.92 & 1.00 & WFI-R & 17.05$\pm$0.01 & 16.04$\pm$0.02 & 0.103$^d$ & 2 & 0.12$\pm$0.03$^3$ \\
184 & 3:32:45.96 & -27:53:16.2 & 3:32:45.96 & -27:53:15.7 & 0.40 & 1.45 & 0.88 & WFI-R & 25.47$\pm$0.19 & $>$24.7 &  &  & 2.31$\pm$0.08$^2$ \\
185 & 3:32:46.33 & -27:53:27.3 & 3:32:46.33 & -27:53:27.0 & 0.30 & 20.03 & 0.99 & ACS-i & 25.13$\pm$0.13 & 21.35$\pm$0.03 &  &  & 1.99$\pm$0.08$^2$ \\
186 & 3:32:46.83 & -27:42:15.1 & 3:32:46.77 & -27:42:11.9 & 3.20 & 10.00 & 0.90 & WFI-R & 16.26$\pm$0.01 & 14.98$\pm$0.01 & 0.103$^a$ & 2 & 0.12$\pm$0.00$^1$ \\
187 & 3:32:46.86 & -27:51:21.2 & 3:32:46.84 & -27:51:21.2 & 0.20 & 38.90 & 0.99 & ACS-i & 24.20$\pm$0.04 & 21.93$\pm$0.06 & 2.292$^d$ & 1 & 1.33$\pm$0.08$^2$ \\
188 & 3:32:46.97 & -27:39:02.3 & 3:32:46.95 & -27:39:03.0 & 0.80 & 156.17 & 1.00 & WFI-R & 17.60$\pm$0.01 & 16.18$\pm$0.01 & 0.152$^b$ & 2 & 0.17$\pm$0.01$^1$ \\
189 & 3:32:47.35 & -27:42:47.5 & 3:32:47.29 & -27:42:48.3 & 1.10 & 50.10 & 1.00 & ACS-i & 21.92$\pm$0.01 & 19.38$\pm$0.04 & 0.730$^d$ & 2 &  \\
190 & 3:32:47.88 & -27:42:32.4 & 3:32:47.91 & -27:42:32.9 & 0.70 & 32.84 & 0.99 & WFI-R & 21.65$\pm$0.01 & 18.92$\pm$0.04 & 0.979$^a$ & 2 & 0.99$\pm$0.02$^1$ \\
191 & 3:32:48.30 & -27:56:26.3 & 3:32:48.30 & -27:56:26.9 & 0.70 & 140.33 & 1.00 & WFI-R & 19.20$\pm$0.01 & 17.35$\pm$0.01 & 0.346$^d$ & 2 & 0.34$\pm$0.01$^1$ \\
192 & 3:32:48.59 & -27:49:34.4 & 3:32:48.57 & -27:49:34.3 & 0.10 & 25.00 & 0.95 & ACS-i & 24.64$\pm$0.11 & 21.07$\pm$0.02 & 1.120$^c$ & 2 & 1.06$\pm$0.08$^2$ \\
193 & 3:32:49.20 & -27:40:49.8 & 3:32:49.22 & -27:40:50.4 & 0.70 & 21.91 & 0.99 & WFI-R & 22.61$\pm$0.02 & 19.63$\pm$0.05 & 0.546$^e$ & 1 &  \\
194 & 3:32:49.35 & -27:58:44.4 & 3:32:49.36 & -27:58:45.8 & 1.50 & 0.82 & 0.80 & WFI-R & 24.23$\pm$0.08 & $>$21.4 &  &  &  \\
195 & 3:32:49.45 & -27:42:35.1 & 3:32:49.43 & -27:42:35.1 & 0.20 & 23.09 & 0.99 & WFI-R & 23.86$\pm$0.05 & 20.55$\pm$0.21 & 0.981$^e$ & 2 &  \\
196 & 3:32:49.89 & -27:34:46.0 & 3:32:49.95 & -27:34:45.7 & 0.90 & 88.85 & 1.00 & WFI-R & 19.23$\pm$0.01 & $>$21.4 &  &  & 0.25$\pm$0.01$^1$ \\
197 & 3:32:49.96 & -27:34:32.4 & 3:32:49.97 & -27:34:32.6 & 0.40 & 268.07 & 1.00 & WFI-R & 18.40$\pm$0.01 & $>$21.4 & 0.251$^b$ & 2 & 0.24$\pm$0.01$^1$ \\
198 & 3:32:51.71 & -27:39:35.1 & 3:32:51.67 & -27:39:36.8 & 1.90 & 1.18 & 0.85 & WFI-R & 23.33$\pm$0.02 & 21.70$\pm$0.27 & 0.780$^b$ & 1 &  \\
199 & 3:32:51.78 & -27:59:56.5 & 3:32:51.80 & -27:59:56.2 & 0.40 & 27.86 & 0.99 & WFI-R & 22.50$\pm$0.02 & $>$21.4 &  &  & 0.55$\pm$0.05$^1$ \\
200 & 3:32:51.81 & -27:44:36.6 & 3:32:51.84 & -27:44:36.8 & 0.60 & 98.15 & 1.00 & WFI-R & 20.56$\pm$0.01 & 18.81$\pm$0.03 & 0.522$^a$ & 2 & 0.52$\pm$0.01$^1$ \\
201 & 3:32:52.08 & -27:44:25.2 & 3:32:52.08 & -27:44:25.1 & 0.10 & 146.99 & 1.00 & WFI-R & 20.78$\pm$0.01 & 18.49$\pm$0.03 & 0.534$^b$ & 2 & 0.46$\pm$0.02$^1$ \\
202 & 3:32:52.54 & -27:59:42.9  &  &  &  &  &  &  &  &  &  &  &   \\
203 & 3:32:52.91 & -27:38:38.2 & 3:32:52.89 & -27:38:38.0 & 0.20 & 31.42 & 0.99 & WFI-R & 23.57$\pm$0.04 & $>$21.4 & 1.226$^b$ & 1 & 1.19$\pm$0.16$^1$ \\
204 & 3:32:53.34 & -28:01:59.6 & 3:32:53.34 & -28:01:58.9 & 0.60 & 24.24 & 0.99 & WFI-R & 23.24$\pm$0.03 & $>$21.4 &  &  & 1.13$\pm$0.01$^1$ \\
205 & 3:32:54.70 & -27:34:21.3 & 3:32:54.76 & -27:34:21.4 & 0.90 & 124.67 & 1.00 & WFI-R & 18.35$\pm$0.01 & $>$21.4 &  &  & 0.15$\pm$0.01$^1$ \\
206 & 3:32:56.46 & -27:58:47.3 & 3:32:56.45 & -27:58:48.1 & 0.90 & 103.77 & 1.00 & WFI-R & 16.72$\pm$0.01 & $>$21.4 &  &  & 0.15$\pm$0.01$^1$ \\
207 & 3:32:57.08 & -28:02:09.7 & 3:32:57.16 & -28:02:09.9 & 1.20 & 30.49 & 0.99 & WFI-R & 20.37$\pm$0.01 & $>$21.4 &  &  & 0.64$\pm$0.01$^1$ \\
208 & 3:32:59.25 & -27:48:58.2 & 3:32:59.31 & -27:48:58.7 & 1.00 & 38.30 & 0.99 & SOFI-K & 23.37$\pm$0.03 & 20.31$\pm$0.09 &  &  & 1.28$\pm$0.98$^3$ \\
209 & 3:32:59.25 & -27:43:24.2 & 3:32:59.21 & -27:43:25.1 & 1.10 & 2.22 & 0.92 & WFI-R & 24.19$\pm$0.06 & 20.48$\pm$0.08 &  &  &  \\
210 & 3:32:59.30 & -27:35:34.1 & 3:32:59.33 & -27:35:34.1 & 0.50 & 33.83 & 0.99 & WFI-R & 22.63$\pm$0.02 & $>$21.4 & 1.128$^b$ & 2 & 1.14$\pm$0.01$^1$ \\
211 & 3:33:01.14 & -27:58:16.1 & 3:33:01.12 & -27:58:17.2 & 1.20 & 7.18 & 0.97 & WFI-R & 23.18$\pm$0.04 & $>$21.4 &  &  &  \\
212 & 3:33:02.01 & -27:49:54.0 & 3:33:02.08 & -27:49:54.3 & 1.00 & 27.30 & 0.99 & WFI-R & 21.17$\pm$0.01 & 18.89$\pm$0.03 & 0.581$^b$ & 2 & 0.57$\pm$0.02$^1$ \\
213 & 3:33:02.69 & -27:56:41.4 & 3:33:02.62 & -27:56:39.8 & 1.80 & 0.30 & 0.70 & GEMS-z & $>$25.5 & $>$21.4 &  &  &  \\
214 & 3:33:03.00 & -27:51:46.0 & 3:33:03.05 & -27:51:45.8 & 0.80 & 2.03 & 0.91 & ACS-i & 25.33$\pm$0.19 & $>$24.7 &  &  & 3.69$\pm$0.57$^3$ \\
215 & 3:33:03.22 & -27:53:06.1 & 3:33:03.18 & -27:53:06.6 & 0.80 & 14.70 & 0.74 & WFI-R & 23.14$\pm$0.02 & 19.52$\pm$0.06 &  &  & 0.73$\pm$0.10$^1$ \\
216 & 3:33:03.30 & -27:53:28.0 & 3:33:03.31 & -27:53:28.0 & 0.50 & 10.00 & 0.90 & SPITZER & $>$25.5 & $>$21.4 &  &  &  \\
217 & 3:33:03.73 & -27:36:10.2 & 3:33:03.74 & -27:36:11.0 & 0.90 & 58.46 & 1.00 & WFI-R & 20.17$\pm$0.01 & $>$21.4 & 0.623$^b$ & 2 &  \\
218 & 3:33:03.87 & -27:50:26.2 & 3:33:03.87 & -27:50:26.2 & 0.10 & 34.86 & 0.99 & WFI-R & 21.99$\pm$0.01 & 19.44$\pm$0.04 & 0.890$^e$ & 2 & 0.86$\pm$0.05$^1$ \\
219 & 3:33:04.50 & -27:38:01.3 & 3:33:04.51 & -27:38:02.5 & 1.30 & 2.33 & 0.92 & WFI-R & 23.58$\pm$0.04 & $>$21.4 & 1.084$^b$ & 1 & 0.99$\pm$0.06$^1$ \\
220 & 3:33:05.18 & -27:40:26.1 & 3:33:05.14 & -27:40:27.4 & 1.50 & 12.90 & 0.98 & WFI-R & 20.36$\pm$0.01 & 18.51$\pm$0.06 &  &  & 0.30$\pm$0.00$^1$ \\
221 & 3:33:05.54 & -27:54:15.0 & 3:33:05.53 & -27:54:16.4 & 1.50 & 19.94 & 0.99 & WFI-R & 20.62$\pm$0.01 & 18.84$\pm$0.03 & 0.325$^d$ & 2 & 0.31$\pm$0.04$^1$ \\
222 & 3:33:05.67 & -27:33:28.9 & 3:33:05.67 & -27:33:27.5 & 1.30 & 2.12 & 0.91 & WFI-R & 24.00$\pm$0.08 & $>$21.4 &  &  &  \\
223 & 3:33:06.21 & -27:48:42.0 & 3:33:06.16 & -27:48:42.0 & 0.80 & 0.30 & 0.70 & WFI-R & 26.51$\pm$0.30 & 24.45$\pm$0.10 &  &  &  \\
224 & 3:33:06.85 & -27:42:05.2 & 3:33:06.86 & -27:42:05.5 & 0.50 & 40.36 & 1.00 & WFI-R & 21.78$\pm$0.01 & 18.66$\pm$0.04 &  &  & 0.56$\pm$0.04$^1$ \\
225 & 3:33:07.32 & -27:44:32.1 & 3:33:07.33 & -27:44:32.6 & 0.70 & 228.79 & 1.00 & WFI-R & 17.66$\pm$0.01 & 16.29$\pm$0.01 &  &  & 0.21$\pm$0.02$^1$ \\
226 & 3:33:07.51 & -27:51:20.2 & 3:33:07.41 & -27:51:21.0 & 1.50 & 0.10 & 0.33 & WFI-R & 25.96$\pm$0.39 & $>$21.4 &  &  &  \\
227 & 3:33:07.75 & -27:53:51.0 & 3:33:07.81 & -27:53:50.9 & 1.00 & 6.20 & 0.80 & WFI-R & 23.74$\pm$0.05 & 21.20$\pm$0.40 &  &  &  \\
228 & 3:33:08.18 & -27:50:33.0 & 3:33:08.17 & -27:50:33.2 & 0.30 & 54.77 & 1.00 & WFI-R & 21.92$\pm$0.01 & 18.95$\pm$0.10 & 0.732$^d$ & 2 & 0.73$\pm$0.04$^1$ \\
229 & 3:33:09.75 & -27:48:01.1 & 3:33:09.74 & -27:48:01.0 & 0.00 & 216.10 & 1.00 & WFI-R & 19.10$\pm$0.01 & 17.57$\pm$0.08 & 0.180$^d$ & 2 & 0.17$\pm$0.02$^1$ \\
230 & 3:33:10.21 & -27:48:42.1 & 3:33:10.20 & -27:48:41.9 & 0.10 & 46.58 & 1.00 & WFI-R & 22.97$\pm$0.02 & $>$21.4 & 1.029$^d$ & 2 & 0.81$\pm$0.06$^3$ \\
231 & 3:33:10.30 & -27:33:06.0 & 3:33:10.31 & -27:33:06.2 & 0.40 & 198.11 & 1.00 & WFI-R & 19.06$\pm$0.01 & $>$21.4 &  &  &  \\
232 & 3:33:11.80 & -27:41:37.9 & 3:33:11.79 & -27:41:38.1 & 0.30 & 16.29 & 0.99 & WFI-R & 23.96$\pm$0.06 & $>$21.4 &  &  & 1.06$\pm$0.12$^1$ \\
233 & 3:33:11.90 & -27:53:46.8 & 3:33:11.88 & -27:53:46.8 & 0.20 & 163.42 & 1.00 & WFI-R & 20.33$\pm$0.01 & $>$21.4 & 0.532$^d$ & 2 & 0.54$\pm$0.02$^1$ \\
234 & 3:33:12.58 & -27:53:22.0 & 3:33:12.60 & -27:53:22.4 & 0.60 & 8.81 & 0.98 & WFI-R & 24.19$\pm$0.08 & $>$21.4 &  &  &  \\
235 & 3:33:13.15 & -27:49:30.9 & 3:33:13.11 & -27:49:30.4 & 0.50 & 38.86 & 0.99 & SOFI-K & $>$25.5 & $>$21.4 &  &  &  \\
236 & 3:33:14.50 & -27:47:47.9 & 3:33:14.52 & -27:47:47.6 & 0.40 & 12.92 & 0.98 & WFI-R & 24.02$\pm$0.05 & $>$21.4 &  &  &  \\
237 & 3:33:14.85 & -28:04:31.8 & 3:33:15.03 & -28:04:29.4 & 3.40 & 0.01 & 0.01 & WFI-R & 23.59$\pm$0.05 & $>$21.4 &  &  &  \\
238 & 3:33:14.98 & -27:51:50.9 & 3:33:14.99 & -27:51:51.1 & 0.30 & 37.59 & 0.99 & WFI-R & 23.19$\pm$0.03 & $>$21.4 & 1.998$^d$ & 1 & 1.10$\pm$0.06$^1$ \\
239 & 3:33:16.37 & -27:47:25.0 & 3:33:16.35 & -27:47:24.8 & 0.20 & 43.48 & 1.00 & WFI-R & 22.94$\pm$0.02 & $>$21.4 &  &  & 1.03$\pm$0.02$^1$ \\
240 & 3:33:16.49 & -27:50:39.8 & 3:33:16.52 & -27:50:39.5 & 0.50 & 232.28 & 1.00 & WFI-R & 17.44$\pm$0.01 & $>$21.4 & 0.086$^d$ & 2 & 0.09$\pm$0.04$^1$ \\
241 & 3:33:16.75 & -27:56:29.8 & 3:33:16.75 & -27:56:30.0 & 0.30 & 111.95 & 1.00 & WFI-R & 21.11$\pm$0.01 & $>$21.4 &  &  & 0.57$\pm$0.02$^1$ \\
242 & 3:33:16.78 & -28:00:15.9 & 3:33:16.77 & -28:00:16.4 & 0.60 & 1.21 & 0.86 & WFI-R & 25.53$\pm$0.18 & $>$21.4 &  &  &  \\
243 & 3:33:16.94 & -27:41:20.9 & 3:33:16.94 & -27:41:21.5 & 0.70 & 199.26 & 1.00 & WFI-R & 18.03$\pm$0.01 & $>$21.4 & 0.148$^b$ & 2 & 0.14$\pm$0.01$^1$ \\
244 & 3:33:17.38 & -27:49:48.8 & 3:33:17.44 & -27:49:47.1 & 1.80 & 0.01 & 0.05 & WFI-R & 25.47$\pm$0.13 & $>$21.4 &  &  &  \\
245 & 3:33:17.75 & -27:49:42.8 & 3:33:17.84 & -27:49:44.8 & 2.40 & 0.01 & 0.01 & WFI-R & 24.81$\pm$0.07 & $>$21.4 &  &  &  \\
246 & 3:33:17.76 & -27:59:06.8 & 3:33:17.77 & -27:59:06.2 & 0.60 & 29.44 & 0.99 & WFI-R & 22.74$\pm$0.02 & $>$21.4 &  &  & 1.13$\pm$0.03$^1$ \\
247 & 3:33:18.32 & -27:34:40.6 & 3:33:18.29 & -27:34:39.8 & 0.80 & 0.89 & 0.82 & WFI-R & 25.36$\pm$0.09 & $>$21.4 &  &  &  \\
248 & 3:33:18.73 & -27:49:40.7 & 3:33:18.71 & -27:49:39.8 & 0.80 & 7.63 & 0.97 & WFI-R & 23.98$\pm$0.05 & $>$21.4 &  &  &  \\
249 & 3:33:19.06 & -27:35:31.0 & 3:33:19.04 & -27:35:30.9 & 0.20 & 287.32 & 1.00 & WFI-R & 18.11$\pm$0.01 & $>$21.4 &  &  & 0.14$\pm$0.01$^1$ \\
250 & 3:33:20.61 & -27:49:10.2 & 3:33:20.60 & -27:49:10.1 & 0.00 & 129.55 & 1.00 & WFI-R & 16.26$\pm$0.01 & $>$21.4 & 0.126$^d$ & 2 & 0.14$\pm$0.00$^1$ \\
251 & 3:33:20.93 & -27:47:56.4 & 3:33:20.86 & -27:47:55.4 & 1.20 & 17.21 & 0.97 & WFI-R & 16.23$\pm$0.01 & $>$21.4 & 0.129$^b$ & 2 & 0.14$\pm$0.00$^1$ \\
252 & 3:33:21.31 & -27:41:37.8 & 3:33:21.32 & -27:41:38.4 & 0.80 & 15.61 & 0.99 & WFI-R & 23.40$\pm$0.06 & $>$21.4 &  &  & 1.15$\pm$0.15$^1$ \\
253 & 3:33:25.85 & -27:43:42.7 & 3:33:25.85 & -27:43:42.5 & 0.10 & 148.17 & 1.00 & WFI-R & 20.59$\pm$0.01 & $>$21.4 & 0.537$^b$ & 2 & 0.57$\pm$0.00$^1$ \\
254 & 3:33:26.54 & -27:44:44.7 & 3:33:26.51 & -27:44:44.8 & 0.30 & 134.54 & 1.00 & WFI-R & 20.20$\pm$0.01 & $>$21.4 & 0.448$^b$ & 2 & 0.44$\pm$0.01$^1$ \\
255 & 3:33:27.55 & -27:57:25.6 & 3:33:27.54 & -27:57:25.8 & 0.30 & 3.40 & 0.94 & WFI-R & 25.11$\pm$0.14 & $>$21.4 &  &  &  \\
256 & 3:33:30.73 & -27:44:02.6 & 3:33:30.69 & -27:44:02.9 & 0.60 & 24.91 & 0.99 & WFI-R & 22.71$\pm$0.02 & $>$21.4 & 0.864$^b$ & 2 & 0.77$\pm$0.02$^1$ \\
257 & 3:33:32.61 & -27:35:39.5 & 3:33:32.57 & -27:35:38.5 & 1.00 & 46.94 & 1.00 & WFI-R & 19.81$\pm$0.01 & $>$21.4 &  &  & 0.56$\pm$0.01$^1$ \\
258 & 3:33:33.46 & -27:53:32.5 & 3:33:33.42 & -27:53:32.6 & 0.40 & 87.05 & 1.00 & SOFI-K & $>$25.5 & $>$21.4 &  &  &  \\
259 & 3:33:34.55 & -27:47:51.4 & 3:33:34.56 & -27:47:51.0 & 0.40 & 32.55 & 0.99 & WFI-R & 22.91$\pm$0.02 & $>$21.4 & 0.860$^b$ & 2 & 0.63$\pm$0.03$^1$ \\
260 & 3:33:35.35 & -27:45:49.4 & 3:33:35.19 & -27:45:50.0 & 2.10 & 3.51 & 0.95 & WFI-R & 17.42$\pm$0.01 & $>$21.4 &  &  & 0.20$\pm$0.01$^1$ \\
261 & 3:33:36.31 & -27:44:32.3 & 3:33:36.34 & -27:44:31.9 & 0.50 & 6.70 & 0.97 & WFI-R & 24.42$\pm$0.07 & $>$21.4 &  &  &  \\
262 & 3:33:36.45 & -27:43:55.5 & 3:33:36.48 & -27:43:56.5 & 1.20 & 0.45 & 0.45 & WFI-R & 25.22$\pm$0.12 & $>$21.4 &  &  &  \\
263 & 3:33:36.85 & -27:36:40.4 & 3:33:36.88 & -27:36:40.5 & 0.60 & 6.65 & 0.97 & WFI-R & 24.30$\pm$0.09 & $>$21.4 & 3.678$^b$ & 1 &  \\
264 & 3:33:38.38 & -28:00:31.2 & 3:33:38.32 & -28:00:30.4 & 1.00 & 37.09 & 0.99 & WFI-R & 20.85$\pm$0.01 & $>$21.4 &  &  &  \\
265 & 3:33:41.30 & -27:38:08.4 & 3:33:41.29 & -27:38:08.4 & 0.10 & 389.68 & 1.00 & WFI-R & 17.41$\pm$0.01 & $>$21.4 & 0.102$^b$ & 2 &  \\
266 & 3:33:42.38 & -27:47:37.2 & 3:33:42.36 & -27:47:36.9 & 0.30 & 39.05 & 0.99 & WFI-R & 22.95$\pm$0.03 & $>$21.4 & 0.776$^b$ & 2 &  \\
\enddata
\tablenotetext{a}{Spectroscopic redshift from \cite{szokoly04}. The average redshift uncertainty is $\Delta$z$=$0.005.  }
\tablenotetext{b}{Spectroscopic redshift from Silvermann et al., in preparation.}
\tablenotetext{c}{Spectroscopic redshift from \citealt{vanzella05,vanzella06,vanzella08}. The average redshift uncertainty is $\Delta$z$=$0.00055. }
\tablenotetext{d}{Spectroscopic redshift from Popesso et al. (2008).}
\tablenotetext{e}{Spectroscopic redshift from \cite{lefevre04}. The average redshift uncertainty is $\Delta$z$=$0.0012.}
\tablenotetext{f}{Spectroscopic redshift from \cite{mignoli05}.}
\tablenotetext{g}{Spectroscopic redshift from \cite{ravikumar07}.}
\tablenotetext{1}{Photometric redshift from \cite{wolf04}.}
\tablenotetext{2}{Photometric redshift from \cite{grazian06}.}
\tablenotetext{3}{Photometric redshift from \cite{zheng04}.}
\tablenotetext{\star}{The optical photometry could be contaminated by a close-by bright star.}
\tablenotetext{\ast}{The optical photometry could be contaminated by a close-by ($\approx 1.3$ arcsec) source.}
\end{deluxetable}

\begin{deluxetable}{rrrrrcrrcccccc}
\tabletypesize{\scriptsize}
\rotate
\tablecaption{Optical, near infrared secondary counterparts of the 20 cm sources in the E-CDF-S \label{opt_catalogue_sec}}
\tablewidth{0pt}
\tablehead{
 & \multicolumn{2}{c}{RADIO} & \multicolumn{2}{c}{OPTICAL}  &  &  &  &  &  &  &  &  & \\
(1)  &  \multicolumn{1}{c}{(2)} &  \multicolumn{1}{c}{(3)} &  \multicolumn{1}{c}{(4)} &  \multicolumn{1}{c}{(5)} &  \multicolumn{1}{c}{(6)} &  \multicolumn{1}{c}{(7)} &  \multicolumn{1}{c}{(8)} &  \multicolumn{1}{c}{(9)} &  \multicolumn{1}{c}{(10)} &  \multicolumn{1}{c}{(11)} &  \multicolumn{1}{c}{(12)} & \multicolumn{1}{c}{(13)} & \multicolumn{1}{c}{(14)}  \\
RID  &  \multicolumn{1}{c}{RA} &  \multicolumn{1}{c}{Dec} &  \multicolumn{1}{c}{RA} &  \multicolumn{1}{c}{Dec} &  \multicolumn{1}{c}{dist} &  \multicolumn{1}{c}{LR} &  \multicolumn{1}{c}{Rel} &  \multicolumn{1}{c}{catalogue} &  \multicolumn{1}{c}{R} &  \multicolumn{1}{c}{K} &  \multicolumn{1}{c}{z} & \multicolumn{1}{c}{Qual-z} & \multicolumn{1}{c}{z$_{phot}$}    \\
 & \multicolumn{2}{c}{(J2000)} & \multicolumn{2}{c}{(J2000)}  & ($^{\prime\prime}$) &  &  &  & (AB) & (AB) &  &  & \\
}
\startdata
38 & 3:31:44.48 & -27:42:11.0 & 3:31:44.49 & -27:42:10.0 & 1.00 & 0.74 & 0.11 & WFI-R & 25.15$\pm$0.10 & $>$21.4 &  &  &  \\
71 & 3:32:00.84 & -27:35:56.4 & 3:32:00.91 & -27:35:55.4 & 1.40 & 1.00 & 0.03 & WFI-R & 24.23$\pm$0.04 & $>$21.4 &  &  & 0.88$\pm$0.06$^1$ \\
98 & 3:32:13.19 & -27:57:44.4 & 3:32:13.25 & -27:57:43.7 & 1.10 & 1.13 & 0.12 & WFI-R & 24.77$\pm$0.05 & $>$21.4 &  &  &  \\
101 & 3:32:13.52 & -27:49:52.5 & 3:32:13.46 & -27:49:51.9 & 0.80 & 13.81 & 0.18 & ACS-i & 22.78$\pm$0.02 & 19.80$\pm$0.01 & 0.731$^c$ & 2 & 0.69$\pm$0.08$^2$ \\
110 & 3:32:17.22 & -27:52:21.3 & 3:32:17.27 & -27:52:19.6 & 1.70 & 0.75 & 0.01 & ACS-i & 24.48$\pm$0.04 & 22.00$\pm$0.04 & 1.097$^a$ & 2 & 0.34$\pm$0.08$^2$ \\
113 & 3:32:19.17 & -27:54:07.7 & 3:32:19.11 & -27:54:07.2 & 0.60 & 44.02 & 0.99 & ACS-i & 24.08$\pm$0.03 & 20.40$\pm$0.01 & 0.964$^a$ & 2 & 0.98$\pm$0.08$^2$ \\
118 & 3:32:21.00 & -27:47:06.3 & 3:32:21.05 & -27:47:07.3 & 1.50 & 8.26 & 0.09 & ACS-i & 23.08$\pm$0.01 & 21.34$\pm$0.03 &  &  & 0.62$\pm$0.08$^2$ \\
120 & 3:32:21.29 & -27:44:35.6 & 3:32:21.32 & -27:44:36.4 & 1.10 & 71.01 & 0.14 & ACS-i & 20.53$\pm$0.01 & 18.22$\pm$0.01 & 0.524$^b$ & 2 & 0.49$\pm$0.08$^2$ \\
121 & 3:32:22.04 & -27:42:44.0 & 3:32:22.16 & -27:42:43.8 & 1.70 & 1.94 & 0.14 & ACS-i & 24.15$\pm$0.04 & 22.38$\pm$0.07 & 1.877$^d$ & 1 & 1.78$\pm$0.08$^2$ \\
121 & 3:32:22.04 & -27:42:44.0 & 3:32:22.05 & -27:42:41.9 & 2.00 & 0.70 & 0.05 & ACS-i & 24.98$\pm$0.07 & $>$24.7 &  &  & 0.66$\pm$0.08$^2$ \\
126 & 3:32:22.68 & -27:41:26.1 & 3:32:22.85 & -27:41:24.8 & 2.60 & 0.62 & 0.03 & ISAAC-K & 24.70$\pm$0.10 & 21.78$\pm$0.04 &  &  & 2.02$\pm$0.08$^2$ \\
136 & 3:32:27.97 & -27:46:39.4 & 3:32:28.00 & -27:46:37.7 & 1.60 & 2.21 & 0.00 & ACS-i & 22.72$\pm$0.01 & 22.08$\pm$0.01 &  &  & 0.04$\pm$0.08$^2$ \\
145 & 3:32:31.47 & -27:46:23.3 & 3:32:31.41 & -27:46:21.4 & 1.90 & 0.51 & 0.01 & ACS-i & 23.89$\pm$0.03 & 23.01$\pm$0.09 & 2.223$^a$ & 2 & 0.75$\pm$0.08$^2$ \\
160 & 3:32:37.29 & -27:51:27.4 & 3:32:37.17 & -27:51:27.9 & 1.50 & 17.19 & 0.19 & ACS-i & 22.48$\pm$0.01 & 19.58$\pm$0.01 &  &  & 1.01$\pm$0.08$^2$ \\
161 & 3:32:37.79 & -27:50:00.4 & 3:32:37.76 & -27:50:01.4 & 1.10 & 6.68 & 0.23 & ACS-i & 23.95$\pm$0.05 & 21.83$\pm$0.03 & 1.004$^c$ & 1 & 1.00$\pm$0.08$^2$ \\
162 & 3:32:37.80 & -27:52:12.4 & 3:32:37.92 & -27:52:11.8 & 1.90 & 0.37 & 0.01 & ACS-i & 24.18$\pm$0.05 & $>$24.7 & 1.603$^a$ & 2 & 0.69$\pm$0.08$^2$ \\
164 & 3:32:38.80 & -27:44:49.2 & 3:32:38.66 & -27:44:49.2 & 1.70 & 0.26 & 0.00 & ACS-i & 26.30$\pm$0.10 & $>$24.7 &  &  & 0.06$\pm$0.08$^2$ \\
169 & 3:32:39.47 & -27:53:01.3 & 3:32:39.47 & -27:53:00.5 & 0.70 & 197.29 & 0.46 & ACS-i & 21.25$\pm$0.01 & 19.15$\pm$0.01 &  &  & 0.68$\pm$0.08$^2$ \\
170 & 3:32:39.68 & -27:48:51.4 & 3:32:39.55 & -27:48:51.7 & 1.60 & 1.95 & 0.13 & ACS-i & 24.84$\pm$0.06 & 21.81$\pm$0.03 & 3.064$^a$ & 2 & 3.06$\pm$0.08$^2$ \\
177 & 3:32:43.16 & -27:55:14.3 & 3:32:43.26 & -27:55:15.1 & 1.80 & 0.28 & 0.01 & ACS-i & 23.83$\pm$0.05 & 21.44$\pm$0.31 & 0.579$^d$ & 2 &  \\
192 & 3:32:48.59 & -27:49:34.4 & 3:32:48.51 & -27:49:34.9 & 1.10 & 0.99 & 0.04 & ACS-i & 24.64$\pm$0.11 & 21.09$\pm$0.02 & 1.117$^e$ & 2 & 1.11$\pm$0.08$^2$ \\
215 & 3:33:03.22 & -27:53:06.1 & 3:33:03.29 & -27:53:06.7 & 1.20 & 5.06 & 0.25 & WFI-R & 23.65$\pm$0.02 & 20.06$\pm$0.06 &  &  & 0.82$\pm$0.06$^1$ \\
227 & 3:33:07.75 & -27:53:51.0 & 3:33:07.70 & -27:53:50.2 & 0.90 & 1.32 & 0.17 & WFI-R & 24.87$\pm$0.07 & $>$21.4 &  &  &  \\
251 & 3:33:20.93 & -27:47:56.4 & 3:33:21.20 & -27:47:53.1 & 4.90 & 0.29 & 0.02 & WFI-R & 21.66$\pm$0.01 & $>$21.4 &  &  &  \\
262 & 3:33:36.45 & -27:43:55.5 & 3:33:36.37 & -27:43:53.9 & 1.80 & 0.35 & 0.35 & WFI-R & 23.92$\pm$0.05 & $>$21.4 &  &  & 1.11$\pm$0.09$^1$ \\
\enddata
\tablenotetext{a}{Spectroscopic redshift from \cite{szokoly04}. The average redshift uncertainty is $\Delta$z$=$0.005.  }
\tablenotetext{b}{Spectroscopic redshift from Silvermann et al., in preparation.}
\tablenotetext{c}{Spectroscopic redshift from \citealt{vanzella05,vanzella06,vanzella08}. The average redshift uncertainty is $\Delta$z$=$0.00055. }
\tablenotetext{d}{Spectroscopic redshift from Popesso et al. (2008)}
\tablenotetext{e}{Spectroscopic redshift from \cite{lefevre04}. The average redshift uncertainty is $\Delta$z$=$0.0012.}
\tablenotetext{f}{Spectroscopic redshift from \cite{mignoli05}.}
\tablenotetext{1}{Photometric redshift from \cite{wolf04}.}
\tablenotetext{2}{Photometric redshift from \cite{grazian06}.}
\tablenotetext{3}{Photometric redshift from \cite{zheng04}.}
\end{deluxetable}

\clearpage
\pagestyle{plaintop}

\clearpage

\clearpage
\clearpage 
\begin{center}
\includegraphics[width=8cm]{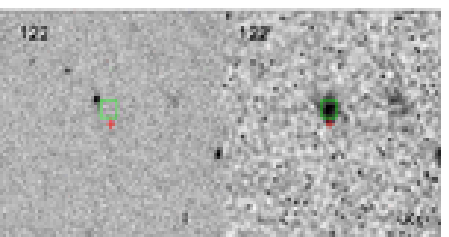}
\includegraphics[width=8cm]{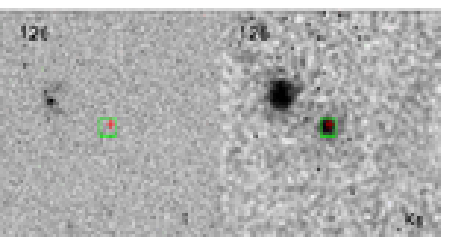}\\
\includegraphics[width=8cm]{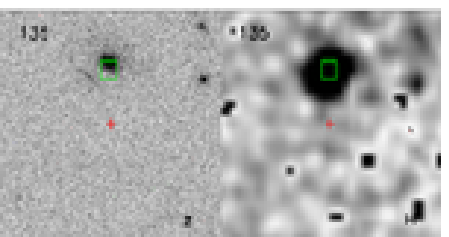}
\includegraphics[width=8cm]{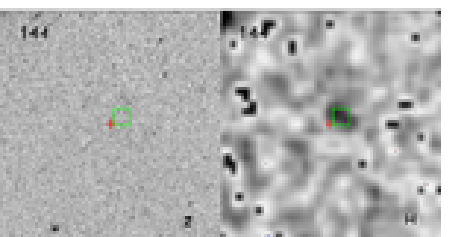}\\
\includegraphics[width=8cm]{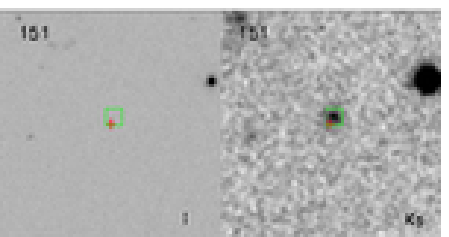}
\includegraphics[width=8cm]{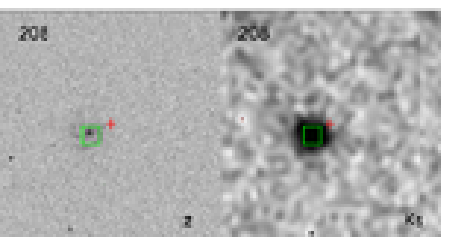}\\
\includegraphics[width=8cm]{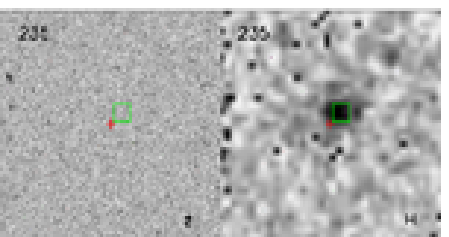}
\includegraphics[width=8cm]{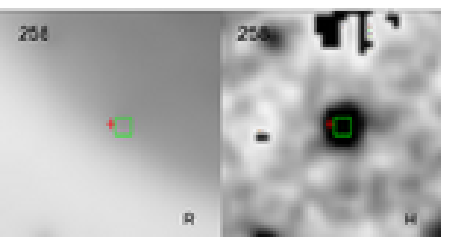}\\
{Fig. 6. --- Continued.}
\end{center}

\clearpage

\clearpage
\clearpage
\begin{center}
\includegraphics[width=5cm]{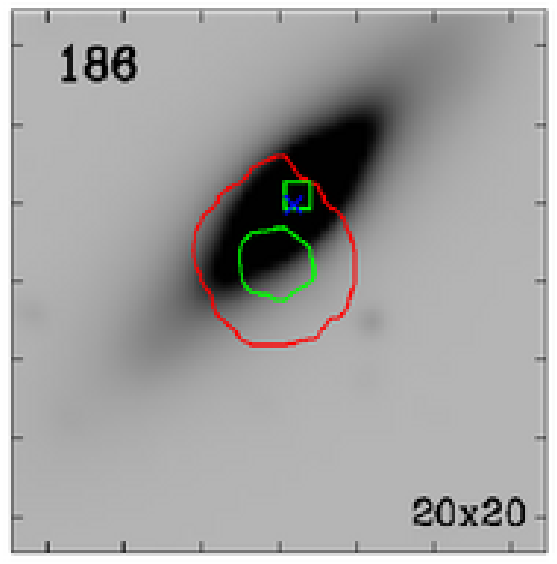}
\includegraphics[width=5cm]{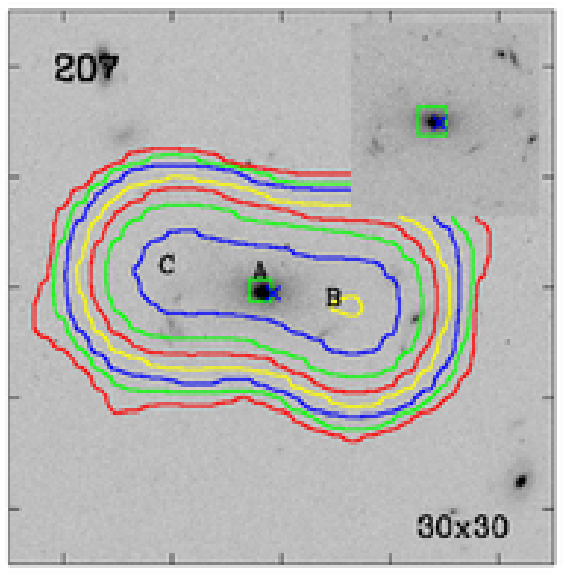}
\includegraphics[width=5cm]{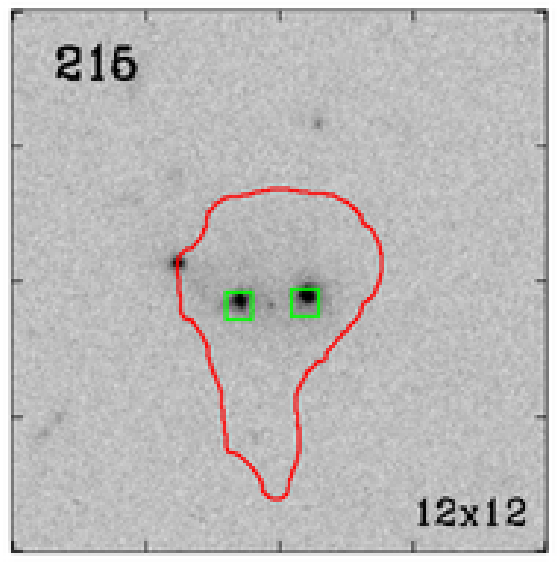}\\
\includegraphics[width=5cm]{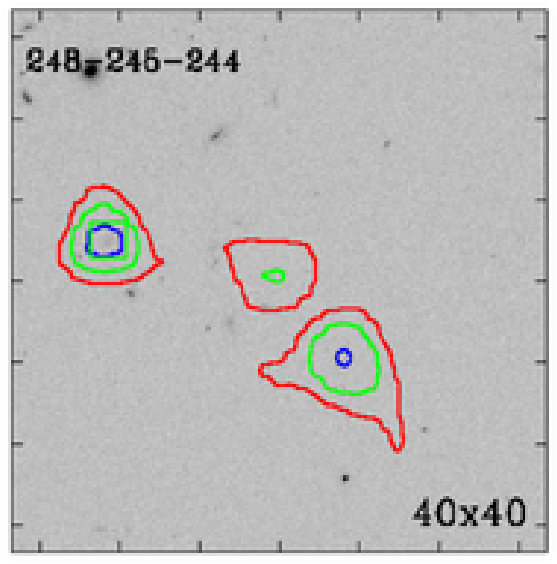}\\
{Fig. 9. --- Continued.}
\end{center}
\clearpage

%\clearpage
%\begin{figure}
%\centering
%\includegraphics[width=16cm]{figures/stack_spec_bright.ps}\\
%\includegraphics[width=16cm]{figures/stack_spec_faint.ps}\\
%\caption{Stacked spectrum of radio sources: S$>$0.15 mJy(top, 6 spectra) and S$<$0.15 mJy(bottom, 16 spectra).}
%\end{figure}

%\clearpage
%\begin{figure}
%\centering
%\includegraphics[width=16cm]{figures/stack_spec_faint_bright.ps}
%\caption{Stacked spectrum of radio sources: S$>$0.15 mJy(red) and S$<$0.15 mJy(blue).}
%\end{figure}

\end{document}